\documentclass[pdflatex,oneside,sn-nature]{sn-jnl}% Style for submissions to Nature Portfolio journals
\newcommand{\beginED}{%
        \setcounter{table}{0}
        \renewcommand{\tablename}{Extended Data Tab.}%
        \setcounter{equation}{0}
        \renewcommand{\theequation}{S\arabic{equation}}%
        \setcounter{figure}{0}
        \renewcommand{\figurename}{Extended Data Fig.}%
         }

\newcommand{\beginSI}{%
        \setcounter{table}{0}
        \renewcommand{\thetable}{S\arabic{table}}%
        \setcounter{equation}{0}
        \renewcommand{\theequation}{S\arabic{equation}}%
        \setcounter{figure}{0}
        \renewcommand{\thefigure}{S\arabic{figure}}%
         }
         
\usepackage{graphicx}%
\usepackage{multirow}%
\usepackage{amsmath,amssymb,amsfonts}%
\usepackage{amsthm}%
\usepackage{mathrsfs}%
\usepackage[title]{appendix}%
\usepackage{xcolor}%
\usepackage{textcomp}%
\usepackage{manyfoot}%
\usepackage{booktabs}%
\usepackage{algorithm}%
\usepackage{algorithmicx}%
\usepackage{algpseudocode}%
\usepackage{listings}%
\begin{document}

%\title[Article Title]{The Evolution of Occurrence and Architecture of Ultra-Short-Period Planet Systems}
\title[Article Title]{Planet Across Space and Time (PAST). VII. The origin and tidal evolution of hot Jupiters constrained by a broken age–frequency relation}
%{A non-monotonic age-frequency relation of hot Jupiters simultaneously constrains their origin and tidal evolution}


\author[1,2,3,4]{\fnm{Di-Chang} \sur{Chen}}\email{dcchen@nju.edu.cn}

\author*[1,2]{\fnm{Ji-Wei} \sur{Xie}}\email{jwxie@nju.edu.cn}

\author[1,2]{\fnm{Ji-Lin} \sur{Zhou}}\email{zhoujl@nju.edu.cn}

\author[5]{\fnm{Fei} \sur{Dai}}\email{fdai@hawaii.edu}

\author[3,4]{\fnm{Bo} \sur{Ma}}\email{mabo8@mail.sysu.edu.cn}

\author[6]{\fnm{Songhu} \sur{Wang}}\email{sw121@iu.edu}

\author[7]{\fnm{Chao} \sur{Liu}}\email{liuchao@nao.cas.cn}

\affil*[1]{\orgdiv{School of Astronomy and Space Science}, \orgname{Nanjing University}, \orgaddress{\postcode{210023}, \state{Nanjing}, \country{China}}}

\affil[2]{\orgdiv{Key Laboratory of Modern Astronomy and Astrophysics}, \orgname{Ministry of Education}, \orgaddress{\postcode{210023}, \state{Nanjing}, \country{China}}}

\affil[3]{\orgdiv{School of Physics and Astronomy}, \orgname{Sun Yat-sen University}, \orgaddress{\postcode{519082}, \state{Zhuhai}, \country{China}}}

\affil[4]{\orgdiv{Center of CSST in the great bay area}, \orgname{Sun Yat-sen University}, \orgaddress{\postcode{519082}, \state{Zhuhai}, \country{China}}}

\affil[5]{\orgdiv{Institute for Astronomy}, \orgname{University of Hawai'i}, \orgaddress{\postcode{HI 96822}, \state{Honolulu}, \country{USA}}}

\affil[6]{\orgdiv{Department of Astronomy}, \orgname{Indiana University}, \orgaddress{\postcode{IN 47405-7105}, \state{Bloomington}, \country{USA}}}

\affil[7]{\orgdiv{National Astronomical Observatories}, \orgname{Chinese Academy of Sciences}, \orgaddress{\postcode{100012}, \state{Beijing}, \country{China}}}


\abstract{The discovery of hot Jupiters has challenged the classical planet formation theory.
Although various formation mechanisms have been proposed, the dominant channel and relative contributions remain unclear. Furthermore, hot Jupiters offer a unique opportunity to test tidal theory and measure the fundamental tidal quality factor $Q^{'}_{*}$, which is yet to be well-constrained.
In this work, based on a hot Jupiter sample around single Sun-like stars with kinematic properties, {we find that the declining trend of their frequency is broken with a ridge at $\sim$ 2 Gyr, providing direct evidence that hot Jupiters are formed with multiple origins of different timescales.}
By fitting with the theoretical expectations, we provide a constraint of $\log Q^{'}_{*} \sim 5.7^{+0.4}_{-0.3}$ for Sun-like stars, which aligns well with the detected number of hot Jupiters with orbital decay.
Moreover, we simultaneously constrain the relative importance of different channels: although the majority of hot Jupiters are formed early, within several tenths of Gyr via `Early' models (e.g., in-situ formation, disk migration, planet-planet scattering and Kozai-Lidov interaction), a significant portion ($38^{+16}_{-14}\%$) should be formed late on a relatively long timescale extending up to several Gyr mainly via the secular chaos mechanism, further supported by the obliquity distribution of ‘late-arrived’ hot Jupiters.
Our findings provide a unified framework that reconciles hot Jupiter demographics and long-term evolution with multichannel formation.
}


\maketitle

The discovery and peculiarities of hot Jupiters present significant challenges to the classical planet formation theories based on our Solar system \citep{1995Natur.378..355M,2018ARA&A..56..175D}.
To explain the origin of hot Jupiters, previous studies have proposed various models, e.g., in-situ formation \citep{2016ApJ...829..114B,2016ApJ...817L..17B}, disk migration \citep{1996Natur.380..606L,2014prpl.conf..667B} and high-eccentricity migration 
\citep{1996Sci...274..954R,2003ApJ...589..605W,2011Natur.473..187N,2011ApJ...735..109W,2015ApJ...805...75P}. 
However, none of these models alone can satisfactorily explain all the observational evidence, and the origin of hot Jupiters remains puzzling (see the review by \citep{2018ARA&A..56..175D}). 
An outstanding question is: what is the predominant channel and how much do different models contribute to the formation of hot Jupiters?

Furthermore, as hot Jupiters are close to the host stars, their orbits are expected to decay with age due to tidal interaction \citep{2008ApJ...678.1396J,2012MNRAS.423..486L}.
Previous studies have looked for observational evidence for the tidal decays from individual \citep{2020ApJ...888L...5Y,2021AJ....161...72T,2024ApJS..270...14W} and ensemble properties of hot Jupiters \citep{2008ApJ...678.1396J,2017AA...602A.107B,2018MNRAS.476.2542C,2019AJ....158..190H,2020AJ....160..138H} as well as the properties (e.g., age, spin-up) of their host stars \citep{2019AJ....158..190H,2023PNAS..12004179C,2023AJ....166..209M,2024AJ....168....7B,2018AJ....155..165P}.
However, the derived stellar modified tidal quality factor $Q^{'}_{*}$ from different works vary by several orders of magnitude \citep{2008ApJ...678.1396J,2017AA...602A.107B,2018MNRAS.476.2542C,2019AJ....158..190H,2020ApJ...888L...5Y,2021AJ....161...72T,2022AJ....163..208B}, ranging from $\sim 10^4-10^9$ (Extended Data Tab. \ref{tab:Qparamodel}).
Therefore, obtaining a well-constrained stellar tidal quality factor is vital to resolve this discrepancy and to achieve a fundamental understanding of tidal phenomena.

The age-frequency relationship of hot Jupiters offers novel insights into their origin and tidal evolution. 
Specifically, different formation mechanisms operate on timescales ranging from millions to billions of years \citep{2018ARA&A..56..175D}, resulting in distinct arrival time distributions.
After hot Jupiters arrive at $\lesssim 0.1$ AU and have been largely circularized, the orbits of hot Jupiters are expected to decay under tidal forces, with the decay rate primarily determined by $Q^{'}_*$ \citep{2008ApJ...678.1396J,2012MNRAS.423..486L}.
Some of them will eventually pass into the Roche limit and be disrupted within the lifetime of their host stars \citep{2012MNRAS.423..486L,2023PNAS..12004179C}.
Consequently, hot Jupiters formed through different origin models, under the tidal forces of stars with varying $Q^{'}_{*}$, will exhibit distinct evolutionary patterns over time.

Based on the kinematic method for estimating age and a sample of hot Jupiters with kinematic properties, {in a previous study \citep{2023PNAS..12004179C}, we derived the age-frequency relation and revealed a generally declining trend using a mathematical exponential function. 
In this work, we perform detailed analyses of the age-frequency trend and focus on the physical interpretation of the observational results by quantitatively fitting with the theoretical expectations of the tidal evolution of hot Jupiters originating from different channels.}
This not only allows us to obtain a precise constraint on the stellar tidal quality factor but also enables the determination of the relative importance of different models in forming hot Jupiters across varying timescales.

\section*{Sample selection}
Our work builds upon the planetary sample selected and the stellar parent sample constructed in our previous study \citep[for details, see the \S~1 and 3 of Supporting information of][]{2023PNAS..12004179C}.
The planetary sample was collected from the kinematic catalogs of the Planets Across Space and Time series \citep[PAST;][]{2021ApJ...909..115C,2021AJ....162..100C} and composed of giant planets orbiting single Sun-like stars in the Galactic disks, which was further divided into three subsamples based on the discovery method and facility, i.e., radial velocity (RV), space-based transit (ST), and ground-based transit (GT).
In this work, we further expand the GT subsample by utilizing the Gaia DR2 catalog.
The final sample consists of 123 hot Jupiters (17 RV, 14 ST and 92 GT) and 936,815 (1662 RV, 19,228 ST and 915,925 GT) parent stars (Methods, \S~1.1).

\section*{Analyses and Results}
\subsection*{A broken Age-frequency relation of hot Jupiters}
\begin{figure}[!h]
\centering
\includegraphics[width=0.8\textwidth]{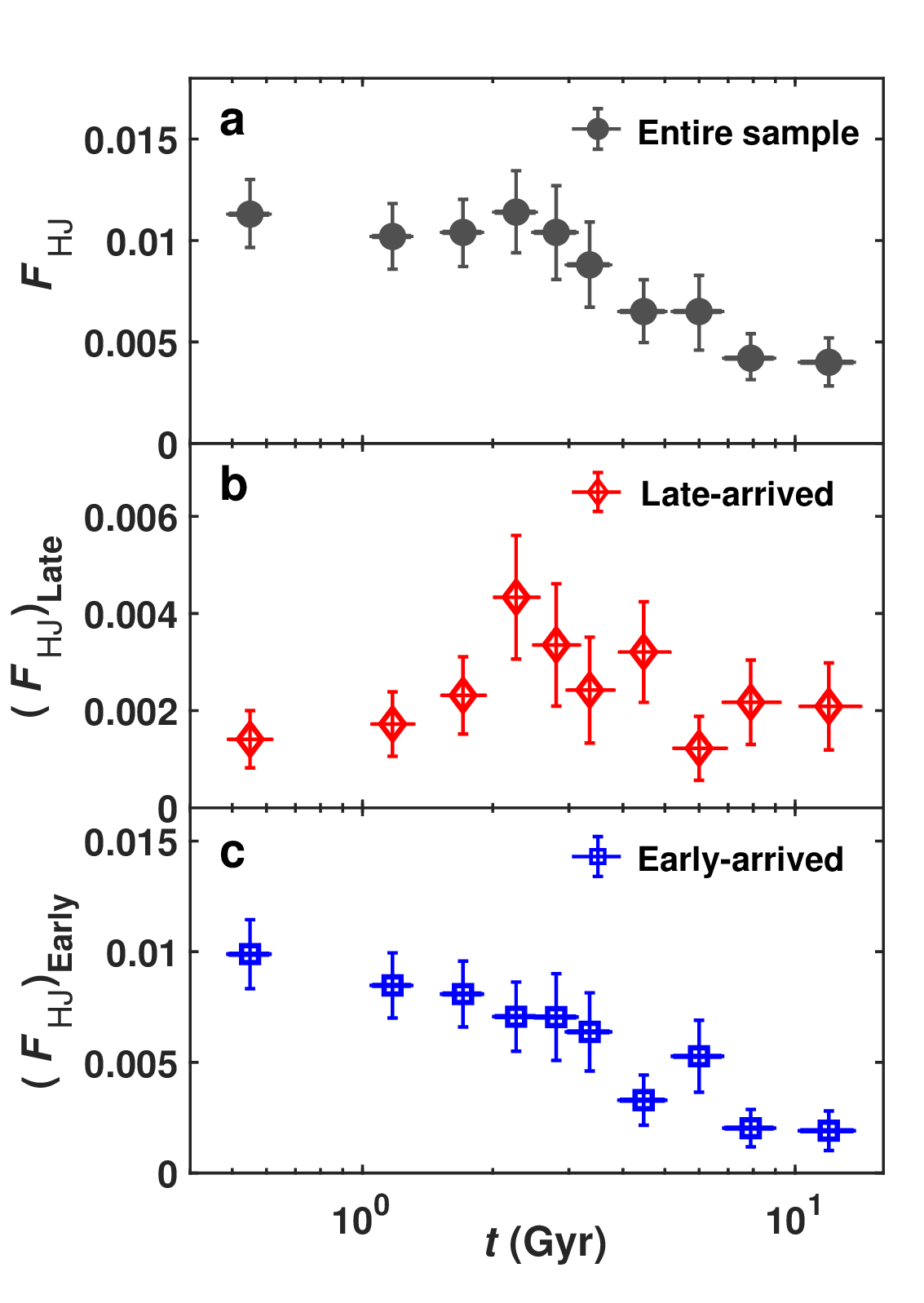}
\caption{\textbf{The observed age-frequency relation of hot Jupiters.}
Frequency of hot Jupiters as a function of kinematic age for the entire hot Jupiter sample (Top), and separately for the two populations: `late-arrived' (Middle) and `early-arrived' (Bottom).
The vertical and horizontal errorbars represent the uncertainties of Frequencies of hot Jupiters and kinematic ages, respectively.
}
\label{figfHJAgecombine}
\end{figure}

We derive a `combined' frequency of hot Jupiters $F_{\rm HJ}$ (i.e., also termed as 'occurrence', defined as the fraction of stars with a hot Jupiter) as a function of kinematic age by jointly using the data from the three subsamples discovered by different methods/facilities (Methods, \S~1.2).
The detection efficiencies and geometric effects are corrected with the same procedures described in \S~3 of Supporting information of \citep{2023PNAS..12004179C}.
As can be seen in the top panel of Figure \ref{figfHJAgecombine}, $F_{\rm HJ}$ generally declines with increasing age (Methods, \S~1.3.1), which is consistent with previous studies \citep{2023PNAS..12004179C,2023AJ....166..209M}.

Beyond the generally declining trend, there are some more intriguing characteristics: a ridge occurs at $\sim 2$ Gyr, and the declining trend of $F_{\rm HJ}$ is mild in the early stage, but becomes steeper in the late stage.
{To evaluate the existence of the ridge, we adopt the findpeaks function from  from the scipy.signal module and returns a ridge at $1.94^{+0.72}_{-0.50}$ Gyr.
We also resample the observation data (e.g., kinematic ages, observed number of planets and stars, detection efficiencies, geometric effects) for 10,000 times from their uncertainties and perform the sequences and reversals test to the resampled data.
In 9,608 out of 10,000 instances, a ridge was observed, corresponding to a confidence level of 96.08\%.
Moreover, on the two sides of the ridge, the decline trend in the late stage is steeper with a smaller exponential index (exponential index $\gamma = -0.18^{+0.04}_{-0.04}$) compared to the early stage ($\gamma = -0.07^{+0.10}_{-0.03}$) with a confidence level of 94.10\% (see (Methods, \S~1.3.2).}
It is worth noting that the kinematic age represents
the average age of a group of stars and the uncertainty denotes the uncertainty of the average age rather than the width of true age distribution. 
To further verify the above result, we also consider a Kepler sample of single Sun-like stars with reliable individual age estimations from gyrochronology and isochrone fitting and {find a broken declining trend of $F_{\rm HJ}$ with a ridge at $\sim 2$ Gyr (Figure S11), which is generally consistent with the kinematic results.}
{We also investigate the impact of age uncertainty on the frequency–age relation and demonstrate that it is insufficient to produce a broken relation as prominent as the observation (Figure S17-S18).}
These features are not anticipated from the tidal decay of hot Jupiters' orbits if they all formed at a very early stage after the stars have formed, and thus suggest that some hot Jupiters may have formed later.

\subsection*{Two hot Jupiter populations with different arrival timescales}
To demonstrate the above inference, we further categorize the hot Jupiter sample into two distinct populations.
Specifically, 54 hot Jupiters with tidal evolution timescales smaller than the ages of host stars at 2-$\sigma$ confidence are selected as `late-arrived' because if they had reached their current orbits soon after stars formed, they would have been tidally disrupted/circularized.
{The $Q^{'}_*$ value used here comes from the best-fits obtained by comparing the observational age-frequency relation with theoretical predictions (as discussed later).}
The other 69 hot Jupiters are selected as `early-arrived'.
We then derive the frequencies of the two populations as a function of kinematic age (Methods, \S~1.3.3).
As shown in the middle panel of Figure \ref{figfHJAgecombine}, {the frequency of the `late-arrived' population first increases ($\gamma=0.44^{+0.46}_{-0.21}$, confidence level of 96.37\%) and then decreases ($\gamma=-0.11^{+0.04}_{-0.05}$, confidence level of 97.83\%), naturally forming a ridge at $\sim 2$ Gyr with a confidence level of 99.87\%. 
%and inducing a significant non-monoticity (with a $p-$value of $4.8 \times 10^{-7}$).
After removing the contribution of the `late-arrived' population, the ridge at $\sim 2$ Gyr disappears and the declining trends of the remaining `early-arrived' population in the early and late stages become nearly identical (middle panel of Figure \ref{figfHJAgecombine}).}
{The above results are based on the combined analysis of the three subsamples. 
For further validation, we also performed the analyses of the temporal evolution using the three subsamples separately, which show similar results though the uncertainties are larger due to the reduced sample sizes (Figure S5, S7, and S9).}
The above results demonstrate that the local ridge and the broken pattern of $F_{\rm HJ}$ are caused by the `late-arrived' population, providing direct evidence that the hot Jupiters do not have a single origin.
As shown below, we interpret them as the tidal evolution of hot Jupiters that formed through different mechanisms on different timescales.
%which simultaneously constrains the relative importance of various hot Jupiter formation mechanisms as well as the stellar tidal quality factor.

\subsection*{Constraints on the formation and tidal evolution}
Various mechanisms have been proposed to explain the origin of hot Jupiters \citep{2018ARA&A..56..175D}.
Considering formation timescales as well as age uncertainties ($\gtrsim$ tenth of Gyr), in this work, these mechanisms are divided into two categories: `Early' model and `Late' model (Methods, \S~2.1).
For the `Early' model, most hot Jupiters form within about several tenths of Gyr via in-situ formation \citep{2016ApJ...829..114B,2016ApJ...817L..17B}, disk migration \citep{1996Natur.380..606L,2014prpl.conf..667B}, planet-planet scattering\citep{1996Sci...274..954R}/Kozai \citep{2011Natur.473..187N}/coplanar secular \citep{2015ApJ...805...75P}.
Whereas, for `Late' model, hot Jupiters are delivered to $\lesssim 0.1$ AU on a long timescale up to several Gyrs. 
Although the planet-star Kozai mechanism can potentially deliver hot Jupiters over Gyr \citep{2003ApJ...589..605W,2007ApJ...669.1298F}, 
we do not consider it as the main channel for the following reasons: (1) the fraction of binary companions capable of inducing Kozai–Lidov oscillations is $\lesssim 16\% \pm 5\%$, and the upper limit will further decrease to $6\% \pm 2\%$ assuming an isotropic distribution for the stellar companion inclinations \citep{2016ApJ...827....8N}; 
(2) the typical timescales for planet-star Kozai cycles are predominantly (99.71\%) less than 1 Gyr (Extended Data Fig. 1); 
(3) potential binaries have been largely excluded in our sample. 
Instead, secular chaos can be a promising candidate for the `Late' model, serving as a robust mechanism consistently capable of producing hot Jupiters around single stars over Gyr timescales.
Previous numerical studies have predicted the formation rate of hot Jupiters through secular chaos as a function of stellar age \citep{2017MNRAS.464..688H}. 
{We also compared the arrival timescales of the secular chaos model with the estimated arrival timescales of the observed`late-arrived' population, which are statistically consistent (with KS $p-$value of 0.3499) after considering the impact of selection criteria (Extended Data Fig. 2).}
In the following, we test the secular chaos mechanism as the `Late' model.

\begin{figure}[!t]
\centering
\includegraphics[width=\textwidth]{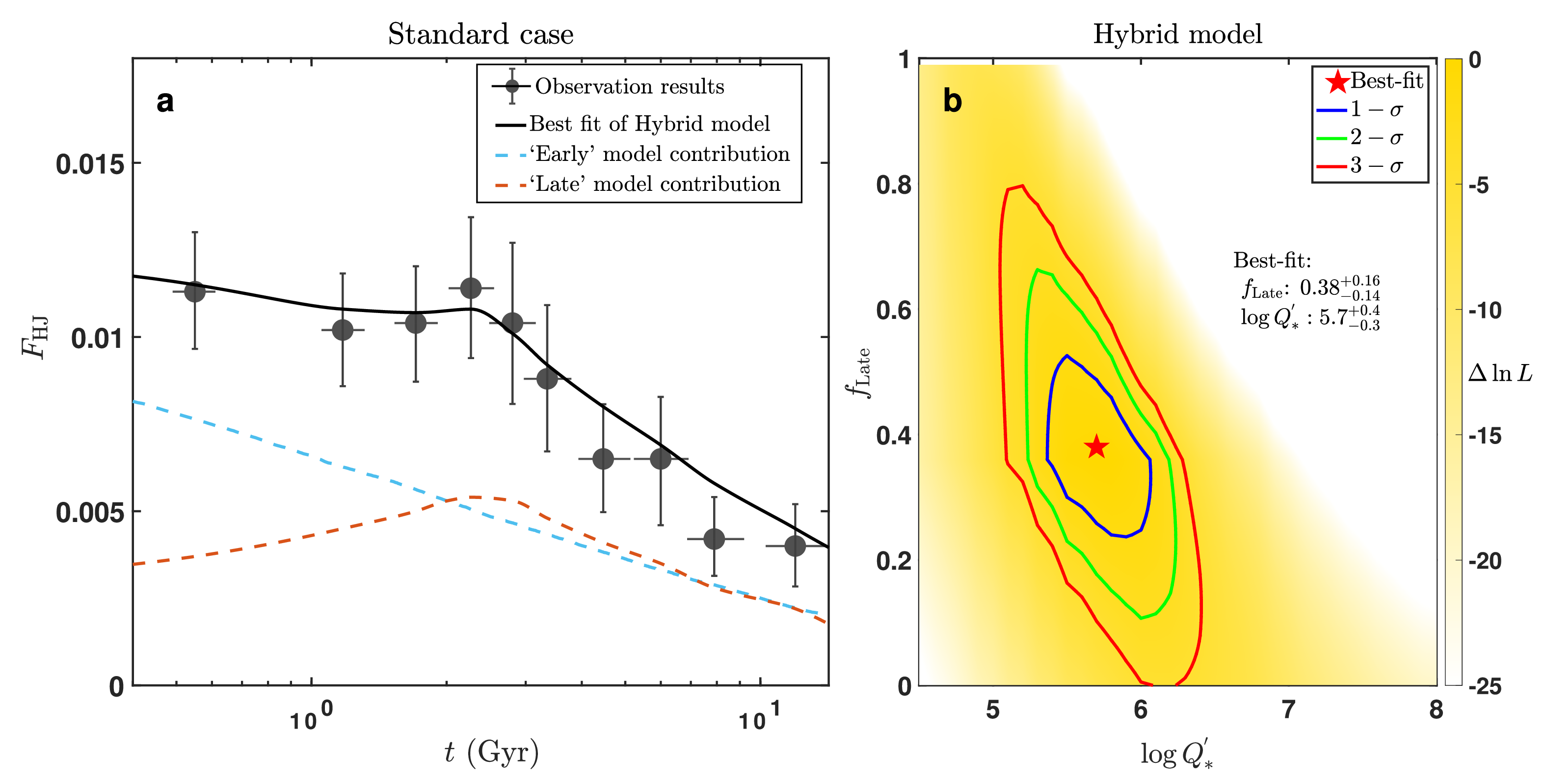}
\caption{\textbf{Fitting the observed age-frequency relation of hot Jupiters with a hybrid model.}
Left panel: The frequency of hot Jupiters $F_{\rm HJ}$ as a function of age. 
The observation data is plotted as solid black points and line segments denote the 1-$\sigma$ interval.
The solid line denotes best-fit of the hybrid model with 38\% hot Jupiter contributed from the `Late' model (red dashed) and 62\% from the `Early' models (blue dashed).
Right panel: Relative likelihood in logarithm as a function of $Q^{'}_{*}$ and $f_{\rm Late}$. 
The blue, green, and red lines indicate the boundary of the 1-$\sigma$, 2-$\sigma$, and 3-$\sigma$ confidence levels.
\label{figFHJmodelStandard}}
\end{figure}

Under the tidal forces of host stars, the temporal evolution of the frequency of hot Jupiters $F_{\rm HJ}$ from different formation models varies significantly (Methods, \S~2.2-2.3).
Specifically, $F_{\rm HJ}$ will decrease for the `Early' model (cyan line of Extended Data Fig. 3).
Whereas, for the `Late' model, $F_{\rm HJ}$ will follow an increasing-decreasing curve (orange line of Extended Data Fig. 3) and
the turning point is around $\sim 2$ Gyr for a typical stellar tidal quality factor ($\sim 10^5-10^7$).
The expectations of the `Early' model and `Late' model coincide well with the observational features of the `early-arrived' and `late-arrived' hot Jupiter populations, respectively (middle and bottom panels of Figure \ref{figfHJAgecombine}).
We are therefore motivated to fit the observed age-frequency relation of hot Jupiters with a hybrid physical model by combining the `Early' and `Late' models (Methods, \S~2.3).

In the hybrid model, there are two free parameters: the tidal quality factor of host stars ($Q^{'}_{*}$) and the fraction of hot Jupiters formed via `Late' model ($f_{\rm Late}$).
Figure \ref{figFHJmodelStandard}
shows the fitting result (the solid black curve) for the hybrid model in the standard case, which matches well with the observational data and provides constraints on the two parameters: $\log Q^{'}_{*} = 5.7^{+0.4}_{-0.3}$ and $f_{\rm Late} = 0.38^{+0.16}_{-0.14}$.
In Figure \ref{figFHJmodelStandard}, we also show the relative contributions of `Early' model (dashed cyan curve) and `Late' model (dashed orange curve).
As can be seen, `Early' model contributes the majority ($62^{+14}_{-16}\%$) of hot Jupiters, leading to the global declining trend.
Whereas, `Late' model plays a complementary role ($38^{+16}_{-14}\%$), forming the local ridge substructure at $\sim 2$ Gyr.
Moreover, the increasing-decreasing curve in $F_{\rm HJ}$ induced by the `Late' model contribution counteracts the declining trend induced by the `Early' model contribution at the early stage but strengthens it at the late stage, naturally explaining the observational broken trend (Methods, \S~3.1).

%The above results are robust.
We also fit the observations using either the `Early' model or the `Late' model alone (Extended Data Fig. \ref{figFHJmodel1a2016MCJ}, \ref{figFHJmodel22016MCJ} and \ref{figFHJmodel12a2016MCJ}) and find that the hybrid model is significantly favored over both the single-`Early' model and the single-`Late' model, with $\rm \Delta AIC$ values of 9.6 and 35.9 respectively.
This corresponds to confidence levels of 98.55\% ($>$2-$ \sigma$) and 99.99\% ($>$3-$ \sigma$), when accounting for uncertainties in the observed frequency and stellar age (Extended Data Tab. 2).
We also investigated various conditions (e.g., initial distributions of $a$ and planetary mass) and found that all the results are consistent within 1-$\sigma$ errorbars (Figure S1-S4, Extended Data Tab. \ref{tab:FQAICdifferentmodels}).
Furthermore, we fit the observational data using the three subsamples separately and all the results show that the hybrid model is preferred (Figure S6, S8, and S10).
The derived constraints on $f_{\rm Late}$ and $Q^{'}_{*}$ from the three subsamples are all consistent with those of the whole sample within 1-$\sigma$ uncertainties (Extended Data Tab. \ref{tab:FQAICdifferentmodels}).

{In addition, previous studies have suggested that the stellar $Q^{'}_{*}$ may depend on the stellar mass and the evolutionary stage \citep{2017ApJ...849L..11W,2018AJ....155..165P,2020MNRAS.498.2270B} as well as the orbital period $P$ of hot Jupiters \citep{2018AJ....155..165P}.
Since we only study single Sun-like stars, these stellar properties are similar across different age bins (Figure S12). 
To account for this potential $Q^{'}_*-P$ dependence, we also fit the observational data using an empirical formula and find that the hybrid model remains preferred compared to the single-`Early' model and single-`Late' model, with confidence levels $>4$-$\sigma$ (Figure S13-15).
Besides, we also adopted the secular chaos model from \citep{2019MNRAS.486.2265T} as the `late' model, and the derived best-fits of $f_{\rm Late}$ and $Q^{'}_{*}$ are highly comparable (Figure S16), suggesting that our results are not sensitive to the choice of secular chaos model.
}

\begin{figure}[!t]
\centering
\includegraphics[width=\linewidth]{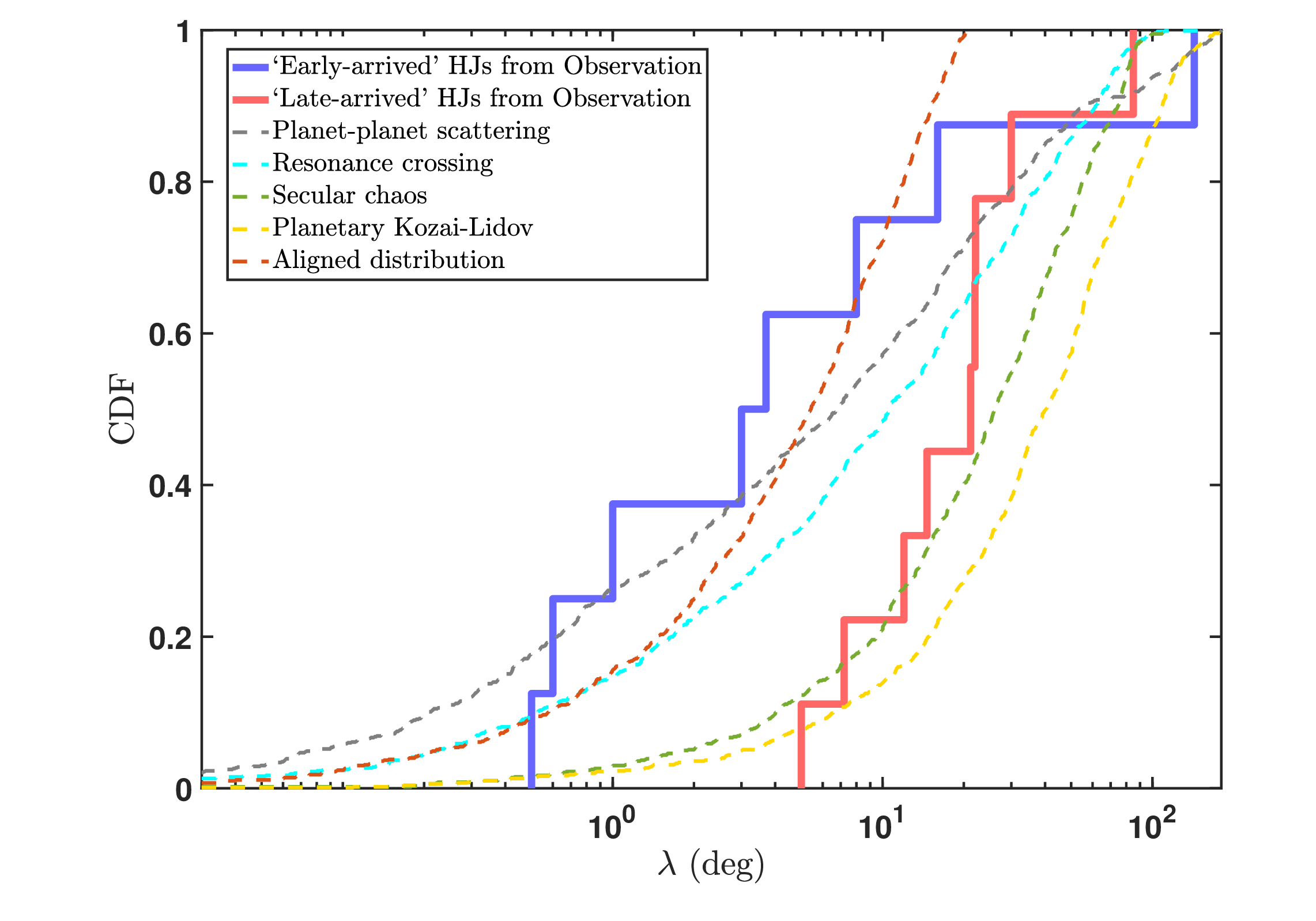}
\caption{\textbf{Additional supporting evidence for the secular chaos as the `Late' model.}
The cumulative distributions of the sky-projected stellar obliquities $\lambda$ for the `early-arrived’ (blue) and `late-arrived’ (red) hot Jupiter systems with long-alignment timescales and young kinematic ages. 
The dashed lines represent the predicted $\lambda$ distributions inferred from the stellar obliquity distributions predicted by different mechanisms and the aligned distribution randomly in $0-20 ^\circ$.}
\label{figObliquityFeHconstraints}
\end{figure}

\subsection*{Additional evidence supporting secular chaos}

The agreement between the observed age-frequency relation and the hybrid model implies that secular chaos \citep{2011ApJ...735..109W,2017MNRAS.464..688H,2019MNRAS.486.2265T} is the dominant mechanism to form the late arrived hot Jupiter.
We further explore other properties of hot Jupiter systems and find some additional evidence supporting this inference.
%We first investigate the stellar obliquities of the `early-arrived' and `late-arrived' hot Jupiter systems (See \S~3.2 of SI).
As shown in Figure \ref{figObliquityFeHconstraints}, we find that the `late-arrived' hot Jupiter systems have a stellar obliquity distribution mainly within $10-90^\circ$, which is consistent with the prediction of the secular chaos mechanism \citep{2019MNRAS.486.2265T} with a KS $p-$value of 0.229, but it deviates from the predictions of all other mechanisms with KS $p-$values$<0.05$ (Methods, \S~3.2, Extended Data Tab. \ref{tab:pvalue_obliquity}). 
%For the `early-arrived' systems are mainly aligned ($\lambda \lesssim 10^\circ$) but with a small fraction in misaligned systems, which matches well with the predictions of the planet-planet scattering, the resonance cross, and the alignment distribution, but is significantly smaller than the prediction of planetary Kozai-Lidov mechanism with a KS $p-$value $<0.003$ (Table \ref{tab:pvalue_obliquity}).
{In addition, since secular chaos mechanism occurs in systems with multiple giant planets, one would expect that the `late-arrived' hot Jupiters are more likely to have outer companions and are preferentially hosted by metal-rich stars. 
For consistency check, we investigate their metallicity distribution and companion presence. 
As expected, }the `late-arrived' hot Jupiter systems are metal-richer ($\rm \overline{[Fe/H]} = 0.12^{+0.02}_{-0.02}$ dex) (Methods, \S~3.3, Extended Data Fig. 7) and show a larger fraction ($F_{\rm OG} = 40^{+26.7}_{-23.3}\%$; Methods, \S~3.4) to have outer giant companions compared to the `early-arrived' hot Jupiters ($\rm \overline{[Fe/H]} = 0.06^{+0.02}_{-0.03}$, $F_{\rm OG} = 8.3^{+7.6}_{-7.0}\%$).

\subsection*{Further validation on $Q^{'}_{*}$}

\begin{figure}[!t]
\centering
\includegraphics[width=0.75\textwidth]{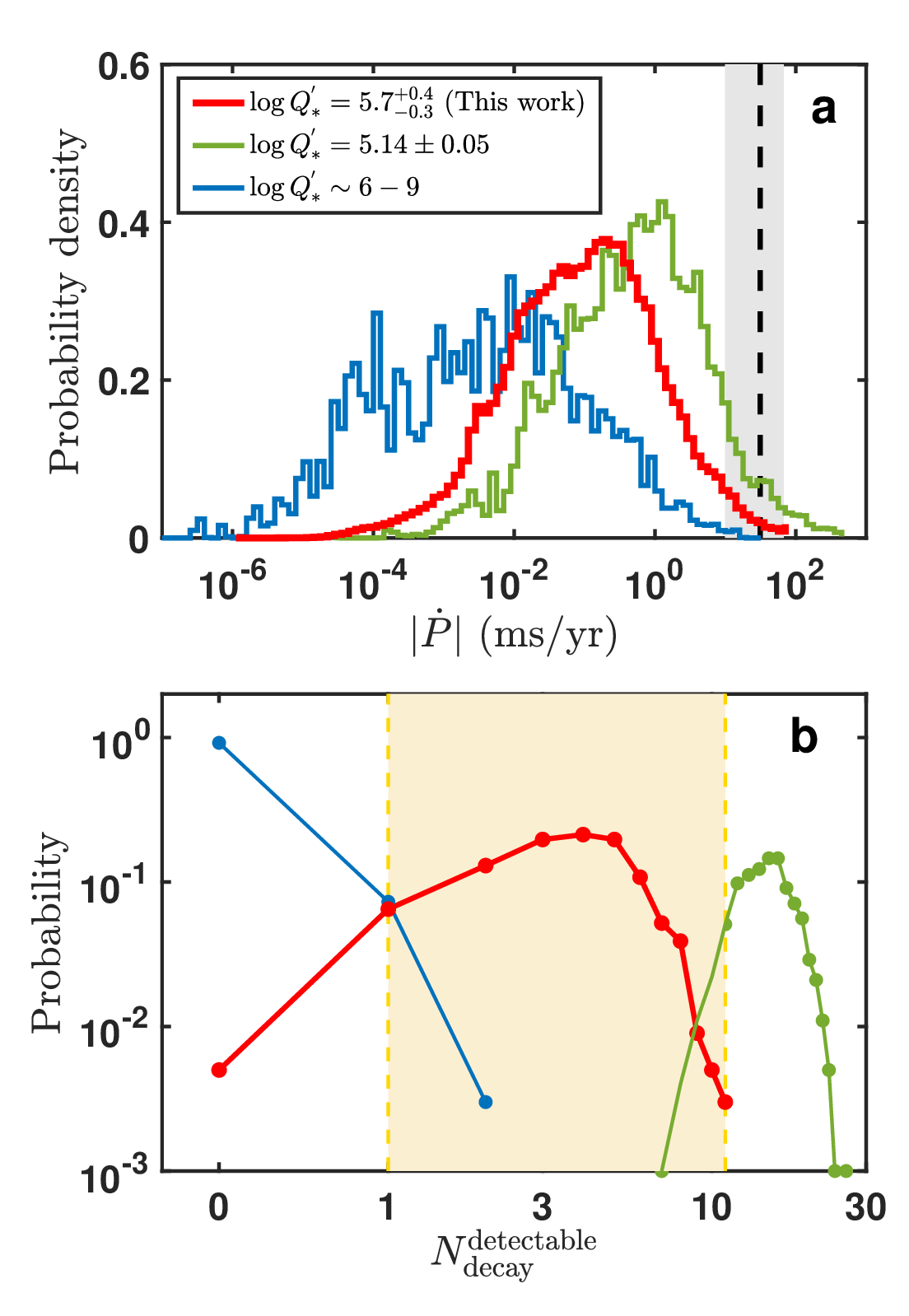}
\caption{\textbf{Evaluation of the constraints on $\log Q^{'}_{*}$ from our work and previous studies.}
The probability density distributions for the absolute decay rates $|\dot{P}|$ of orbital period (Top) and {the probability of} the expected numbers of hot Jupiters with detectable orbital decay (Bottom).
Solid lines in different colors represent results for different stellar tidal quality factors: this work (red), WASP 12 b (green) and previous ensemble studies (blue).
In the top panel, the vertical dashed line and grey region denote the median value and 1-$\sigma$ interval of the typical detection threshold of $|\dot{P}|$ for orbital period decay.
In the bottom panel, the yellow lines (region) indicate the lower and upper limits (range) of the observed numbers with orbital decay.}
\label{figNumberofdecayQ}
\end{figure}

To further validate the constraints on $\log Q^{'}_{*}$ derived in our work, we calculate the absolute decay rates $|\dot{P}|$ of orbital period and the expected numbers of hot Jupiters with detectable orbital period decay, $N^{\rm detectable}_{\rm decay}$ (see Figure \ref{figNumberofdecayQ}), and then compare these with the observation number, i.e., 1 confirmed case +10 candidates identified in Wang et al. (2024) (\citep{2024ApJS..270...14W}, Methods, \S~3.5). 
As can be seen, the expected value $N^{\rm detectable}_{\rm decay} = 4 \pm 2$ (red lines), which lies within the observed range of $1-11$ with a confidence level of 99.5\%, aligning well with the observational results. 
Previous studies have also provided constraints on $Q^{'}_{*}$ (Extended Data Tab. \ref{tab:Qparamodel}) based on the evolution of hot Jupiters in the individual systems \citep{2020ApJ...888L...5Y,2024ApJS..270...14W} and the ensemble statistics \citep{2008ApJ...678.1396J,2017AA...602A.107B,2018MNRAS.476.2542C,2019AJ....158..190H,2020AJ....160..138H}.
For comparison, we tested $Q^{'}_{*}$ derived from previous studies by repeating the above procedure and found that the resulted $N^{\rm detectable}_{\rm decay}$ do not match the observations.
Specifically, when taking the constraints of $\log Q^{'}_{*} = 5.14 \pm 0.05$ derived from individual system WASP 12 (the only one with conclusive evidence of orbital decay; green lines), we found that $N^{\rm detectable}_{\rm decay} = 15 \pm 3$, which is significantly ($\gtrsim 4$-$\sigma$) higher than the number of confirmed system and remains overestimated even if all 10 candidates are eventually confirmed.
In contrast, using the constraints of $\log Q^{'}_{*} \sim 6-9$ derived from previous ensemble studies (blue lines), the orbital decay of hot Jupiters is expected to be unobservable yet, with a confidence level of 92.5\%.

\subsection*{Discussions and conclusions}
One of the longest-standing puzzles in exoplanet research is how hot Jupiters form.
Previous studies have attempted to constrain the origin of hot Jupiters from the observed distribution of the semi-major axis/eccentricity/age/configurations for hot Jupiters \citep[see the review by][]{2018ARA&A..56..175D}.
For example, Wu et al. (2023) \citep{2023AJ....165..171W} reporetd that $\gtrsim 12 \pm 6\%$ of hot Jupiters have nearby planetary companions and thus form via quiescent dynamical mechanisms (i.e., in-situ formation and disk migration).
Hamer et al. (2022) \citep{2022AJ....164...26H}
found that high-stellar-obliquity hot Jupiter systems are kinematically younger than low-stellar-obliquity hot Jupiter systems, implying that some hot Jupiters in misaligned systems may arrive late.
However, it remains unclear which of these formation mechanisms actually plays a role and how much they quantitatively contribute \citep{2018ARA&A..56..175D}.
In this study, we analyze the age-frequency relation of hot Jupiters orbiting single Sun-like stars and provide direct evidence that hot Jupiters form through multiple channels operating on different timescales (Figure \ref{figfHJAgecombine}). 
We further quantitatively determine the contributions of different channels, revealing that $38^{+16}_{-14}\%$ of hot Jupiters should arrive late with timescales up to several Gyrs via the `Late' model (Figure \ref{figFHJmodelStandard}).
Additionally, we found more evidence from obliquity that supports secular chaos \citep{2011ApJ...735..109W,2017MNRAS.464..688H,2019MNRAS.486.2265T} as a promising mechanism for producing such a `late-arrived' population (Figure \ref{figObliquityFeHconstraints}).

The evolutionary patterns of hot Jupiters have long served as a critical framework for testing tidal theories and constraining stellar tidal factors (Extended Data Tab. \ref{tab:Qparamodel}).
However, $Q^{'}_{*}$ remain poorly constrained, with notable discrepancies between results obtained from the orbital decay of hot Jupiter/spin-up of host stars in individual systems \citep[$\sim 4.5-7$;][]{2020ApJ...888L...5Y,2024ApJS..270...14W,2018AJ....155..165P} and the observed distributions of hot Jupiter ensembles \citep[$\sim 6-9$;][]{2008ApJ...678.1396J,2017AA...602A.107B,2018MNRAS.476.2542C,2019AJ....158..190H,2020AJ....160..138H}.
Moreover, the $Q^{'}_{*}$ values from previous studies do not match the observed number of hot Jupiters with orbital decay (Figure \ref{figNumberofdecayQ}).
This is likely due to the limitations of their methodologies.
Specifically, the limited sample size of individual systems with precise orbital decay rates or stellar spin periods/ages hampers the ability to represent the overall distribution, potentially introducing a bias toward a larger $|\dot{P}|$ and lower tail of $Q^{'}_{*}$.
On the other hand, {previous ensemble studies on the distribution/evolution of the hot Jupiter population \citep{2008ApJ...678.1396J,2017AA...602A.107B,2018MNRAS.476.2542C,2019AJ....158..190H,2020AJ....160..138H} assumed that hot Jupiters were all `early-arrived’ and neglected the contribution of `late-arrived' population, leading to overestimate in $Q^{'}_{*}$ and thus an underestimate in $N^{\rm detectable}_{\rm decay}$.
Hansen 2010 \citep{2010ApJ...723..285H} and 2012 \citep{2012ApJ...757....6H}  constrained $\log Q^{'}_{*} \sim 7-9$ from the eccentricity distribution of short-period exoplanets, but the analysis involved host stars near the 
convective-radiative envelope transition that may have less efficient
dissipation and larger $Q^{'}_{*}$ \citep{2018AJ....155..165P}.
In this work, by considering the effects of not only early formed but also late-arrived hot Jupiters, we obtain a more precise constraint on $\log Q^{'}_{*} = 5.7^{+0.4}_{-0.3}$ for Sun-like stars from a large sample analysis, aligning well with both the observed age-frequency relation (Figure \ref{figFHJmodelStandard}) and the detected number of hot Jupiters with orbital decay (red lines in Figure \ref{figNumberofdecayQ}).
Very recently, during the peer-review period of this paper, we noted that Millholland et al. (2025) \citep{2025ApJ...981...77M} analyzed the distribution of insprial timescales ($\sim 0.001-0.3$ Gyr) of short-period hot Jupiters and provided a constraint of $\log Q^{*}_{'} \sim 5.5-7.0$, consistent with our results. 
Such a consistence is expected since these hot Jupiters are probably in the very late stages of tidal evolution, which is primarily dominated by $Q^{'}_{*}$ rather than formation channels.}
Furthermore, our derived constraint is well consistent with those derived from binary interactions ($\log Q^{'}_{*} \sim 5.5$; \citep{2005ApJ...620..970M,2014AJ....148...38M}).
In addition, our results emphasizes the necessity of considering the formation pathways and timescales of hot Jupiters when constraining $Q^{'}_{*}$.

In summary, we find that the declining of the frequency of hot Jupiters with age is broken, which simultaneously puts constraints on the stellar tidal quality factors and the relative importance of the `Early' and `Late' models to form hot Jupiters. 
Although the secular chaos mechanism \citep{2011ApJ...735..109W,2017MNRAS.464..688H,2019MNRAS.486.2265T}  is identified as the dominant channel in the `Late' model, the situation regarding the `Early' model (timescales within tenths of Gyr) remains unclear, highlighting the need for a larger sample of hot Jupiters orbiting young stars.
{Moreover, our results focus primarily on understanding formation channels in single-star systems. Investigating planetary systems in binary star samples can offer further insights into constraining the formation mechanisms (e.g., binary Kozai) of hot Jupiters \citep[e.g.,][]{2024ApJ...977L..11S,2025ApJ...980L..31W}.}
%, e.g., hot Jupiters around young stars \citep{2023Univ....9..192F} and/or with atmospheric element abundance measurements \citep{2021Natur.598..580L},
Future observations and analyses of hot Jupiters from TESS \citep{2015JATIS...1a4003R}, PLATO \citep{2014ExA....38..249R}, JWST \citep{2006SSRv..123..485G} and ET 2.0 \citep{2024ChJSS..44..400G} will test our results and further provide more insights on their origin and evolution.

\clearpage

{\beginED
\begin{center}
{ \LARGE  Methods}\\[0.5cm]
\end{center}

\subsubsection*{1.Observational Analyses}
\label{sec.obs}

\subsubsection*{1.1 Sample}
\label{sec.sample.GT.enlarge}
%Enlarging the Ground-based transit subsample by constructing parent sample based on Gaia data
In a previous paper Chen et al. 2023 (hereafter as Chen2023; \citep{2023PNAS..12004179C}), we selected giant planets around single Sun-like stars in the Galactic disk from the planet host catalog of Planet Across Space and Time (PAST) \uppercase\expandafter{\romannumeral1} \citep{2021ApJ...909..115C} and divide the planetary sample into three subsamples, i.e., radial velocity (RV), ground-based transit (GT) and space-based transit (ST).
Then we constructed parent samples for the three subsamples (see \S~3 of supporting information (SI) of Chen2023 \citep{2023PNAS..12004179C}).
Specifically, the RV parent sample were constructed by using data from Keck, Lick, HAPRS \citep{2014ApJS..210....5F,2017AJ....153..208B,2020A&A...636A..74T}.
The ST parent sample were constructed by using data from LAMOST-Gaia-Kepler catalog \citep{2021AJ....162..100C}.
For the GT subsample, we conducted their parent sample primary based on the information from (Super)WASP survey \citep{2003ApJ...585.1056G,2007ASPC..366..187W}, which searched stars with V-band magnitudes in the range of $\sim 8-13$ \citep{2003ApJ...585.1056G}.
For homogeneity, we only adopted Tycho-2 catalog, which is only 99\% complete to V-band magnitude of $V \lesssim 11$ \citep{2000A&A...355L..27H}.

%For the GT subsample, planets were discovered through various surveys. However, most of these surveys have not released their parent stellar samples or the selection criteria required to reconstruct the parent samples from stellar catalogs.  The exception is the (Super)WASP survey, which has provided such information \citep{2007ASPC..366..187W,2007MNRAS.381..851C,2010MNRAS.404.1849B}. Thus 

In this work, to cover the entire magnitude range of SuperWASP and further expand the GT planetary subsample, we utilize the Gaia DR2 catalog, which provides parallax, celestial coordinates (RA and DEC), proper motions, and magnitudes.
We then select stars from the Gaia DR2 catalog using similar criteria as SuperWASP \citep{2007ASPC..366..187W}: \\
(1) V-band magnitude in the range $8 \leq V \leq 13$; \\
(2) Dwarf stars by excluding giant stars with $\log g <4.0$.
Note this criterion is not exactly same as the reduced proper motion (RPM) method but it offers a more direct approach with the same physical meaning.
We then collect RV data from APOGEE, RAVE, LAMOST and Gaia with the same criteria as PAST \uppercase\expandafter{\romannumeral1}.
Using the astrometric and RV data, we calculate their kinematic properties (i.e., Galactic position, velocity, and component).
We make the quality control and keep Sun-like stars in the Galactic disk with the same criteria  
as described in \S~1 of the SI of Chen2023 \citep{2023PNAS..12004179C}.
The final parent GT stellar sample contains 915,925 stars.
We then crossmatch the parent stellar sample with planets discovered by transit method and ground-based facilities to obtain the planetary sample.
To ensure uniformity in the RV follow-up of SuperWASP transit candidates, we retain only hot Jupiters with high signal-to-red noise ratios (see \S~3.2.1 of SI of Chen2023 \citep{2023PNAS..12004179C} for more details).
The number of hot Jupiters in the GT sample increases from 43 (in Chen 2023) to 92.

In sum, based on the kinematic catalogs of PAST series \citep{2021ApJ...909..115C,2021AJ....162..100C}, we select giant planets around single Sun-like stars and construct their parent stellar samples. 
The final sample consists of:
\begin{enumerate}
    \item RV: 17 hot Jupiters and 85 warm/cold Jupiters from 1662 stars;
    \item ST: 14 hot Jupiters and 19 warm/cold Jupiters from 19,228 stars;
    \item GT: 92 hot Jupiters from 915,925 stars;
\end{enumerate}

\subsubsection*{1.2 Deriving the $F_{\rm HJ}$ by combining RV, GT and ST data}
\label{sec.sample.combine}
Using the planetary and parent stellar samples, by correcting the detection efficiency and geometric effect, one can derive the frequencies of giant planets for the three subsamples with the following formula:
\begin{equation}
    F = \frac{\sum\limits_{i=1}^{N_{\rm p}}{\frac{1}{\rm GE}}}{\sum\limits_{i=1}^{N_*}{{\rm DE}}} \equiv \frac{{N_{\rm p}}}{{N_{\rm eff}}},
\end{equation}
where $N_{\rm p}$ and $N_{\rm *}$ are the numbers of planets and stars.
$\rm DE$ denotes the planetary detection efficiency detected by a given star, regardless of
whether the star has actually detected planets or not.
GE is the geometric effect of a given planet.
$N_{\rm eff}$ is the effective detection number of targets after correcting the effects of geometric effect and detection efficiency.

%Finally, there are 1662 (17 and 85), 915,925 (92), and 19,268 (14 and 19) stars (hot Jupiters and warm/cold Jupiters) in the RV, GT, and ST subsamples, respectively.

In this subsection, for a given sample, we aim to derive a joint frequency of hot Jupiters by combining the three subsamples and eliminating the influence of metallicity.
The detailed procedure is as follows:
\begin{enumerate}   
    \item We calculate the detection efficiencies and geometric effects for stars and planets  for the three subsamples with the same procedure described in \S~3.1-3.3 of SI of Chen2023 \citep{2023PNAS..12004179C}.
    We then obtain the ${N_{\rm p}}$ and ${N_{\rm eff}}$ of the ten bins for the three subsamples.

    \item We eliminate the effect of stellar $\rm [Fe/H]$ by unifying to solar metallicities (i.e., $\rm [Fe/H] =0$).
    It has been found that the frequencies of giant planets are positively correlated with stellar $\rm [Fe/H]$ as $F_{\rm HJ} \propto 10^{\beta \times \rm [Fe/H]}$ \citep{2023PNAS..12004179C}.
    Thus the stellar $\rm [Fe/H]$ induce a difference in $F_{\rm HJ}$ by a factor of:
    \begin{equation}
     \rm Factor_{[\rm Fe/H]} = 
    \frac{\sum 10^{\beta \times [\rm Fe/H]}}{\sum 10^{\beta \times 0}}
    = \frac{\sum 10^{\beta \times [\rm Fe/H]}}{N_*},
    \end{equation}
    Thus we eliminate the effect of $\rm [Fe/H]$ by dividing the frequency of each bin by $\rm Factor_{[\rm Fe/H]}$ for the three subsamples.
    
    \item $F_{\rm HJ}$ derived from the GT subsample are likely underestimated since we do not correct the effect of weather window, which is challenging to quantify as key parameters required for such an estimation are either incomplete or unpublished.
    To address the weather window effect, we assume that the GT subsample have an identical $F_{\rm HJ}$ with the RV and ST subsamples.
    Then we re-normalize its amplitude to the intersection of RV and ST subsamples, which further reduce the effective size of GT parent sample by a factor of $F_{\rm norm} = 0.73$. 

    \item We finally obtain the combined $F_{\rm HJ}$ as follows:
    \begin{equation}
        F_{\rm HJ} = \frac{N_{\rm p}^{\rm Com}}{N^{\rm Com}_{\rm eff}},
    \end{equation}
\end{enumerate}
where $N_{\rm p}^{\rm Com}$ and $N^{\rm Com}_{\rm eff}$ are the `combined' numbers of planets and effective star in each bin:
\begin{equation}
        {N_{\rm p}^{\rm Com}} = N^{\rm RV}_{\rm p} + N^{\rm ST}_{\rm p} + N^{\rm GT}_{\rm p},
\end{equation}

\begin{equation}
\begin{aligned}
{N^{\rm Com}_{\rm eff}} = & N^{\rm RV}_{\rm eff} * {\rm Factor}^{\rm RV}_{\rm [Fe/H]} + N^{\rm ST}_{\rm eff}* {\rm Factor}^{\rm ST}_{\rm [Fe/H]} \\
& + N^{\rm GT}_{\rm eff} * {\rm Factor}^{\rm RV}_{\rm [Fe/H]}*F_{\rm norm}.
\end{aligned}
\end{equation}    

To access the uncertainties of the `combined' frequencies $F_{\rm HJ}$, we assume that the observed number $N_{\rm HJ}, N_{\rm *}$ obey the Poisson distribution.
For $\rm DE$ and $\rm GE$, we take a Monte Carlo method by resampling the stellar and planetary properties based on their uncertainties.
For $X_{\rm [Fe/H]}$, we take $\beta$ as $1.6 \pm 0.3$ \citep{2023PNAS..12004179C} and calculate the uncertainties of $X_{\rm [Fe/H]}$ with Equation S2 by means of error propagation.
The uncertainties of $F_{\rm HJ}$ are set as the 50$\pm$34.1 percentiles in the resampled distributions.

\subsubsection*{1.3 Statistical analyses of the age-frequency trend of hot Jupiter}
\label{sec.obs.analy}
In this subsection, we conduct mathematical analyses to quantitatively characterize the properties of the observational data.

\subsubsection*{1.3.1 $F_{\rm HJ}$ is generally declining with age}
\label{sec.obs.analy.decline}
We divide the whole planetary sample into ten bins according to the relative membership probability for the thick disk to thin disk $TD/D$.
Then we calculate the frequencies of hot Jupiters by adopting the methods described in \S~1.2.
For each bin, the kinematic age $t$ is calculated from the vertical velocity dispersion using the refined age-velocity dispersion relation (AVR) of PAST \uppercase\expandafter{\romannumeral1} \citep{2021ApJ...909..115C}, which gives
\begin{equation}
t = \left(\frac{\sigma}{k\rm \, km \ s^{-1}}\right)^{\frac{1}{b}}\, \rm Gyr,
\label{Eq_S6}
\end{equation}
where $t$ is stellar age, $\sigma$ is the velocity dispersion.
$k$ and $b$ are two coefficients for AVR (listed in Table 5 of \citep{2021ApJ...909..115C}). 
The uncertainties of kinematic ages $t$ are calculated by means of error propagation considering the uncertainties of vertical velocity dispersion and the AVR coefficients.

In the top panel of Figure \ref{figfHJAgecombine} in the main text, we show the evolution of $F_{\rm HJ}$ with age.
As can be seen, there is an obvious declining trend for $F_{\rm HJ}$ with increasing age.
To quantify this declining trend, we fit the relation between $F_{\rm HJ}$ and $t$ with two models: a constant model and an exponential model.
The Formula forms for the two models are:
\begin{eqnarray}
F_{\rm HJ} = C_0, 
\label{eqS1}
\end{eqnarray}

\begin{eqnarray}
F_{\rm HJ} = C_{1} \times {\rm exp}({\gamma} \times \bar{t}),
\label{eqS3}
\end{eqnarray}

For each model, we fit the relation between $F_{\rm HJ}$ and $\bar{t}$ (Gyr) with the Levenberg-Marquardt algorithm (LMA). 
In order to avoid being trapped in some local best fits, we randomly generate 1,000 sets of initial starting guesses of the fitting parameters before LMA fitting. 
We then select the one with lowest residual sum of squares (RSS) from the 1,000 local best-fits as the global best fit.
In order to compare the global best fits of the two models,  we calculate the Akaike information criterion (AIC) \citep{CAVANAUGH1997201}  for each model.
The resulting AIC values are $-93.2$ and $-113.5$ for the constant and exponential model, respectively. 
Therefore a constant model can be confidently ruled out compared with the exponential model with AIC score difference $\rm \Delta AIC>10$.

To consider the effect of uncertainties and obtain the confidence level, we resample $F_{\rm HJ}$ and $t$ from the given distributions and refit the data. 
We repeat this procedure 10,000 times and obtain 10,000 sets of best fits and their respective AIC scores. 
Of the 10,000 sets of resampled data, the exponential decay model is preferred with a smaller AIC score comparing to the constant model for 9,986 sets, corresponding to a confidence level of 99.86\%. 
Therefore, we conclude that $F_{\rm HJ}$ generally declines with age.
The best fit is mathematically expressed as
\begin{equation}
    F_{\rm HJ} = 1.3^{+0.2}_{-0.2} \% \times {\rm exp}({-0.16^{+0.03}_{-0.04} \times \bar{t}}),
\label{eqfHJbestfit}
\end{equation}
which is consistent with the results derived from the three subsamples in Chen2023 \citep{2023PNAS..12004179C}.

\subsubsection*{1.3.2 The declining of $F_{\rm HJ}$ is {broken with a ridge at $\sim 2$ Gyr}}
Besides the global declining, we also find some local features.
Specifically, we note that there exists a local ridge at $\sim 2$ Gyr. 
Moreover, on the two sides of the ridge, $F_{\rm HJ}$ seems to follow different evolution trends.
%, suggesting that the declining trend of $F_{\rm HJ}$ is not monotonic.
%To evaluate the significance of the above features, we adopt the sequences and reversals metric, which is defined as follows:
%\begin{eqnarray}
%I_{ij} (i<j)= \begin{cases} {1}&\mbox{if $F_{i}>F_{j}$}\\ {0}&\mbox{if $F_{i} \le F_{j}$}. \end{cases}
%\end{eqnarray}
%For ten data points, $I$ is an array of 45 elements.
%For a monotonically declining function, all the 45 elements of $I_0$ will be equal to 1.
%Thus, we perform the sign-test between $I_0$ and $I_{\rm HJ}$ derived from $F_{\rm HJ}$.
%The resulting $p-$value is 0.0156, suggesting that $F_{\rm HJ}$ is statistically not monotonic.
%Out of the 10,000 sets, the resulting $p-$values are smaller than 0.05 for 9,786 times, corresponding to confidence levels of 97.86\%.
{To evaluate the existence of the ridge, we also adopt the findpeaks function, showing a ridges at 1.94 Gyr.
To consider the effect of the uncertainties of observational data, we resample the observation data (e.g., kinematic ages, observed number of planets and stars, detection efficiencies, geometric effects) for 10,000 times from their uncetainties.} 
For each set of resampled data, we adopt the findpeaks function to determine the existence and position of ridges.
Out of the 10,000 sets, the findpeaks function returns a ridge for 9,608 times, corresponding to confidence levels of 96.08\%.
{Moreover, we also fit $F_{\rm HJ}$ as a function of $t$ at two sides of the ridge using the exponential model.
The exponential indexes $\gamma$ of the early and the late stages are $-0.07^{+0.10}_{-0.03}$ and $-0.18^{+0.04}_{-0.04}$, respectively.
Of the 10,000 times of resampled data, $\gamma$ of the late stage is smaller than that of the early stage for 9,410 times, corresponding to a confidence level of 94.10\%.}

%\item To evaluate the existence of the ridges, we adopt a findpeaks function, showing a ridges at 1.94 Gyr. To quantify its significance, we resample the observation data (e.g., kinematic ages, observed number of planets and stars, detection efficiencies, geometric effects) for 10,000 times from their uncetainties.
   %For each set of resampled data, we adopt the findpeaks function to determine the existence and position of ridges.
   %Out of the 10,000 sets, the findpeaks function returns a ridge for 9,608 times, corresponding to confidence levels of 96.08\%, respectively. 
   %The 1-$\sigma$ interval of ridge is located at $(1.44 ,\ 2.66)$ Gyr.

\subsubsection*{1.3.3 The two populations of hot Jupiters of different origin channels of different timescales}
\label{sec.obs.analy.twopop}
One natural explanation for the broken $F_{\rm HJ}$-age function is that some hot Jupiters are late-arrived, causing some local increase in hot Jupiter frequency at late stage (see the orange line of Figure \ref{figHJTSofDM}).
To further validate this explanation, we select hot Jupiters as `late-arrived' if they satisfy at least one of the following conditions:
\begin{enumerate}
    \item Kinematic age $t$ larger than their in-spiral timescale $t_{\rm in}$ at 2-$\sigma$ confidence;
    \item eccentricity $e>0$ at 2-$\sigma$ but kinematic age larger than the circularization timescale $t_{\rm cir}$ at 2-$\sigma$ confidence.
\end{enumerate}
$t_{\rm in}$ and $t_{\rm cir}$ are estimated using the following equations \citep{2009MNRAS.395.2268B,2012MNRAS.423..486L}:
\begin{equation}
    t_{\rm in} = \frac{2}{13}  \frac{2Q^{'}_{*}}{9}  \frac{M_*}{M_{\rm p}}  \left(\frac{a}{R_*}\right)^5  \frac{P}{2\pi},
\end{equation}
\begin{equation}
    t_{\rm cir} = 1.6 \ {\rm Gyr} \times \frac{Q_{\rm p}}{10^6}  \frac{M_{\rm p}}{M_{\rm J}} \left(\frac{M_*}{M_\odot}\right)^{-\frac{3}{2}} \left(\frac{R_{\rm p}}{R_{\rm J}}\right)^{-5} \left(\frac{a}{\rm 0.05 AU}\right)^{\frac{13}{2}},
\end{equation}
where $M_*$ and $M_{\rm p}$ are the masses of the host star and the planet. 
$R_*$ and $R_{\rm p}$ are the radii of the host star and the planet.
$a$ and $P$ are the orbital semi-major axis and period of the planet.
The subscripts $\odot$ and $\rm J$ denote the Sun and Jupiter.
$Q_{\rm p}$ is the planetary tidal quality factor and we here adopt a fiducial value often used for close Jovian planets as $10^6$ referring to \citep{2004ApJ...610..477O,2014ApJ...787...27Q}.
$Q^{'}_{*}$ is the stellar tidal quality factor and we adopt the best fits from the observational results by comparing with the theoretical predictions (see later analyses in \S~3.1).

{In Extended Data Fig. 2, we also plot the cumulative distribution of the difference between the kinematic age and tidal evolution timescale, $t-t_{\rm tide}$ for the `late-arrived' population.
As can been, the majority of `late-arrived' hot Jupiters should take $\gtrsim 1$ Gyr to reach their current orbits since the formation of their host stars, which provides a strong constraint to the `late' model.
As discussed later, we adopt the secular chaos model from \citep{2017MNRAS.464..688H} as the `late' model.
Compared to the arrived timescale via secular chaos model \citep{2017MNRAS.464..688H}, $t-t_{\rm tide}$ from the observational sample seems to be larger. 
This is expected since our selection criteria for the `late-arrived' observational sample are rather strict. 
As we mentioned before, a portion of ‘late-arrived’ hot Jupiters (e.g., $t<t_{\rm tide}$) may not be identified. 
Thus, we generated a synthetic population of 10,000 hot Jupiters with the arrival timescale distribution as \citep{2017MNRAS.464..688H}. 
We then calculate their inspiral timescales. 
Subsequently, we assigned ages and uncertainties to these synthetic hot Jupiters by randomly sampling from the observed `late-arrived' hot Jupiter population. 
Then, we selected survived synthetic hot Jupiters that can be classified as `late-arrived'. 
Finally, we conducted a two-sample KS test between the arrival timescales of synthetic and observed hot Jupiters that can be selected as `late-arrived', yielding a p-value of 0.3499. 
The above consistence further supports the secular chaos as the ‘Late’ model.}

%\begin{figure}[!t]
%\centering
%\includegraphics[width=\linewidth]{FHJ-Age_RVGTSTcombine_Late.eps}
%\caption{The frequency of `late-arrived' hot Jupiters  by combining RV, GT and ST subsamples as a function of kinematic age. We also print the $p-$ value of the monotonicity.}
%\label{figfHJAgecombined_Late}
%\end{figure}

%\begin{figure}[!t]
%\centering
%\includegraphics[width=\linewidth]{FHJ-Age_RVGTSTcombine_Early.eps}
%\caption{The frequency of `early-arrived' hot Jupiters by combining RV, GT and ST subsamples as a function of kinematic age. We also print the $p-$ value of the monotonicity.}
%\label{figfHJAgecombined_Early}
%\end{figure}

The selected `late-arrived' population consists of 54 hot Jupiters (5 RV, 6 ST and 43 GT).
The other hot Jupiters are classified as `early-arrived'.
It is worth noting that the selected `early-arrived' population may also contain a portion of `late-arrived' hot Jupiters, which could not be identified using the aforementioned criteria due to the current age precision.
Then we divide the two populations into ten bins with the same intervals as the whole sample and derived their `combined' frequencies with the same procedures as described in \S~1.2.
The frequencies of`late-arrived' and `early-arrived' populations after correcting detection biases are $0.23^{+0.04}_{-0.03}\%$ and $0.62^{+0.08}_{-0.07}\%$, which contribute to $\sim 27.3^{+5.3}_{-4.5}\%$ and $\sim 72.7^{+4.5}_{-5.3}\%$ of the whole hot Jupiter frequency ($0.85^{+0.09}_{-0.07}\%$), respectively.

In the middle and bottom panels of Figure \ref{figfHJAgecombine}, we show the `combined' frequencies of the `late-arrived' and `early-arrived' hot Jupiter populations as a function of kinematic age, respectively.
{As can be seen, for the `late-arrived' hot Jupiters, their frequency first increases and then decreases, forming a ridge at $\sim$ Gyr.
To quantify the above feature, we adopt the findpeaks function, returning a ridge at $1.90^{+0.61}_{-0.49}$ Gyr with a confidence level of 99.87\%.
At the early stage, as star ages, $F_{\rm HJ}$ increase with $\gamma = 0.44^{+0.46}_{-0.22}$ and the confidence level (i.e., the fraction of $\gamma>0$) is 96.37\%.
In contrast, in the late stage, $F_{\rm HJ}$ becomes to decrease with age and the fitted $\gamma$ is $-0.11^{+0.05}_{-0.04}$, correspond to a confidence level of 97.83\%.
%We calculate their the sequences and reversals metric $I$ and perform the sign-test, resulting a $p-$value of $4.8 \times 10^{-7}$.
%We also resample the data considering the uncertainties and the resulting $p-$values are smaller than $3 \times 10^{-3}$ for 9,992 times out of 10,000 sets of resampling, corresponding to a confidence level of 99.92\% and suggesting a significant monotonicity.
After removing the contribution of the  `late-arrived' population, the frequency of the remaining `early arrived' population shows a significantly declining trend with a $\rm \Delta AIC = 24.5$.
Moreover, the ridge at $\sim 2$ Gyr disappear, and the fitted $\gamma$ before and after $\sim 2$ Gyr are $-0.18^{+0.12}_{-0.37}$ and $-0.16^{+0.07}_{-0.18}$ respectively, which are well consistent within $1-\sigma$ interval.}

The above results support the proposed explanation that the observed broken age-frequency trend is a consequence of multiple origins of hot Jupiters. 
Specifically, the `early-arrived' population dominates the global declining trend, while the `late-arrived' population induces the local ridge.
As discussed later in \S~2.3, we put constraints on the origin and tidal evolution of hot Jupiters by comparing the observational results with the theoretical predictions. 

%such a monotinitically declining trend of $F_{\rm HJ}$ is not expected of the tidal evolution of hot Jupiters if hot Jupiters are  formed via different origins of different timescales.

\subsubsection*{2. The Age-Frequency Relation of Hot Jupiters from Theory}
\label{sec.theory}

\subsubsection*{2.1 The origin of hot Jupiters}
\label{sec.theory.origin}
At present, there are mainly three classes of origin theories for the hot Jupiters: in-situ formation, disk migration, and high-eccentricity migration \citep{2018ARA&A..56..175D}. 
In the in-situ formation scenario, hot Jupiters form at their current locations through gravitational instability \citep{2007prpl.conf..607D} and/or core accretion \citep{1996Natur.380..606L,2014prpl.conf..619C}.
It was believed that in-situ formation is difficult to operate at close-in orbit \citep{2005ApJ...621L..69R,2006ApJ...648..666R}, but some recent studies suggest that it may be feasible under some certain conditions \citep{2016ApJ...829..114B,2016ApJ...817L..17B}, which however could not explain the vast majority of hot Jupiter population.
Thus some form of migrations must have occurred for most of hot Jupiters after their initial formation.
To date, two primary migration mechanisms have been proposed: disk migration and high-eccentricity migration. 
In the disk migration scenario, hot Jupiters form several astronomical units (AU) away and migrate inward to $\lesssim 0.1$ AU under torques from the gaseous protoplanetary disk 
\citep{1996Natur.380..606L,2008ApJ...685..584I,2014prpl.conf..667B}.
For the high-eccentricity migration scenario, hot Jupiters are driven into high-eccentricity orbits under close encounters between planets \citep{1996Sci...274..954R}, Kozai–Lidov interactions \citep{1962AJ.....67..591K,1962P&SS....9..719L} induced by a distant binary companion star/massive planet on an inclined orbit \citep{2003ApJ...589..605W,2011Natur.473..187N}, secular chaos in multi-planetary systems in mildly inclined and eccentric orbits \citep{2011ApJ...735..109W,2017MNRAS.464..688H} or planet-planet coplanar secular \citep{2015ApJ...805...75P} and then moved from a large separation to several hundredths of AU \citep{2014A&A...562A..71B,2019A&A...631A..47K}.

Hot Jupiters of different origins are expected to have distinct formation timescales. 
For in-situ formation and gas disk migration, hot Jupiters must form or migrate to their current locations before the dissipation of the gas disk, i.e., $\lesssim 1-10$ Myr \citep{1986ApJ...309..846L}. 
In contrast, for the high-eccentricity migration, the timescale varies significantly for different mechanisms.
Specifically, in the case of planet-planet scattering, the eccentricities of hot Jupiters increase via a random walk process, with their periapsides shrinking to $\sim 0.1$ AU over a timescale of thousands of years \citep{2012ApJ...751..119B}.
For hot Jupiters formed via secular Kozai–Lidov interactions by inclined massive planets \citep{2011Natur.473..187N},
most of them are tidally ensnared on their first close approach to the host star with a timescale ranging from $10^4-10^8$ years. Afterward, they are tidally circularized to orbits of $\lesssim 0.1$ AU within $\sim 1$–100 Myr (summarized in Fig. 3 and \S~2.3.1.3 of \citep{2018ARA&A..56..175D}).
Whereas, under the secular chaos in multi-planetary systems, the migration timescale of hot Jupiter can span the entire lifetime of the host star, i.e., $\sim 0.1-10$ Gyr  \citep{2011ApJ...735..109W,2017MNRAS.464..688H}.

Alternatively, the planet-star kozai effect can also deliver hot Jupiters under certain conditions \citep{2003ApJ...589..605W,2007ApJ...669.1298F}.
The planet eccentricity and periapse oscillate with a typical timescale as \citep{1998MNRAS.300..292K}:
\begin{equation}
    \tau_{\rm kozai} = \frac{m_1+m_2+m_3}{m_3} \frac{2P_{\rm out}^2}{3\pi P_{\rm in}} (1-e_{\rm out}^3)^{3/2},
\end{equation}
where $m_1$, $m_2$, $m_3$ are the masses of central stars, planets and outer companions, respectively.
$P_{\rm in}$ and $P_{\rm out}$ are the orbital periods of the giant planet and companion.
$e_{\rm out}$ is the eccentricity of the companion.
We then exclude planets that are likely dynamically
unstable according to the following formula: $\frac{a_{\rm in}}{a_{\rm out}} > 0.330-0.417 e_{\rm out}+0.069 e_{\rm out}^2$ \citep{2007ApJ...670..820W}.
Furthermore, because Kozai cycles are driven by weak gravitational forces from the outer binary, they can be easily suppressed by general relativity pericenter precession in the inner binary \citep{2007ApJ...669.1298F}.
Therefore, the Kozai effects can only occur when their timescale is shorter than the pericenter precession timescale (following Equation (23) in \citep{2007ApJ...669.1298F}).
To obtain the distributions of the Kozai timescales, we adopt the following conditions: the distributions of orbital period, eccentricity and mass ratio of outer companions referring fig. 13, 15 and 16 of \citep{2010ApJS..190....1R}, and the distribution of orbital semi-major axes of giant planets between 1 and 5 au based on the power law distribution from \citep{2008PASP..120..531C}.
Then we calculate with Equation S13 and keep those shorter than the corresponding pericenter precession timescale.
Extended Data Fig. 1 shows the distribution of the planet-star Kozai timescale.
As shown, the Kozai timescales are mainly between $10^4-10^8$ years, which is comparable to the results of previous studies \citep{2007ApJ...670..820W}.
The majority (94.28\%) of the Kozai timescales are less than 0.1 Gyr, with nearly all (99.71\%) being less than 1 Gyr.

Here we classify these mechanisms into two types according to their timescales:
\begin{itemize}
   \item \textbf{`Early' model}, i.e., in-situ foramtion, disk migration and high-eccentricity migration via planet-planet scattering \citep{1996Sci...274..954R}/Kozai-Lidov interactions \citep{2011Natur.473..187N}/coplanar secular \cite{2015ApJ...805...75P}. 
   Their formation timescales, typically within several tenths of Gyr, are comparable to (or even shorter than) the typical uncertainties in kinematic age estimates and are significantly shorter than the evolutionary timescales of our sample (i.e., $\sim 0.5$–10 Gyr). Therefore, given the current limitations in age accuracy, we collectively group the aforementioned mechanisms into a single category.

   \item \textbf{`Late' model}, i.e., High-eccentricity migration via secular chaos in multi-planetary systems, which could form hot Jupiters over a relatively long timescale extending to several Gyrs.
   Previous studies show that the stellar Kozai effect can also deliver hot Jupiters over Gyr under certain conditions \citep{2007ApJ...669.1298F,2011ApJ...735..109W}, but here we do not consider it as the main channel for the following reason: (1) the fraction of hot Jupiter host stars with binary companions capable of inducing Kozai–Lidov oscillations is at most $16\% \pm 5\%$ (when overcoming the general relativity pericenter precession) and will further decrease to $6\% \pm 2\%$  assume an isotropic distribution of the inclinations of stellar companions \citep{2016ApJ...827....8N}.
   (2) The stellar kozai would mainly deliver hot Jupiters with timescales $\lesssim$ tenths of Gyr.
   (3) the potential binaries have been excluded in our sample.
   
\end{itemize}
Without considering the subsequent tidal evolution, the frequency of hot Jupiters via `Early' model is expected to have no (significant) dependence on stellar age.
In contrast, if `Late' model is dominating, the frequency of hot Jupiters would grow with increasing age (See Extended Data Fig. 3).

\subsubsection*{2.2 The tidal decay of hot Jupiters}
\label{sec.theory.tidal}

Because hot Jupiters are close to their host stars, the tidal interactions can reduce the orbital eccentricity ($e$) of close-in planets \citep{1996Sci...274..954R,2008ApJ...678.1396J}.  
%This is one probably reason that lead to the observed dichotomy between the e-values for hot Jupiters ($<0.1$) and cool giant planets ($\sim 0.3$). 
Tides can also transfer orbital energy and angular momentum from the hot Jupiters to their host stars, resulting in the spin-up of the stars  \citep{2009MNRAS.396.1789P,2010ApJ...719..602S,2013ApJ...775L..11M,2014ApJ...786..139T,2015A&A...577A..90M}.
As a consequence, the semi-major axis ($a$) of hot Jupiters would shrink with age.
Part of them may have already passed into the Roche limit and been tidally disrupted by the host star \citep{2008ApJ...678.1396J,2009ApJ...692L...9L}. 

In order to quantitatively determine the tidal reduction of the semi-major axis $a$, we adopt the classical equations of equilibrium 
tidal evolution: \citep{GOLDREICH1966375,2012MNRAS.423..486L} 

\begin{equation}
\begin{aligned}
\frac{1}{a} \frac{{\rm d}a}{{\rm d}t} = & -[\frac{63}{2}(GM^3_*)^{1/2}\frac{R^5_{\rm p}}{Q^{'}_{\rm p}M_p}e^2+\frac{9}{2}(G/M_*)^{1/2}\frac{R^{5}_{*}M_{\rm p}}{Q^{'}_{*}} \\
&\times (1+\frac{57}{4}e^2))] a^{-13/2},
\label{eqdadt}
\end{aligned}
\end{equation}

\begin{equation}
\begin{aligned}
\frac{1}{e} \frac{{\rm d}e}{{\rm d}t} = & -[\frac{63}{4}(GM^3_*)^{1/2}\frac{R^5_*}{Q^{'}_{\rm p}M_{\rm p}}e^2+\frac{225}{16}(G/M_*)^{1/2} \\
&\frac{R^{5}_{*}M_{\rm p}}{Q^{'}_{*}}] a^{-13/2},
\label{eqdedt}
\end{aligned}
\end{equation}
where $G$ is the gravitational constant, $R$ and $M$ are the radius and mass, and $Q^{'}$ is the modified tidal dissipation factor \citep{GOLDREICH1966375}.
The subscripts $*$ and $\rm p$ refer to the star and planet, respectively.
As can been, due to the tides, $a$ decays with the increase of age.
The tidal decay rate is dependent on various parameters (e.g., the initial $a$ and $Q^{'}_{*}$).

Plenty of previous studies have estimated the stellar tidal quality factor $Q^{'}_{*}$,  using various samples and methodologies.
Specifically, Jackson et al.  (2008) \citep{2008ApJ...678.1396J}, Bonomo et al. (2017) \citep{2017AA...602A.107B}, Penev et al. (2018) \citep{2018AJ....155..165P} and Labadie-Bartz et al. (2019)  \citep{2019ApJS..240...13L} provided constraints based on the evolution of planetary semi-major axis ($a$), eccentricity ($e$), and stellar spin-up.
Whereas, Meibom et al. (2005) \citep{2005ApJ...620..970M} and Milliman et al. (2014) \citep{2014AJ....148...38M} restrict the $Q^{'}_{*}$ from the tidal interaction between stellar binaries. 
Collier Cameron \& Jardine (2018) \citep{2018MNRAS.476.2542C}estimated the tidal quality factor by reproducing the observed orbital separation distribution, assuming an initially uniform distribution inferred from Doppler surveys.
Hamer \& schlaufman (2019) \citep{2019AJ....158..190H} obtains a upper limit of $\log Q^{'}_{*} \lesssim 10^7$ by assuming that hot Jupiters were destroyed before the end of the main sequence lifetime of their host stars.
Recently, Yee et al. (2020) \citep{2020ApJ...888L...5Y} and Turner et al. (2021) \citep{2021AJ....161...72T} constrained $Q^{'}{*}$ by analyzing the orbital decay of WASP-12b, a transiting hot Jupiter with a 1.09-day orbit around a late-F star.
Additionally, some other studies have detected hints of orbital decay for tens of hot Jupiters (e.g., WASP-4b, 43b) and provided estimations for $Q^{'}_{*}$ \citep{2020AJ....159..150P,2022AJ....163..281T}.
We have summarized the constraints on the value of $Q^{'}_{*}$ from the literature in Extended Data Tab. 1.

\subsubsection*{2.3 Numerical simulation}
\label{sec.theory.simu}

In this section, we conduct numerical simulations for the tidal decay of hot Jupiters of different origin to obtain the age evolution of hot Jupiters from theory.

For the origin of hot Jupiters, we consider the following three models (as shown in Extended Data Fig. 3): 

\begin{enumerate}
    \item `Early' model within a short time after stars formed. 
    As mentioned before, the formation timescale is significantly shorter than the evolutionary timescale in our sample, thus the simulation simplistically approximates it as 0.
    %由于现在的年龄误差较大，模拟中简单等效于$t=0$
    \item `Late' model. 
    The formation timescale obeys a distribution referring to Fig. 9 of Hamers et al. (2017) \citep{2017MNRAS.464..688H}. 
    \item Hybrid model (`Early' plus `Late'). 
    We assume $f_{\rm Late}$ of hot Jupiters are formed via the `Late' model and the other $1-f_{\rm Late}$ are via the `Early' model.
    In each simulation, $f_{\rm Late}$ is set as a free parameter range from 0 to 1 with the interval of 0.01. 
\end{enumerate}

For the subsequent tidal evolution, we use a simplified model to perform numerical simulations by setting the orbital eccentricity $e=0$ since the circularization timescale is order of magnitudes less than the in-sprial timescale for hot Jupiters \citep{2008ApJ...678.1396J,2012MNRAS.423..486L}.
That is to say, the conditions for a hot Jupiter's `arrival' defined in this work refer to it entering within 0.1 AU of its star, with its orbit having been largely circularized.
The simulations are carried out by adopting the eighth-order Runge-Kutta method and the precision is set as $10^{-12}$. 
As shown in Equation \ref{eqdadt}, the initial conditions consist of the stellar tidal dissipation factor $Q^{'}_{*}$, the masses and radii of star and planet, and the initial period distribution of hot Jupiters.

The initial conditions are set as follows:
For the stellar mass and radius, we adopt the same distributions as hot Jupiter host stars in our observed sample.
The initial distributions of planetary mass, radius and orbital period are set as follows: 
\begin{enumerate}
    \item `Early' model: For the planetary mass and radius, we adopt the same distributions as the observed warm/cold Jupiters.
    For the initial orbital period distribution, we here adopt a distribution inferred from the confirmed planets in Kepler field \citep{2016A&A...587A..64S}.

    \item `Late' model: mass and period as the results of Fig. 8 and 3 of Hamers et al. (2017) \citep{2017MNRAS.464..688H}.
    The planetary radius are estimated with the planetary mass-radius relation, $\frac{R_{\rm p}}{R_{\rm J}} = 1.2 \left( \frac{M_{\rm p}}{M_{\rm J}} \right)^{0.27}$ \citep{2016A&A...589A..75M}.
\end{enumerate}
The above initial condition is hereafter referred to as the standard case.
We also perform simulations on different initial conditions and compare to the observations. 
The influence of initial conditions will be discussed later in detail in \S~4.1 and 4.2.

In each set of simulation, we put 10,000 hot Jupiters with the aforementioned initial conditions.
During the simulations, the hot Jupiters were removed from the system when they fell into the Roche limit, $a_R = (R_p/0.462)(M_*/M_p)^{1/3}$ \citep{2005Icar..175..248F}. 
$\log Q^{'}_{*}$ is set at the interval of 0.1 from 4 to 9.
Upon completion of the simulations, the returned information consist of the simulation time, semi-major axis, and the lifetimes of each hot Jupiter. 
%In the context that follows, the subscripts \emph{obs} and \emph{sim} refer to observation and numerical simulation, respectively.

\subsubsection*{3. Constraints on the Formation and Tidal Evolution of Hot Jupiters}
\label{sec.constraints}

\subsubsection*{3.1 Constraints from the age-frequency relation}
\label{sec.theory.comparison}

To explore how the origin mechanisms and tides shape the observation, we calculate the number fraction of survived hot Jupiters to the total number in simulations,  $f_{\rm HJ}^{\rm sim}$. 
To compare the observation data with theoretical models, we transfer it to ${F^{\rm sim}_{\rm HJ}}$ with the following formula:
\begin{equation}
   {F^{\rm sim}_{\rm HJ}} = F_0 \times f_{\rm HJ}^{\rm sim}, 
\end{equation}
where $F_0$ is the scaling factor for each set of simulation. 
The $F_0$ factor is calibrated by equating the observational data to the simulation results, using the average age of the observed hot Jupiter population as the value of $t$ in the simulation.

To obtain the best match, we calculate the likelihood $L$ as a function of $f_{\rm Late}$ and $Q^{'}_{*}$ in the following formula:
\begin{equation}
L = \prod^{10}_{i=1} \frac{1}{\sqrt{2\pi} \sigma_i} {\rm exp} \left( -\frac{{\Delta F_{\rm HJ}}^2}{2{\sigma_i}^2}\right),
\label{eqlikelihood}
\end{equation}
%\begin{equation}
%\ln L = \sum^{10}_{i=1} \left( -\ln {\sqrt{2\pi} \sigma_i} -  \frac{{\Delta \left(\frac{{N_{\rm HJ}}}{{N_{\rm J}-N_{\rm HJ}}}\right)}^2_{\rm model-obs}}{2{\sigma_i}^2}\right),
%\label{eqlnlikelihood}
%\end{equation}
where $\Delta F_{\rm HJ}$ is the difference between the observational and model results.
$\sigma_i$ denotes the uncertainties of $F_{\rm HJ}$ for each age bin shown in Figure 1.
The best-fits of $Q^{'}_{*}$ and $f_{\rm Late}$ have the maximum $L$ and the confidence intervals
(1-$\sigma$: 68.3\%, 2-$\sigma$: 95.4\% and 3-$\sigma$: 99.7\%) are estimated using a spline function.

In what follows, we show the best fits for the three origin models of hot Jupiters:
\begin{itemize}
\item \textbf{`Early' model  ($f_{\rm Late}=0$)}

Extended Data Fig. 4 illustrates the best match of the $F_{\rm HJ}$ from the simulation (cyan line) along with the distribution of likelihood as a function of $Q^{'}_{*}$ when all Jupiters are formed via the `Early' model.
As can be seen, $F_{\rm HJ}$ derived from the simulations decreases with age, which is generally consistent the observational results but can not explain the ridge at $t \sim 2$ Gyr and the broken relation.
By comparing to the observational results, we obtain the best fit of $\log Q^{'}_{*}$ is $6.2^{+0.3}_{-0.3}$ for the standard case.
The AIC score for the best-fit comparing to the observational results is -106.9.

\item \textbf{`Late' model ($f_{\rm Late}=1$)}

If hot Jupiter are all formed via the `Late' model, their frequency first increases and then decreases with age due to the combined effects of continuous supplement and tidal destruction. 
Extended Data Fig. 5 shows the best match ($\log Q^{'}_{*} = 4.8^{+0.3}_{-0.2}$) of the numerical simulation via the `Late' model.
Due to the strong tidal effect, the turning point occurs before $\sim 0.6$ Gyr and $F^{\rm sim}_{\rm HJ}$ decreases within the age range of the observational sample, which barely agree with the observational decay.  Nevertheless, significant deviations remain between the best match of the numerical simulations and the observational results.
The AIC score for the best-fit is -80.6.

\item \textbf{Hybrid model}

Extended Data Fig. 6 compares the numerical simulation of the hybrid model with the observational results and shows the distribution of likelihood as a function of $Q^{'}_{*}$ and $f_{\rm Late}$. 
As can be seen, the best fit of theoretical result matches well with the observation data (i.e., the long-term decay as well as the broken age-frequency relation and the local ridge at $\sim 2$ Gyr).
The good match between the observational results and theoretical models allows us to provide constraints on the stellar tidal quality factor and relative importance of different origin mechanisms of hot Jupiters.
For the standard case, the best fits for $f_{\rm Late}$ and $\log Q^{'}_{*}$ are $0.38^{+0.16}_{-0.14}$ and $5.7^{+0.4}_{-0.3}$, respectively.
The AIC score for the best-fit of the hybrid model is -116.5. 
\end{itemize}

In order to compare the fitting quality between the three model, we calculate the AIC scores for their best fits.
As shown in Extended Data Tab. 2, the AIC values are -106.9, -80.6, and -116.5 for the `Early' model, `Late' model and Hybrid model, respectively.
%To quantify the significance, we resample the observed numbers $N_{\rm HJ}$, $N_{*}$ from Poisson distributions. The detection efficiency, geometric effect, kinematic age are resampled from normal distribution given their values and uncertainties for 10,000 times.
In the 10,000 sets of resampled data, comparing to the single origin of `Early' model and `Late' model, the AIC scores of the hybrid model are smaller for 9,855 and 9,999 times, corresponding to confidence levels of 98.55\% and 99.99\% respectively.
Therefore, we can confidentially conclude that the hybrid model is more preferred.

\subsubsection*{3.2 Additional evidence from the obliquity distributions}
\label{sec.Obliquity}
Stellar obliquity refers to the angle between a planet’s orbital axis and its host star’s spin axis, which can provide important clues to the formation and dynamic evolution of hot Jupiters.
Specifically, the initial distribution of stellar obliquity is determined by the formation and
evolution of a protoplanetary disk \citep{2010MNRAS.401.1505B,2011MNRAS.412.2790L}.
Subsequently, the stellar obliquity can be further excited by the dynamic mechanisms via interactions caused by planetary and/or binary companions, e.g., planet–planet scattering %citep{1996Sci...274..954R,2008ApJ...686..580C,2008ApJ...678..498N}
, stellar/planetary Lidov–Kozai (LK) effects %\citep{2003ApJ...589..605W,2007ApJ...669.1298F,2011Natur.473..187N,2012ApJ...754L..36N,2015ApJ...799...27P}, and
secular chaos. %\citep{2011ApJ...735..109W,2017MNRAS.464..688H}.
The predicted stellar
obliquity distributions of the hot Jupiters via different mechanisms are different (see fig. 22 of \citep{2022PASP..134h2001A}).
For example, the majority of hot Jupiter systems caused by the planet-planet scattering mechanism would be aligned but a small fraction of systems would have a diverse range of stellar obliquities \citep{2012ApJ...751..119B}.
In contrast, the secular chaos mechanism will result in misaligned systems with obliquities $\lesssim 90^\circ$ \citep{2019MNRAS.486.2265T}.
Lastly, the tidal interaction could realign the host star’s spin axis with the planet’s orbital axis and thus reduce the stellar obliquity \citep{2022PASP..134h2001A}.
Therefore, investigating the distribution of the stellar obliquities of hot Jupiter systems can help to constrain their formation channel and tidal evolution \citep{2022AJ....164...26H,2023AJ....166..112D}.

From observations, typically only the sky-projected
stellar obliquity $\lambda$ (the projections onto the plane of the sky) can be mainly measured through the Rossiter–McLaughlin
effect \citep{1924ApJ....60...15R,1924ApJ....60...22M}.
%as well as starspot crossings \citep{2011ApJ...740L..10N,2013AN....334..180S}, $v\sin{i}$ technique \citep{2010ApJ...719..602S} and asteroseismology \citep{2013Sci...342..331H}. 
To date, the stellar obliquities have been measured for hundreds of hot Jupiters.
Previous studies have found observational evidence that many hot Jupiters in misaligned systems are late-arrived within $\lesssim 0.1$ AU long after their parent protoplanetary disks dissipated \citep{2022AJ....164...26H}. 
Dong \& Foreman-Mackey (2023) \citep{2023AJ....166..112D} reported that most hot Jupiter system are aligned and the other misaligned systems have nearly isotropic stellar obliquities with no strong clustering near 90 degrees.

We initialize the sample of hot Jupiters with $\lambda$ measurements based on the Table A of \citep{2022PASP..134h2001A}.
We also compare with the TEPCat orbital obliquity catalog (https://www.astro.keele.ac.uk/jkt/tepcat/obli-
quity.html) and remove hot Jupiters with the differences in $\lambda$ larger than three times of uncertainties.
Then we cross-match with our planetary sample and yield 39 hot Jupiters with $\lambda$ measurements.

As mentioned before, the tidal interaction would reduce the observed $\lambda$.
The timescale of tidal-realignment is not well determined yet \citep{2014ARA&A..52..171O}.
In the equilibrium tidal theory, the timescale for realignment is comparable to the in-sprial timescale \citep{2022PASP..134h2001A}, which is dependent on the stellar mass, radius and planetary mass, radius, orbital semi-major axes.
Referring to \citep{2022PASP..134h2001A}, we adopt the dimensionless tidal dissipation parameter,
\begin{equation}
    \left(\frac{M_{\rm p}}{M_*}\right)^{-2} \left(\frac{a}{R_*}\right)^6, 
\end{equation}
where $M_*$ and $R_*$ are the stellar mass and radius.
$M_{\rm p}$ and $a$ are the planetary mass and semi-major axes.
Then we select hot Jupiter systems with long-alignment timescales using the following criterion: $\left(\frac{M_{\rm p}}{M_*}\right)^{-2} \left(\frac{a}{R_*}\right)^6>10^{12}$ or $T_{\rm eff}>6250$ K \citep{2022PASP..134h2001A}.
Furthermore, as shown in Fig. 11 of \citep{2022PASP..134h2001A}, hot Jupiter systems older than $\sim 4$ Gyr have relatively smaller $\lambda$, thus we also only keep hot Jupiter systems with kinematic ages $<4$ Gyr.
We also make Kolmogorov–Smirnov (KS) tests and check the resulted $p$ values to quantify the age evolution.
%As shown in Figure \ref{figLambda_alignment}, hot Jupiter systems with long-alignment timescales and kinematic ages $<4$ Gyr have larger $\lambda$ comparing to that of others (with KS $p-$value of 0.0437), supporting that the tidal forces have realigned some hot Jupiter systems.

In Figure \ref{figObliquityFeHconstraints}, we show the distributions in $\lambda$ for the `early-arrived' and `late-arrived' hot Jupiter systems with long-alignment timescales and young kinematic ages.
As can be seen, the `early-arrived' systems are mainly aligned ($\lambda \lesssim 10^\circ$) but with a small fraction in misaligned systems.
In contrast, the `late-arrived' hot Jupiter systems have a stellar obliquity distribution mainly within $10-90^\circ$.

To further put constraints on the origin of hot Jupiters, we then compare the observed $\lambda$ distribution with the predictions of different formation mechanisms, i.e., planet-planet scattering \citep{2012ApJ...751..119B}, resonance crossing \citep{2018MNRAS.480.1402A}, secular chaos \citep{2019MNRAS.486.2265T}, and planetary Kozai-Lidov effect \citep{2016ApJ...829..132P}.
We also conduct an aligned distribution of stellar obliquity $\phi$ as uniformly distributed in the range of $0-20^\circ$.
Specifically, we first generate 1,000 hot Jupiter systems for each of the above mechanisms with the distributions of stellar obliquity $\phi$.
Then we obtain the sky-projected obliquity $\lambda$ distribution from the $\phi$ distribution with the following formula \citep{2023AJ....166..112D}:
\begin{equation}
    \lambda = \arctan{\frac{\sin{\phi}\sin{\theta}}{-\sin{\phi}\cos{\theta}\cos{i_{\rm orb}}+\cos{\phi}\sin{i_{\rm orb}}}},
\end{equation}
where $\theta$ is the azimuthal angle of the stellar spin
axis relative to the orbital axis, and $i_{\rm orb}$ is the orbital inclination.
Referring to \citep{2023AJ....166..112D}, we take $\theta$ and $\cos{i_{\rm orb}}$ as uniformly distributed between 0 and 1.
Finally, we perform two-sample KS tests between the $\lambda$ distributions of the observed `early-arrived'/`late-arrived' hot Jupiter systems and the prediction of the five mechanisms.

The resulted $p-$values are summarized as Extended Data Tab. 3.
As can be seen, the $\lambda$ distribution of the `early-arrived' population matches well with the predictions of the planet-planet scattering \citep{2012ApJ...751..119B}, the resonance cross \citep{2018MNRAS.480.1402A}, and the alignment distribution, but is significantly smaller than the prediction of planetary Kozai-Lidov mechanism \citep{2016ApJ...829..132P} with a KS $p-$value $<0.003$.
For the `late-arrived' hot Jupiters, their $\lambda$ distribution is consistent with the prediction of the secular chaos mechanism \citep{2019MNRAS.486.2265T} with a KS $p-$value of 0.229, but it deviates from the predictions of all other mechanisms with KS $p-$values$<0.05$.
The most-favored mechanisms (with largest $p-$values, highlighted in Extended Data Tab. 3) for the `early-arrived' and `late-arrived' hot Jupiters are planet-planet scattering and secular chaos, respectively.

\subsubsection*{3.3 Consistency check of stellar metallicity}
Stellar metallicity also has important impacts on the formation and evolution of hot Jupiters \citep{2015ApJ...799..229W}. 
From theory, gentle disk migration is expected to operate in environments with a range of metallicities.
Whereas high eccentricity migration via secular chaos occurs in systems with multiple giant planets, which are more likely to form around metal-rich stars born in disks with more solids.
%From observation, previous studies have shown some supporting evidence: eccentric hot Jupiters tend to be hosted by stars with super-Solar metallicities, whereas circular hot Jupiters are hosted by stars with both super-Solar and sub-Solar metallicities \citep{2013ApJ...767L..24D,2016ApJ...820...93S}.

Stellar metallicity is known to be correlated with
other stellar properties \citep{2022AJ....163..249C}.
Thus we adopt the NearestNeighbors method in scikit-learn to select the nearest neighbor in the stellar mass and $TD/D$ from the `early-arrived' hot Jupiter hosts for every `late-arrived' hot Jupiter host star.
As shown in Extended Data Fig. \ref{figFeHconstraint}, after eliminating the effects of stellar mass, radius and age, the `early-arrived' hot Jupiter host stars have a relative smaller $\rm [Fe/H]$ comparing to the `late-arrived' hot Jupiter hosts.
We generate the 10,000 bootstrapped sample from the $\rm [Fe/H]$ of `late-arrived' and `early-arrived' hot Jupiter hosts (after parameter controlling) and calculate their average metallicities $\overline{\rm [Fe/H]}$. The uncertainties are taken as the $50 \pm 34.1$ percentile. The $\overline{\rm [Fe/H]}$ are $0.06^{+0.02}_{-0.03}$ and $0.12^{+0.02}_{-0.02}$ dex for the `early-arrived' and `late-arrived' hot Jupiter hosts, respectively. Of the 10,000 set of bootstrapping, the $\overline{\rm [Fe/H]}$ of `late-arrived' hot Jupiter hosts are larger for 9,607 times, supporting our inference that the `late-arrived' hot Jupiters are formed via secular chaos.

\subsubsection*{3.4 Consistency check of configurations of hot Jupiter systems}
Secular chaos mechanism occurs in systems with multiple giant planets. 
Thus we expect that the `late-arrived' hot Jupiters are more likely to have outer companions.
Therefore, we adopt the 17 hot Jupiters in the RV sample using data from the Keck, Lick and HAPRS and calculate the fraction to have outer giant companions for the `early-arrived' and `late-arrived' populations. 
Out of the 5 `late-arrived' hot Jupiters, 2 of them ($40^{+26.7}_{-23.3}\%$) have outer giant companions. 
Whereas, for the `early-arrived' hot Jupiters, only 1/12 ($8.3^{+7.6}_{-7.0}\%$) has an outer giant companion. 
It is worth noting that the sample size is currently very small and thus, the observed difference in the fraction of hot Jupiters with outer giant planets necessitates further verification with a larger sample.

%Figure \ref{figConstraintssummary} summarizes the main constraints on the origin of hot Jupiters derived in this work.
%Based on the temporal evolution of hot Jupiters, we simultaneous put constraints on the stellar tidal quality factors and the relative importance of the `Early' and `Late' models to form hot Jupiters. 
%The observed distributions of stellar obliquity and $\rm [Fe/H]$ of provide additional evidence supporting the secular chaos as the formation mechanism of `late-arrived' hot Jupiters.

\subsubsection*{3.5 Evaluation on $\log Q^{'}_{*}$ from the observed number of hot Jupiters with orbital decay.}
Previous studies have detected hints of the orbital decay of hot Jupiters through monitoring of their transit-time variations (TTVs) \citep{2020ApJ...888L...5Y,2020AJ....159..150P}.
Recently, by
combining the high-precision 2 minute cadence transit data
provided by TESS with archival data from previous work, Wang et al. (2024) identified 11 candidates with decreasing orbital periods that also passed the leave-one-out cross validation test from 326 hot Jupiters \citep{2024ApJS..270...14W}.
Out of the 11 candidates, only the orbital decay of WASP 12 b has been confirmed \citep{2020ApJ...888L...5Y,2021AJ....161...72T}.
For the other 10 candidates, the possibility that mechanisms other than tidal interactions (e.g., apsidal precession \citep{2009ApJ...698.1778R}, the Rømer effect \citep{2019AJ....157..217B}, and the Applegate effect\citep{2010MNRAS.405.2037W}) may contribute to the variations in orbital periods cannot be ruled out.
Therefore, 1 and 11 can be taken as the lower and upper limits of the number of hot Jupiters with observable orbital decay.

Under the equilibrium tide theory, the rate of decay in the orbital semi-major axis or period of hot Jupiters is strongly correlated with the stellar tidal quality factor $\log Q^{'}_{*}$. Consequently, the number of hot Jupiters with detectable orbital period decay depending on $\log Q^{'}_{*}$ under the current observational precision. 
Therefore, comparing the theoretically expected numbers with observations can validate the constraints on $\log Q^{'}_{*}$ derived from our work and previous studies.

For the 326 hot Jupiters in Wang et al. (2024) \citep{2024ApJS..270...14W}, we calculate the theoretical prediction of their decay rate of orbital due to tidal interaction.
For the stellar tidal quality factor, we examine three different distributions:
\begin{enumerate}
    \item Results from this work: $\log Q^{'}_{*} = 5.7^{+0.4}_{-0.3}$.
    \item Results from WASP-12 system \citep{2021AJ....161...72T}: 
    $\log Q^{'}_{*} = 5.14 \pm 0.05$.
    \item Results from previous assemble studies: $\log Q^{'}_{*} \sim 6-9$. 
    Here we set $\log Q^{'}_{*}$ as uniformly distributed in $6-9$. 
\end{enumerate}

We adopt the $3\sigma$ errorbars of the observed period change rate as the detection threshold (as shown in Fig. 9 of \citep{2024ApJS..270...14W}).
Hot Jupiters with calculated $\dot{P}$ values exceeding the detection limit are considered to exhibit detectable orbital decay and we count the expected numbers $N^{\rm detectable}_{\rm decay}$ for the above three $\log Q^{'}_{*}$ distributions.
To consider the effect of uncertainties and obtain the confidence level, we resample the data (i.e., stellar/planetary parameters, $Q^{'}_{*}$, detection limit) considering the uncertainties  for 1,000 times and repeat the above procedure.
Then we count the numbers of hot Jupiters with detectable orbital decay $N^{\rm detectable}_{\rm decay}$ by taking different $\log Q^{'}_{*}$ distributions for each set of resampled data.
The 1-$\sigma$ intervals of $N^{\rm detectable}_{\rm decay}$ are taken as their $50 \pm 34.1$ percentiles.

Figure \ref{figNumberofdecayQ} displays the probability density distributions for the absolute decay rates $|\dot{P}|$ of orbital period (top panel) and expected numbers of hot Jupiters with detectable orbital decay (bottom panel).
As can be seen, a smaller $Q^{}_{*}$ leads to a greater $|\dot{P}|$.
Consequently, more hot Jupiters can exceed the detection threshold, making a larger numbers with detectable orbital decay. 
When taking a $\log Q^{'}_*$ as the WASP 12 system, $N^{\rm detectable}_{\rm decay} = 15 \pm 3$, which is larger than the observed lower and upper limits with confidence level of $>99.9\%$ and $92.1\%$, respectively.
When taking a $\log Q^{'}_*$ as previous ensemble studies ($\sim 6-9$), the predicted $|\dot{P}|$ will be too small to be identified as detectable.
Out of 1000 resampled data, only 75 cases resulted in detectable orbital decays of hot Jupiters.
Whereas, when taking our derived results $\log Q^{'}_* = 5.7^{+0.4}_{-0.3}$, the expected $N^{\rm detectable}_{\rm decay} = 4 \pm 2$. 
Out of 1000 resampled data, the resulted $N^{\rm detectable}_{\rm decay}$  fall within the upper and lower limits for 995 times.
Therefore, compared to previous individual and ensemble studies, our derived constraint on the stellar tidal factor aligns better with the observations.

{\subsection*{Data availability} 

\small The data were collected and processed from publicly available datasets and the selection procedures are included in the Main text and Methods.
The data that support the findings of this study are available upon request to the corresponding author: Ji-Wei Xie (jwxie@nju.edu.cn).}

\subsection*{Acknowledgements}

\small We thank Subo Dong and Wei Zhu for helpful discussions and suggestions.
LAMOST is operated and managed by the National Astronomical Observatories, CAS supported by the Chinese NDRC. 
This research has made use of the NASA Exoplanet Archive, which is operated by the California Institute of Technology, under contract with the National Aeronautics and Space Administration under the Exoplanet Exploration Program. 
This paper makes use of data from the first public release of the WASP data as provided by the WASP consortium.

This work is supported by the National Key R\&D Program of China (2024YFA1611803) and the National Natural Science Foundation of China (NSFC; grant No. 12273011, 12150009, 12403071).
We also acknowledge the science research grants from the China Manned Space Project with NO.CMS-CSST-2021-B12. 
J.-W.X. also acknowledges the support from the National Youth Talent Support Program.
D.-C.C. also acknowledges the Cultivation project for LAMOST Scientific Payoff, Research Achievement of CAMS-CAS and the fellowship of Chinese postdoctoral science foundation (2022M711566). 

{\color{black}
\subsection*{Author Contributions Statement}

\small J.-W.X. conceived the project and designed the research; D.-C.C. led the data analyses and numerical simulations; D.-C.C. and J.-W.X. analyzed the results and drafted the manuscript; and all authors contributed to discussing the results, editing, and revising the manuscript.

\subsection*{Competing Interests Statement}

\small The authors declare no competing interests as defined by Nature Portfolio, or other interests that might be perceived to influence the results and/or discussion reported in this paper.}

\newpage

\begin{figure}[!t]
\centering
\includegraphics[width=0.9\textwidth]{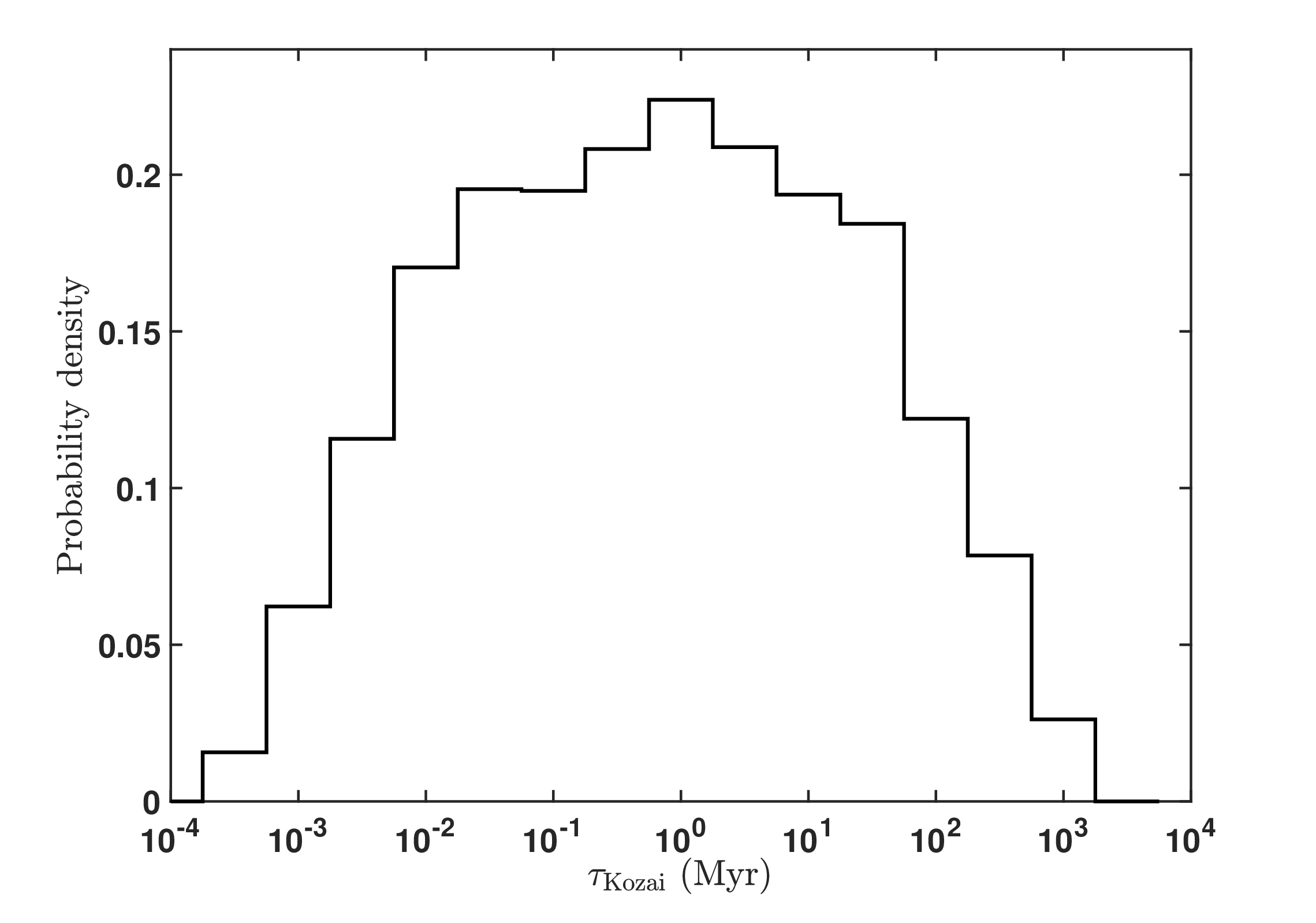}
\caption{\textbf{The probability density distribution for the typical timescale of planet-star Kozai cycles.}
To overcome general relativity pericenter precession, we only keep these Kozai–Lidov oscillations with timescales shorter than the pericenter precession timescales.
\label{figKozaitime}}
\end{figure}

\begin{figure}[!h]
\centering
\includegraphics[width=0.9\textwidth]{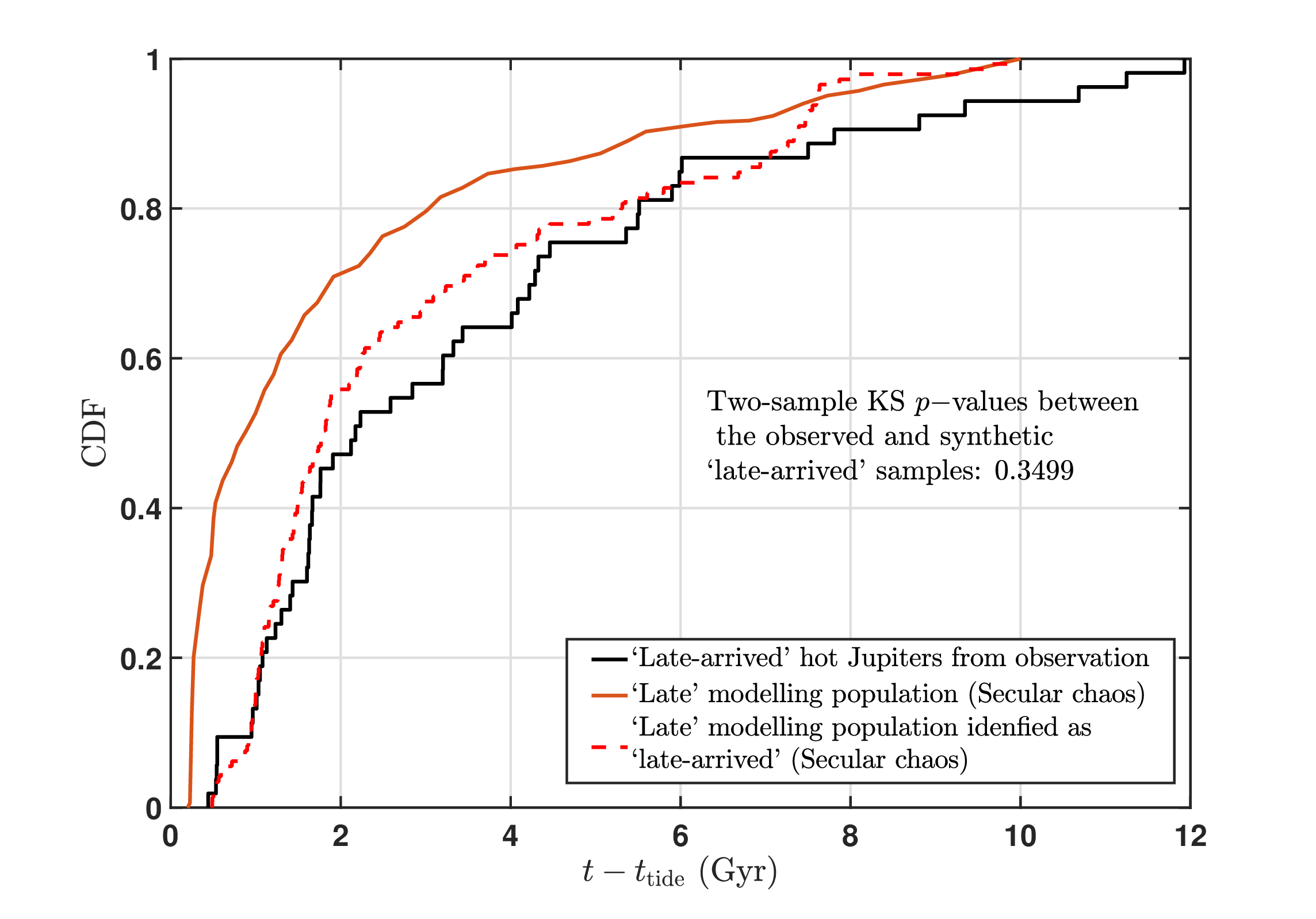}
\caption{\textbf{The cumulative distribution of the difference between the kinematic ages and tidal evolution timescales $t-t_{\rm tide}$.}
The results from the observed  and synthetic hot Jupiter sample that can be classified as `Late-arrived' are plotted as solid black line and dashed red line, respectively.
We print the two-sample KS test $p-$value.
The solid orange line denotes the arrival timescale of hot Jupiters via the secular models derived from simulations \citep{2017MNRAS.464..688H}.
\label{figLate_arrive_time}}
\end{figure}

\begin{figure}[!t]
\centering
\includegraphics[width=0.75\textwidth]{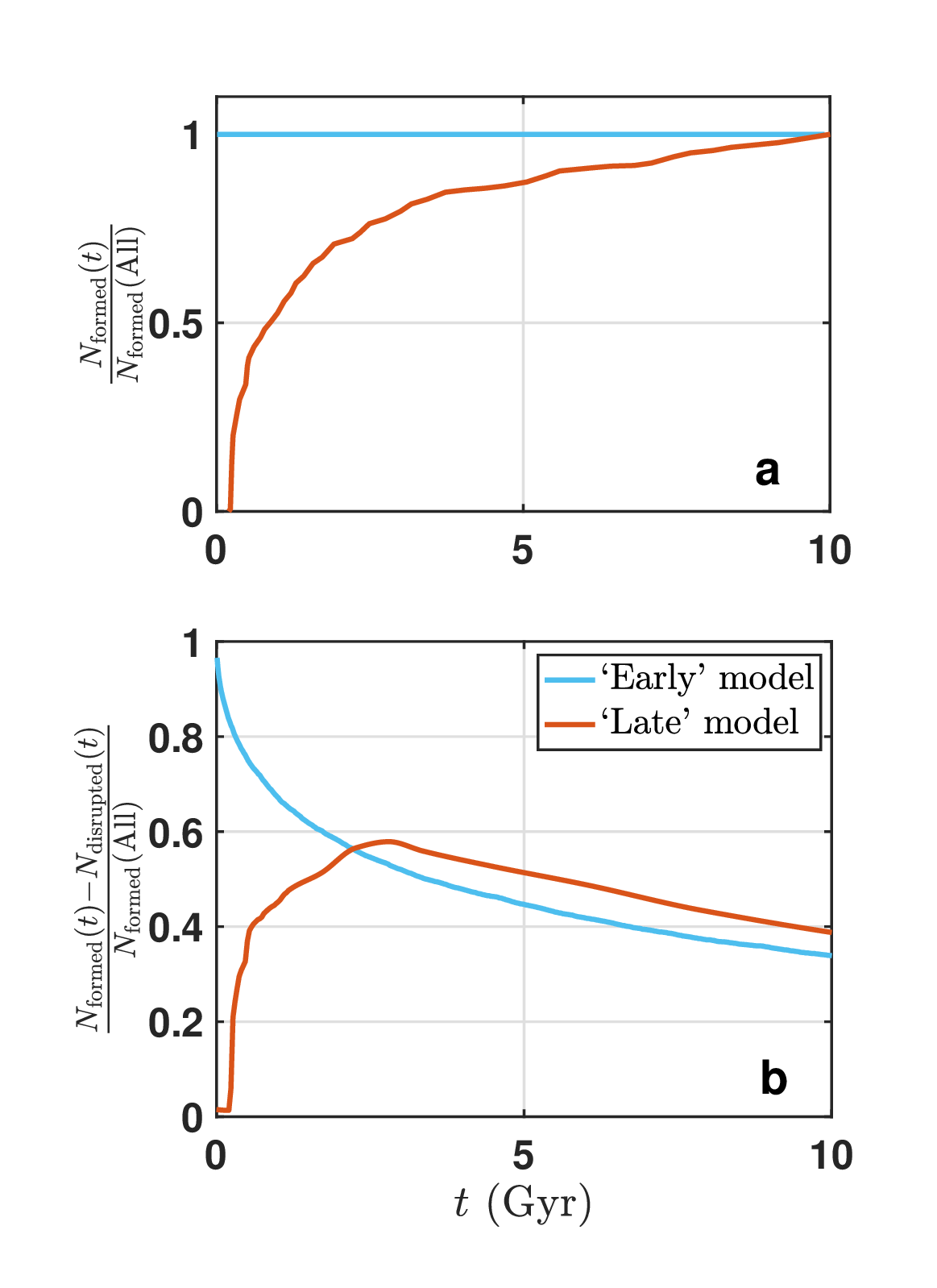}
\caption{\textbf{The evolutionary patterns of hot Jupiters formed via different origin models.}
Top panels: The number ratio of the formed hot Jupiters before $t$ over the formed hot Jupiters of all times as a function of age for different origin models.
The green and origin lines represent
`Early 'model and `Late' model, respectively.
Bottom panels: The number ratio of the left (formed $-$ tidally disrupted)  hot Jupiters before $t$ over the formed hot Jupiters of all times as a function of age for different origin models.
The initial conditions are set as the standard case, i.e., the a-distribution of hot Jupiters is set as the results inferred from Kepler data \citep{2016A&A...587A..64S} and the initial planetary mass is set as that of cold Jupiters. 
The modified stellar tidal quality factor $Q^{'}_{*} = 10^{6}$.
\label{figHJTSofDM}}
\end{figure}

\begin{figure*}[!t]
\centering
\includegraphics[width=\textwidth]{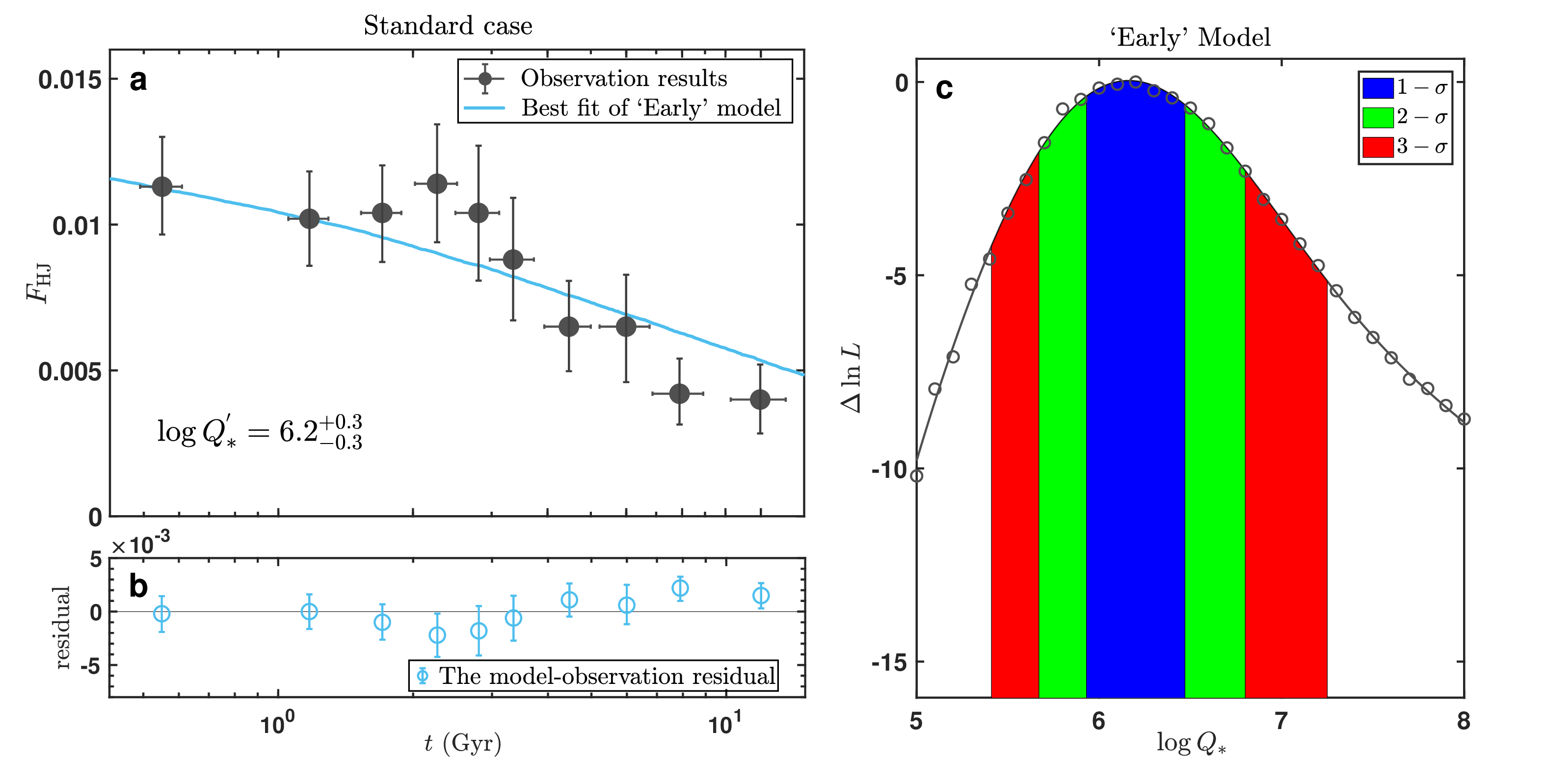}
\caption{\textbf{Fitting the observed age-frequency relation of hot Jupiters with the `Early' model for the standard case.}
The comparison between $F_{\rm HJ}$ obtained from observation data and theoretical simulation. 
\textit{Here the hot Jupiters are formed/migrated all by `Early' model (i.e., $f_{\rm Late}=0$).}
Left-Top panel: 
The observation data is plotted as solid black points and line segments denote the 1-$\sigma$ interval. 
The solid line denotes the best match.
The initial conditions of simulations are as standard case.
The modified stellar tidal quality factor $Q^{'}_{*}$ ranges from $10^4$ to $10^9$.
Left-Bottom panel:
The residual of the best match of numerical simulation to the observational results.
Right panel: Relative likelihood in logarithm as a function of $Q^{'}_{*}$. The blue, green, and red hatched regions indicate the 1-$\sigma$, 2-$\sigma$, and 3-$\sigma$ confidence levels.
\label{figFHJmodel1a2016MCJ}}
\end{figure*}

\begin{figure*}[!t]
\centering
\includegraphics[width=\textwidth]{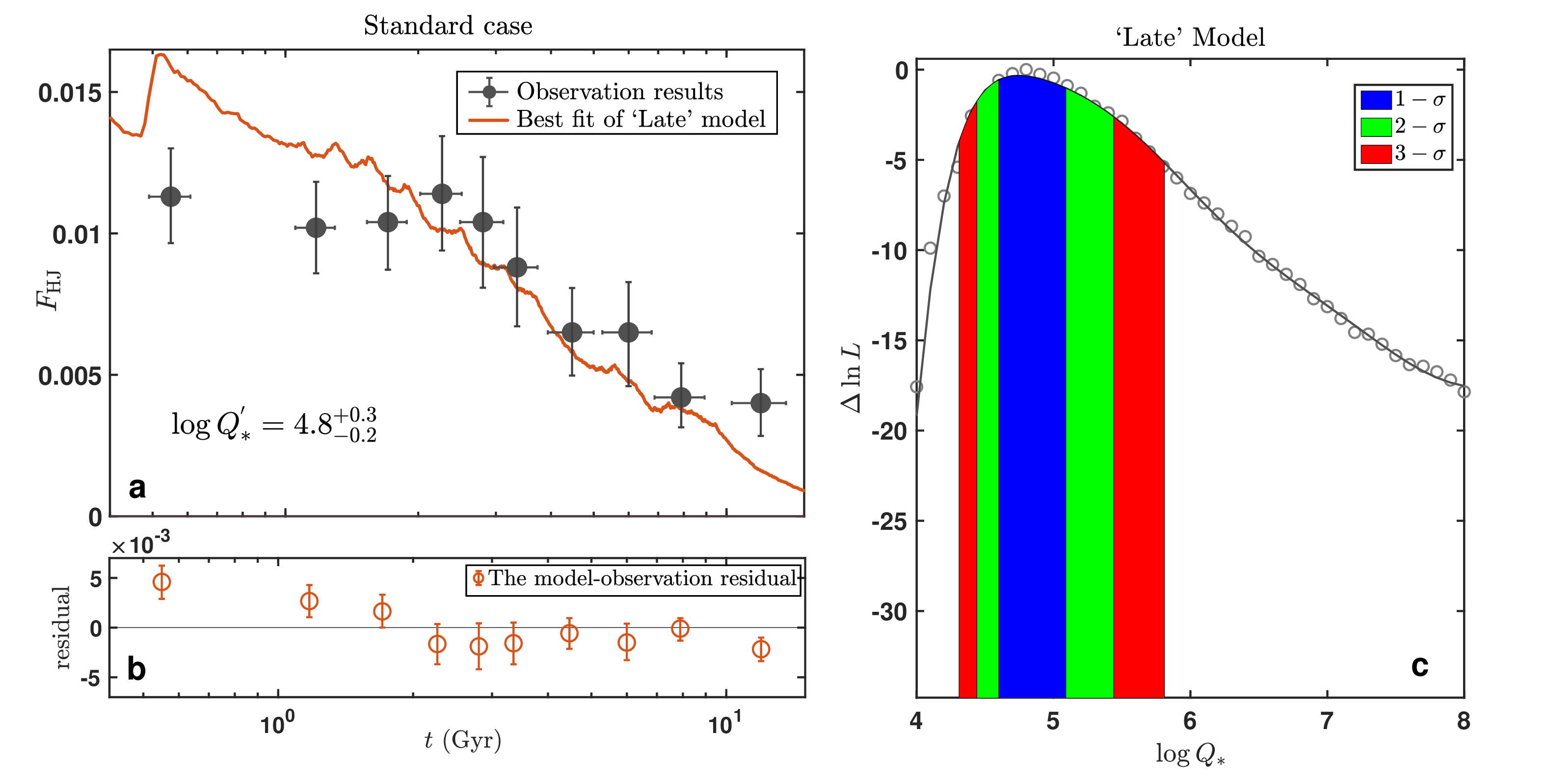}
\caption{\textbf{Fitting the observed age-frequency relation of hot Jupiters with the `Late' model for the standard case.}
Similar to Extended Data Fig. \ref{figFHJmodel1a2016MCJ} but
\textit{here the hot Jupiters are formed/migrated all by `Late' model (i.e., $f_{\rm Late}=1$)}.
\label{figFHJmodel22016MCJ}}
\end{figure*}

\begin{figure}[!t]
\centering
\includegraphics[width=\textwidth]{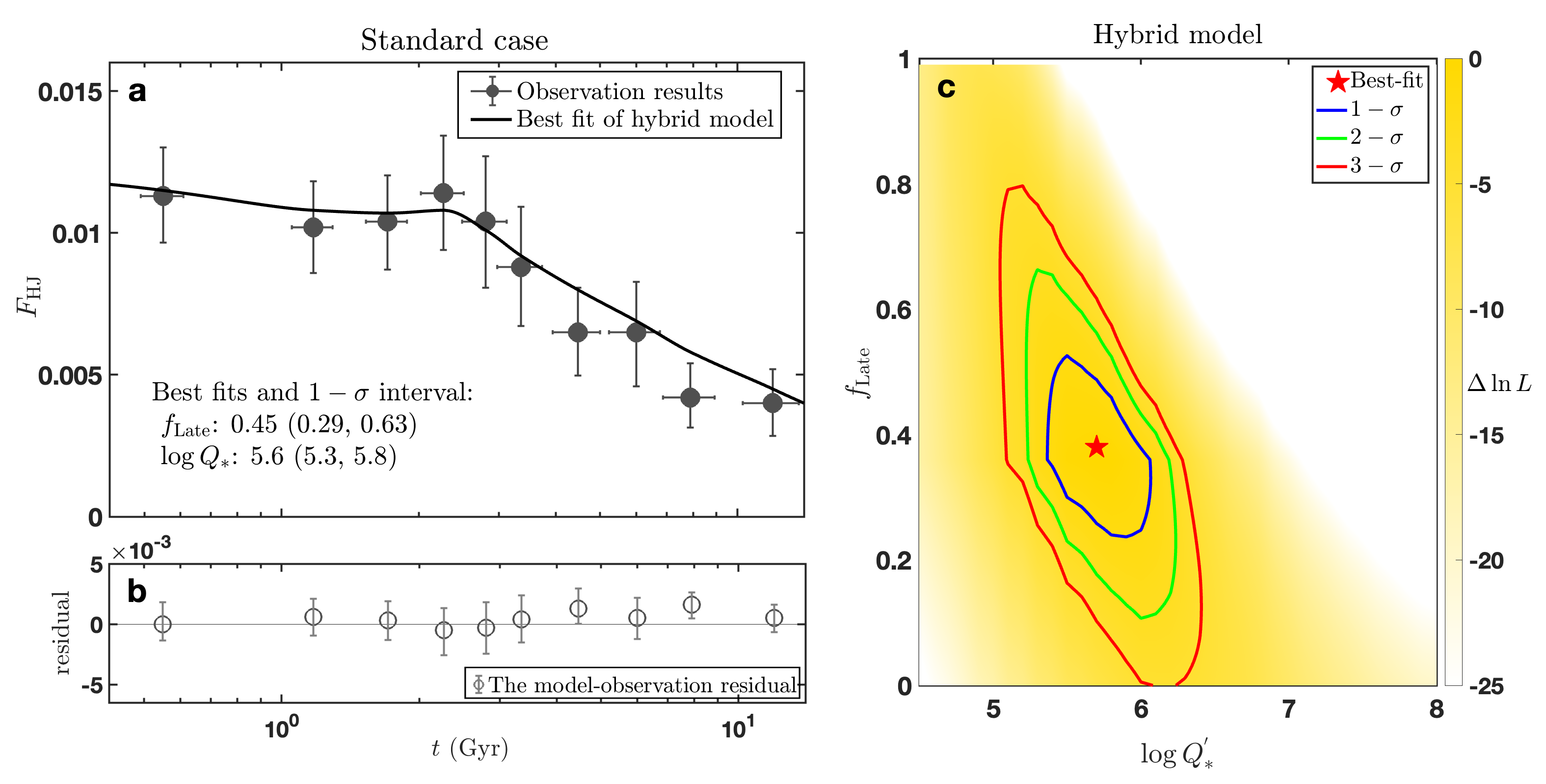}
\caption{\textbf{Fitting the observed age-frequency relation of hot Jupiters with the Hybrid model for the standard case.}
Similar to Extended Data Fig. \ref{figFHJmodel1a2016MCJ} but
\textit{here the hot Jupiters are formed/migrated by two origin mechanisms: `Early' model plus `Late' model.}
$F_{\rm Late}$ is the fraction of hot Jupiters formed by the `Late' model.
The right panel displays the relative likelihood in logarithm as a function of $Q^{'}_{*}$ and $f_{\rm Late}$.
\label{figFHJmodel12a2016MCJ}}
\end{figure}

\begin{figure}[!t]
\centering
\includegraphics[width=0.85\linewidth]{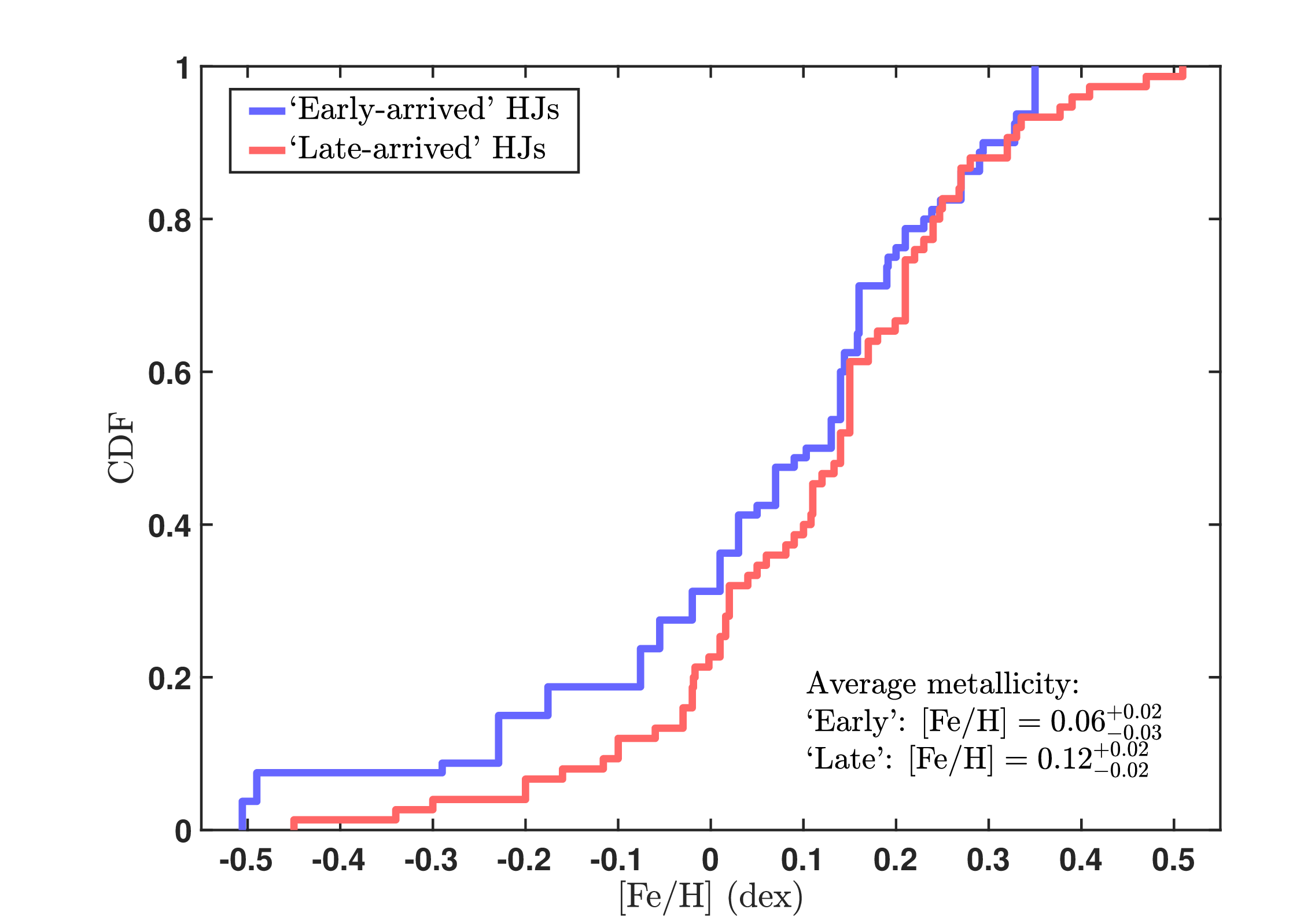}
\caption{
\textbf{The cumulative distributions of stellar metallicity $\rm [Fe/H]$.} 
The red and blue lines denote the `late-arrived’ hot Jupiter hosts and the `early-arrived’ hot Jupiter hosts neighboring to the `late-arrived’ hot Jupiter hosts in stellar mass, radius and $TD/D$.
The average metallicities and corresponding uncertainties from bootstrapping are printed in the top-right corner.}
\label{figFeHconstraint}
\end{figure}

\clearpage

\begin{table*}[!t]
\centering
\renewcommand\arraystretch{1.25}
\caption{\textbf{The constraints of the value of $Q^{'}_{*}$ from previous observational studies.}}
{\footnotesize
\label{tab:Qparamodel}
\begin{tabular}{lcccl} \hline
Literature Source   &  $\log Q^{'}_{*}$ & Sample & Method & Comments  \\ \hline
Meibom\&Mathieu (2005)  &  $\sim 5.5$  & Binaries  & ensemble statistics  & Distribution of orbital  \\
 & & & & $e-P$ in M35\\ 
 %may be dependent on the mass of the secondary \\ 
Milliman et al. (2014)  & $\sim 5.5$ & Binaries & ensemble statistics  & Tidal circularization of  \\
 & & & & binary orbits in NGC 6819 \\ \hline %may be dependent on the mass of the secondary \\
Hansen (2012)  &  $7-9$ & Planet hosts & ensemble studies  & eccentricity distribution of hot  \\
& & & & Jupiters based on stars near \\
& & & & transition between convective  \\  & & & & and radiative envelopes \\
%which may have less efficient dissipation
% Rely on the isochrone ages of host stars, which suffer  \\
% &  &  &  & relatively large errors for main-sequence stars \\
%Did not consider the hot Jupiter formed at late stage \\ 
Jackson et al. (2008) &  $\sim 4-8$ & Hot Jupiters  & ensemble statistics & Distributions of orbital $a-e$  \\ 
& & & &  (loose constraints) \\
Bonomo et al. (2017)   & $\gtrsim 6-7$ & Giant planets & Ensemble statistics & Distribution of orbital $e$ \\
 & & & & versus distance \\
Collier Cameron\&Jardine (2018)  & $8.3 \pm 0.14$ & Hot Jupiters & ensemble statistics & Distribution of orbital distance \\
%Did not consider the hot Jupiter formed at late stage \\
Hamer\&Schlaufman (2019)  & $\lesssim 7$ & hot Jupiters & ensemble statistics & Assuming  hot Jupiters are  \\
 &  & &  & tidally disrupted within \\
 & & & & main-sequence stage \\
%Derived from age difference of hot Jupiter hosts to cold Jupiter hosts and only obtain a upper limit 
Hamer\&Schlaufman (2020) & $\gtrsim 7$ & Planets & ensemble statistics &  Assuming USPs survive \\
& & & &  the main-sequence stage  \\ 
Millholland et al. (2025)  & $\sim 5.5-7.0$ & hot Jupiters & ensemble statistics & Distribution of insprial \\
& & & & timescales (0.001-0.3 Gyr) of \\ 
& & & & very-short-period hot Jupiters \\
\hline
Nielsen et al. (2020)  & $6-9$ & Hot Jupiter & individual measurement & Assuming HIP 65Ab' insprial \\ 
 & & & & timescale of 80 Myr to Gyr \\ \hline
%and only obtain a lower limit
Penev et al. (2018)  & $5-7$ & Hot Jupiter hosts & individual measurement & Spin-up of host stars\\ 
%Assuming that tides have managed to spin up the star\\
% & & & & to the observed rate within the age of the system \\ 
Labadie-Bartz et al. (2019)  &  $\sim 6.2$ & hot Jupiter hosts & individual measurement & Spin-up of host stars \\ \hline
%Individual may not represent typical value of the population \\
Yee et al. (2020)  &  $5.24 \pm 0.03$ & Hot Jupiter & individual measurement & Orbital decay of WASP-12b \\
%Individual may not represent typical value of the population\\
Maciejewski et al. (2020)  & $\gtrsim 6.6$ & Hot Jupiter & individual measurement  & Orbital decay of WASP-18b\\
Patra et al. (2020)  & $\gtrsim 5.0$ & Hot Jupiter & individual measurement & Orbital decay of 11 HJs \\
Davoudi et al. (2021)  & $\gtrsim 5.6 \pm 0.1$ & Hot Jupiter & individual measurement & Orbital decay of WASP-43 b \\
Turner et al. (2021)  & $5.14 \pm 0.05$ & Hot Jupiter & individual measurement & Orbital decay of WASP-12b \\
Turner et al. (2022)  & $4.71^{+0.07}_{-0.09}$ & Hot Jupiter & individual measurement & Orbital decay of WASP-4 b \\
Bai et al. (2022)  & $5.09^{+0.05}_{-0.06}$ & Hot Jupiter & individual measurement & Orbital decay of WASP-12b \\
Akinsanmi et al. (2024)  & $5.23^{+0.03}_{-0.04}$ & Hot Jupiter & individual measurement & Orbital decay of WASP-12b \\

%Individual may not represent typical value of the population \\
%Efroimsky et al. (2021) \citep{2021arXiv211108273E} & $\sim 4.67$ & hot Jupiter WASP-12 b & individual measurement & Individual may not represent typical value of the population \\
\hline
\end{tabular}}
\end{table*}

\begin{table*}[!t]
\renewcommand\arraystretch{1.25}
\centering
\caption{\textbf{The fitting parameters for theoretical models under various conditions for different observational data.}}
{\footnotesize
\label{tab:FQAICdifferentmodels}
\begin{tabular}{c|c|cc|cc|ccc} \hline
  & Initial  &  \multicolumn{2}{c}{`Early' model ($f_{\rm Late}=0$)}   &  \multicolumn{2}{c}{`Late' model  ($f_{\rm Late}=1$)} & \multicolumn{3}{c}{Hybrid Model }\\
    & condition  & $\log Q^{'}_{*}$  & AIC  & $\log Q^{'}_{*}$  & AIC & $f_{\rm Late}$ &  $\log Q^{'}_{*}$ & AIC \\ \hline
    \multicolumn{9}{c}{\S~\ref{sec.theory.comparison}: fitting for the whole sample in the standard case} \\ \hline
    Whole sample & Standard case &  $6.2^{+0.3}_{-0.2}$ & -106.9 
    & {$4.8^{+0.3}_{-0.2}$} & {-80.6} &  {$0.38^{+0.16}_{-0.14}$} &  {$5.7^{+0.3}_{-0.3}$} & {-116.5} \\ \hline
    \multicolumn{9}{c}{fitting for the whole sample for different initial planetary mass} \\ \hline
    Whole sample & Case 1 &  $5.9^{+0.3}_{-0.2}$ & -105.4  &  {$4.8^{+0.3}_{-0.2}$} & {-80.6}  & $0.41^{+0.15}_{-0.12}$ &  $5.4^{+0.2}_{-0.3}$ & -115.0 \\ \hline
    \multicolumn{9}{c}{fitting for the whole sample for different initial orbital period} \\ \hline
    Whole sample & Case 2 &  $6.2^{+0.2}_{-0.2}$ & -104.8 & $4.8^{+0.3}_{-0.2}$ & -80.6 & $0.30^{+0.16}_{-0.12}$  & $5.6^{+0.3}_{-0.4}$ & -113.2 \\
    Whole sample & Case 3 &  $6.4^{+0.3}_{-0.2}$ & -104.6  &  {$4.8^{+0.3}_{-0.2}$} & {-80.6}  & $0.35^{+0.16}_{-0.13}$ &  $5.8^{+0.4}_{-0.4}$ & -115.3 \\
    Whole sample & Case 4 &  $6.5^{+0.4}_{-0.3}$ & -105.5 
    & {$4.8^{+0.3}_{-0.2}$} & {-80.6} &  {$0.45^{+0.15}_{-0.13}$} &  {$5.7^{+0.4}_{-0.3}$} & {-112.3} \\
    \hline
\multicolumn{9}{c}{fitting for the three subsamples in the standard case} \\ \hline
RV subsample & Standard case &  $5.6^{+0.4}_{-0.4}$ & -83.2 & $4.5^{+0.4}_{-0.5}$ & -79.9 & $0.21^{+0.42}_{-0.21}$  & $5.3^{+0.4}_{-0.5}$ & -86.3 \\
ST subsample & Standard case &  $6.0^{+0.5}_{-0.5}$ & -86.9  &  {$4.9^{+0.4}_{-0.4}$} & {-85.3}  & $0.44^{+0.39}_{-0.43}$ &  $5.8^{+0.4}_{-0.7}$ & -91.8 \\
GT subsample & Standard case &  $6.0^{+0.4}_{-0.4}$ & -102.5
& {$4.9^{+0.4}_{-0.3}$} & {-89.8} &  {$0.41^{+0.14}_{-0.14}$} &  {$5.7^{+0.6}_{-0.5}$} & {-111.4} \\
\hline 
\end{tabular}}
\flushleft
{\small Standard case: (a) `Early' model: initial mass and radius distributions as observed warm/cold Jupiters; initial period distribution as the result from Kepler data \citep{2016A&A...587A..64S}; (b) `Late' model: initial distributions referring to Hamers et al. 2017 \citep{2017MNRAS.464..688H}. \\
Case 1: Similar to Standard case but the mass distribution of `Early' model as observed hot Jupiters. \\
Case 2-4: Similar to Standard case but the period distribution of `Early model' is taken as uniform in logarithmic space, the disk migration component, and high-e migration component of Nelson et al. (2017), respectively. 
}
\end{table*}

\begin{table*}[!t]
\renewcommand\arraystretch{1.25}
\centering
\caption{\textbf{The KS $p-$values of the obliquity distributions between the observed hot Jupiter systems and theoretical predictions (Fig. \ref{figObliquityFeHconstraints}).}}
{\footnotesize
\label{tab:pvalue_obliquity}
\begin{tabular}{l|cccccc} \hline
 & Planet scattering & Resonance crossing & Secular chaos & Planetary Kozai & Stellar Kozai & Aligned distribution \\ \hline
`early-arrived' & \textbf{0.808} & 0.317 & $8.0 \times 10^{-3}$ & $8.1 \times 10^{-4}$ & $4.2 \times 10^{-3}$ & 0.719\\ \hline
`late-arrived' & 0.032 & 0.045 & \textbf{0.229} & 0.0138 & $5.1 \times 10^{-3}$ & $2.5 \times 10^{-3}$ \\
 %& Planet scattering & Resonance crossing & Secular chaos & Stellar Kozai & Planetary Kozai \\ \hline
%`early-arrived' & 0.808 & 0.317 & $8.0 \times 10^{-3}$ & $3.1 \times 10^{-4}$ & $8.1 \times 10^{-4}$ \\ \hline
%`late-arrived' & 0.032 & 0.045 & 0.229 & $4.1 \times 10^{-3}$ & 0.0138 \\
\hline 
\end{tabular}}
\end{table*}
}

\clearpage
\beginSI
\begin{center}
{ \LARGE  Supplementary Information}\\[0.5cm]
\end{center}

\subsubsection*{Influence of the initial mass distribution of hot Jupiters}
From theory, the initial mass distribution of hot Jupiters is expected to be similar to that of warm/cold Jupiters if disk migration is dominating. In contrast, under high-eccentricity migration (e.g., secular chaos), hot Jupiters tend to have smaller mass distributions, as smaller planets are more easily driven into high-eccentricity orbits by interactions with other planets \citep{2011ApJ...735..109W,2015ApJ...805...75P}.
Therefore, the intrinsic initial mass distribution of hot Jupiters is hard to be estimated accurately and is expected to lie between those of observed hot and warm/cold Jupiters.

\begin{figure}[!t]
\centering
\includegraphics[width=\textwidth]{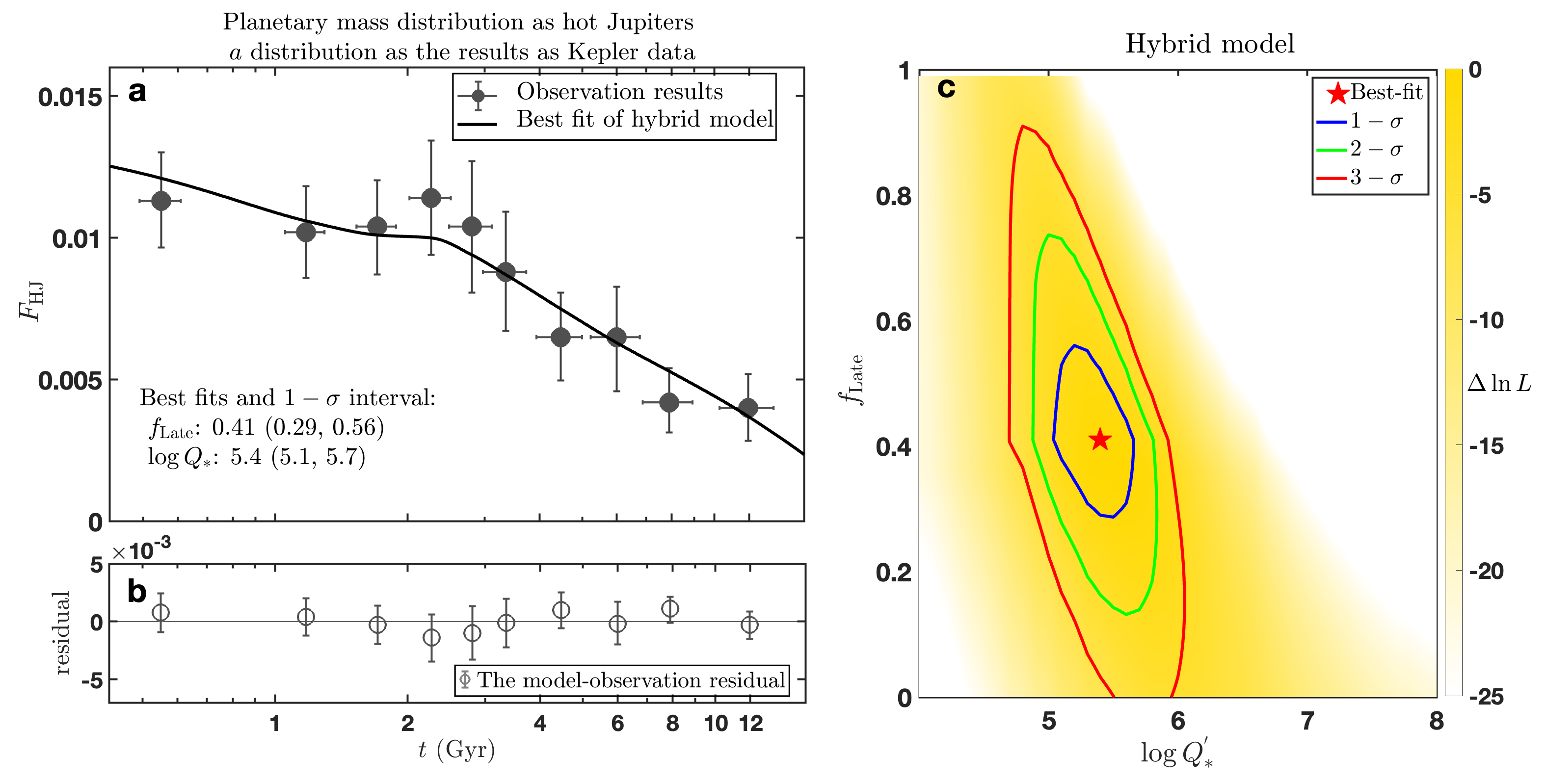}
\caption{{Similar to Extended Data Fig. 6 but the initial planetary mass in simulations is set as the median value of observed hot Jupiters.}
\label{figFHJmodel12a2016MHJ}}
\end{figure}

To evaluate the impact of the initial mass of hot Jupiters, we conduct simulations using the similar initial conditions as \S~2.3 but the planetary mass and radius are set as the distribution of the observed hot Jupiters.
Figure \ref{figFHJmodel12a2016MHJ}  show the best match of the simulation results comparing to the observations.
As shown in Extended Data Tab. 2, the best fits (1-$\sigma$ intervals) of $f_{\rm Late}$ are nearly identical for different initial mass distributions.
For the stellar tidal quality factor, the difference in the initial median planetary mass induce a systematic difference of $\sim 0.3$ in $\log Q^{'}_{*}$ to maintain a similar $\frac{M_{\rm p}}{Q^{'}_{*}}$.
Nevertheless, this systematic difference in $Q^{'}_{*}$ is statistically insignificant considering the uncertainties.

\subsubsection*{Influence of the initial $a-$distribution of hot Jupiters}
As shown in Equation S14, the decay rate and the inspiral timescale of hot Jupiters strongly depend on the initial distribution of orbital period.
Therefore, different assumptions of the initial period distribution may produce different numerical simulation results and affect the constraints on the $Q^{'}_{*}$ and $f_{\rm Late}$.

To quantify the influence of initial period distribution on the $Q^{'}_{*}$ and $f_{\rm Late}$, we perform numerical simulations by adopting different period distributions as following:
\begin{enumerate}
    \item the uniform distribution in logarithmic space inferred from early Doppler survey \citep{2007ARA&A..45..397U}; 
    \item the distribution as the disk migration component derived from Nelson et al. (2017) \citep{2017AJ....154..106N}, 
    $\frac{{\rm d}N}{{\rm d}x} \propto x^{\gamma-1}, \ x \equiv \frac{a}{a_{\rm R}}, \ \gamma = -0.04$;
    \item the distribution as the high-eccentricity migration component derived from Nelson et al. (2017) \citep{2017AJ....154..106N} which have a $\gamma = -1.38$.
\end{enumerate}
Then we compare the simulation with the observational results (as shown in Figure S2-S4).
The fitting parameters by adopting different period distributions are summarized in Extended Data Tab. 2.
As can be seen, the best fits of the $Q^{'}_{*}$ are consistent within 1-$\sigma$ errorbars for different initial period distributions. 

\subsubsection*{Comparisons with results from different subsamples}
\label{sec.subsample.res}
In this subsection, we analyze the temporal evolution of hot Jupiter frequency by using these three sub-samples separately with the same procedures as the whole sample. 
The results are as follows:
\begin{enumerate}
    \item RV subsample. 
    In the top panel of Figure \ref{figfHJAgeRV}, we plot the frequency of hot Jupiters from the RV subsample as a function of age.
    We then fit $F_{\rm HJ}$ with the constant model and the exponential model. 
    The resulting difference in AIC score is 8.6, demonstrating that $F_{\rm HJ}$ is generally declining with age.
    {We also note that the declining trend is relatively mild before $\sim 3$ Gyr and becomes steeper at late stage.}
    %We also perform the sequences and reversals test.
    %The resulting $p-$value is 0.0625, suggesting a weak non-monotonicity.
    We divide the RV hot Jupiter subsample into two populations (i.e., `late-arrived' and `early-arrived') and then obtain their frequencies as a function of age.
    {As shown in the middle panel of Figure \ref{figfHJAgeRV}, the frequency of `late-arrived' RV hot Jupiter population firstly increases and then decreases, forming a ridge at $\sim 2-4$ Gyr.
    After removing these `late-arrived' hot Jupiters, the ridge at $\sim 2$ Gyr disappears and the declining trend of $F_{\rm HJ}-t$ trend becomes more significant (with a larger $\rm \Delta AIC$ of 14.3).}

    We then perform numerical simulations for the tidal decays of hot Jupiters formed via different origins ($f_{\rm Late}$) with various stellar tidal quality factor ($Q^{'}_{*}$).
    %For the stellar mass and radius, we adopt the same distributions as our RV hot Jupiter host star. 
    %The initial condition of planetary properties is set as the standard case.
    Figure \ref{figFHJmodel12a2016MCJRV} shows the best match of $F_{\rm HJ}$ from the hybrid model comparing to the observational results.
    The Hybrid model is preferred comparing to the single `Early' model and single `Late' model with $\rm \Delta AIC$ of 3.1 and 6.4, corresponding to confidence levels of $\sim 1-\sigma$ and $\sim 2-\sigma$, respectively.
    The best-fits for the $f_{\rm Late}$ and $Q^{'}_*$ are $21^{+42}_{-21}\%$ and $5.3^{+0.4}_{-0.5}$, respectively. \\
    
    \item ST subsample. 
    Figure \ref{figfHJAgeST} displays the frequency of hot Jupiters derived from the ST subsample.
    {Similar to the results from the entire sample, $F_{\rm HJ}$ shows a ridge at $\sim 2-3$ Gyr and the declining trend is more mild at the early stage.}
    %We perform the sequences and reversals test
    %The resulting $p-$value is 0.0313, suggesting a non-monotonically declining trend.
    {For the two populations, the frequency of `late-arrived' RV hot Jupiter population firstly increases and then decrease, forming a ridge at $\sim 2-3$ Gyr (shown in the middle panel of Figure \ref{figfHJAgeST}).
    After removing the contribution of `late-arrived' population, the ridge at $\sim 2$ Gyr disappears and the frequency of the `early-arrived' hot Jupiters shows similar declining trends before and after $\sim 2$ Gyr (shown in the bottom panel of Figure \ref{figfHJAgeST}).}

    Figure \ref{figFHJmodel12a2016MCJST} shows the best match of $F_{\rm HJ}$ from the hybrid model comparing to the observational results.
    As can be seen, the hybrid model is preferred comparing to the single `Early' model and single `Late' model with $\rm \Delta AIC$ of 4.9 and 6.5, respectively.
    The best-fits for the $f_{\rm Late}$ and $Q^{'}_*$ are $44^{+41}_{-34}\%$ and $5.7^{+0.5}_{-0.6}$, respectively. \\

    \item GT subsample. 
    The frequency of hot Jupiters derived from the ST subsample is generally declining with age (with a $\rm Delta AIC$ of 6.2).
    {Similar to the results from the entire sample, $F_{\rm HJ}$ shows a ridge at $\sim 2$ Gyr and the declining trend is more mild at the early stage.}
    %We also perform the sequences and reversals test, resulting a $p-$value of monotonicity of $1.2 \times 10^{-4}$, 
    We then divide the ST hot Jupiter subsample into two populations and investigate the temporal evolution of their frequencies.
    {As shown in the middle panel of Fig. \ref{figfHJAgeGT}, the frequency of `late-arrived' hot Jupiter population firstly increases and then decreases, forming a ridge at $\sim 2-4$ Gyr.
    After removing the contribution of `late-arrived' population, the ridge at $\sim 2$ Gyr disappears and the declining trends at early and late stages for the`early-arrived'  hot Jupiter population become well consistent.}

    Figure \ref{figFHJmodel12a2016MCJGT} shows the best match of $F_{\rm HJ}$ from the hybrid model comparing to the observational results.
    As can be seen, the hybrid model is preferred comparing to the single `Early' model and single `Late' model with $\rm \Delta AIC$ of 8.9 and 21.6, corresponding to confidence levels of $\gtrsim 2-\sigma$ and $\gtrsim 3-\sigma$ respectively.
    The best-fits for the $f_{\rm Late}$ and $Q^{'}_*$ are $41^{+14}_{-14}\%$ and $5.7^{+0.6}_{-0.5}$, respectively.
\end{enumerate}

The above analyses demonstrate that all the three subsamples show results similar to those of the whole sample.
The advantage of combining the three samples is that the results from whole sample have smaller uncertainties in $\log Q^{'}_{*}$ and $F_{\rm Late}$
%The two main sections of the supplement can be split up using headings.

% If your supplement is very short you might need to uncomment the following line to avoid
% layout problems with the figures and tables.
%\newpage

{
\subsubsection*{Comparisons with results using ages from other methods}
\label{sec.dis.otherages}
It is worth noting that the kinematic ages represent the average ages of a group of stars.
For a consistency check of the age-frequency relation of hot Jupiters derived from the kinematic method, we collect individual ages obtained through other methods and obtain the evolution trend of $F_{\rm HJ}$:
\begin{enumerate}
    \item Gyrochronology age. 
    Gyrochronology estimates the age of a main-sequence
    star based on its rotation period \citep{2010ApJ...722..222B,2010ApJ...721..675B}.
    By adopting the method provided by Spada \& Lanzafame \citep{2014ApJS..211...24M}, we calculate gyrochronology ages for 31,860 Kepler stars based on the rotation data from \citep{2020A&A...636A..76S}.
    Then we select Sun-like stars with $T_{\rm eff}$ in the range of $4700-6500$ K and remove potential binaries by excluding stars with re-normalized unit
    weight error (RUWE)$>1.2$, leaving 22,570 stars.
    For the planetary sample, we cross-match with Kepler DR25 catalog and yield 16 hot Jupiters.

    To obtain the age-frequency trend, we divide the stellar sample into 5 bins with equal size according to their gyrochronology age and calculate the frequencies of hot Jupiters by correcting the geometric effect/detection efficiency and eliminating the effect of $\rm [Fe/H]$ with same procedure described in \S~3.2 of Chen2023 \citep{2023PNAS..12004179C}.
    Figure \ref{figFHJAge_Rotation_Isochrone} shows the frequency of hot Jupiters as a function of gyrochronology age (pink points).
    {As can be seen, $F_{\rm HJ}$ remains relatively stable  when $t\lesssim 2$ Gyr and then declines sharply.}
    It is also worth noting that the ages of some stars in the last two bins may be underestimated due to stalled spin-down.
    %We also perform the sequences and reversals test and the resulted $p-$value is 0.125, suggesting a weak non-monotonicity. 

    \item Isochrone age.
    \citep{2020AJ....159..280B} provided isotropic ages for
    186,301 Kepler stars with a typical uncertainty of $\sim 56\%$ by fitting the isochrone grid.
    Here we select Sun-like stars with reliable age estimations by applying the following criteria: $\rm GOF>0.99$ \& $TAMS<20$ Gyr \& $t<14$ Gyr \& $\frac{{\rm err}\_t}{t}<1/3$, yielding 17665 Kepler stars.
    For the planetary sample, we cross-match with the Kepler DR25 catalog and obtained 10 hot Jupiters. 
    We then divided the stellar sample into 5 bins and derive the frequency of hot Jupiters of different age bins.
    As shown in Figure \ref{figFHJAge_Rotation_Isochrone} (pink points), $F_{\rm HJ}$ generally decreases when $t>2$ Gyr.
\end{enumerate}

Due to methodological limitations, the Gyrochronology and isochron method can only provide reliable ages for the younger and older ends, respectively. 
{By combining the Kepler sample with individual age estimations from two methods, the declining trend of $F_{\rm HJ}$ is mildly ($\gamma = 0.03^{+0.08}_{-0.10}$) when $t \lesssim 2$ Gyr and becomes steeper ($\gamma = -0.12^{+0.05}_{-0.15}$) at the late stage.}
%We perform the sequences and reversals test and the resulted $p-$ value is 0.0625, suggesting a moderate non-monotonicity.

The above analyses demonstrate that the age-frequency trend of hot Jupiters derived from individual age estimations is generally consistent with the kinematic results, further reinforcing the reliability of the broken pattern.

\subsubsection*{Influence of the dependence of $Q^{'}_{*} $ on stellar/planetary properties}
\label{sec.dis.Qsdependence}
In the above analyses, we fit the observed frequency-age trend of hot Jupiters to theoretical predictions, assuming a common tidal quality factor.
Some previous studies have suggested that $Q^{'}_{*}$ depend on factors such as stellar mass, age, evolutionary stage, and the orbital periods of hot Jupiters \citep{2017ApJ...849L..11W,2018AJ....155..165P,2020MNRAS.498.2270B}.
Thus, in this subsection, we explore the influence of the above dependences of $Q^{'}_{*}$ as follows:
\begin{enumerate}
    \item Stellar properties. 
    $Q^{'}$ may decrease at specific stellar evolutionary stages (e.g., subgiant) due to the change of stellar structures \citep[e.g., the increase of stellar radii and convective envelopes;][]{2009ApJ...705L..81V,2017ApJ...849L..11W}.
    In this work, the stellar sample has been restricted to Sun-like stars, i.e., main-sequence stars with $T_{\rm eff}$ in the range of 4700-6500 K. 
    In Figure \ref{figMass_Age_HJhosts}, we compare the mass distributions for host stars of different kinematic ages, which are statistically indistinguishable (with KS $p-$values $>0.1$).
    Moreover, $Q^{'}_{*}$ does not differ significantly with age during the main-sequence stage \citep[see fig. 3 of][]{2020MNRAS.498.2270B}.
    Therefore, in this work, it is reasonable to fit using a common tidal factor when considering only stellar properties.

    \item Planetary orbital period (tidal forcing period). 
    As suggested by the observational \citep{2018AJ....155..165P} and theoretical analyses \citep{2020MNRAS.498.2270B}, the stellar $Q^{'}_{*}$ may vary with the orbital period of hot Jupiters $P$, with an empirical formula,
    \begin{equation}
        Q^{'}_{*} (P_{\rm tide}) = {\rm max}[\frac{10^6}{P^{3.1}_{\rm tide}}, 10^5], 
    \label{eqQsPorb}
    \end{equation}
    
    \begin{equation}
        P_{\rm tide} \equiv \frac{1}{2(P^{-1}-P^{-1}_{\rm spin})},
    \end{equation}
    where $P_{\rm tide}$ and $P_{\rm spin}$ are the tidal force period and stellar spin period, respectively.
    In the top panel of Figure \ref{figQs_tin_P_Penev2018}, we show the derived $Q^{*}_{*}$ as a function of the orbital period $P$, where $P_{\rm spin}$ is set as a typical value of 18 days for Sun-like stars \citep{2010ApJ...722..222B,2014ApJS..211...24M}.
    As can be seen, $Q^{'}_{*}$ remains constant when $P \gtrsim 4$ days, while for $P \lesssim 3$ day, $Q^{'}_{*}$ increases with decreasing $P$. 
    We also calculate the in-sprial timescale $t_{\rm in}$.
    The stellar and planetary periods are taken as the median values for the `Early' model population in the standard case.
    As shown in the bottom panel, $t_{\rm in}$ become longer but does not differ significantly compared to typical age uncertainties (i.e., 0.2-2 Gyrs when t $\sim 0.5-10$ Gyr).
    This is not unexpected since $t_{\rm in} \propto P^{6.5}$ , while the variation in $Q^{'}_{*}$ primarily occurs at shorter periods (corresponding to $t_{\rm in} \lesssim 1$ Gyr) with a weaker correlation, scaling as $P^{-3.1}$. 

    To further quantify the $Q^{*}_{'}-P$ dependence on our inference (i.e., the multiple origins of hot Jupiters) and the fitted $f_{\rm Late}$, we make numerical simulations in the standard case by adopting $Q^{'}_{*}$ as Equation \ref{eqQsPorb}.
    As shown in the top panel of Figure \ref{figHJTSofDM_QsPorb}, for the `Early' model, the fraction of hot Jupiters survived (proportional to $F_{\rm HJ}$) decreases quickly at first, then decreases more slowly when $t \gtrsim 4$ Gyr. 
    This is expected since the orbital period $P$ decays with increasing age.
    As star ages, $P$ will decrease to $\lesssim 3$ days and lead to an increase in $Q^{'}_{*}$, weakening tidal dissipation and slowing the decline rate of $F_{\rm HJ}$.  
    For the `Late' model, as shown in the bottom panel, $F_{\rm HJ}$ would first increase and then decrease.

    We then fit the observed age-frequency relation with a hybrid model by adopting $Q^{'}_{*}$ as Equation \ref{eqQsPorb}.
    The initial conditions are set as the standard case.
    Figure \ref{figFHJmodel12standard_QsPorb} shows the best match of $F_{\rm HJ}$ from the hybrid model comparing to the observational results.
    As can be seen, the Hybrid model is preferred comparing to the single `Early' model (i.e., $f_{\rm Late} = 0$) and single `Late' model (i.e., $f_{\rm Late} = 1$) with $\rm \Delta AIC$ of 23.2 and 12.3, corresponding to confidence levels of $\gtrsim 4$-$\sigma$.
    The best-fit of $f_{\rm Late}$ is $0.58^{+0.08}_{-0.07}$, which is a bit larger but statistically consistent with the results ($0.38^{+0.16}_{-0.14}$) when taking a constant $\log Q^{'}_{*} = 5.7^{+0.4}_{-0.3}$ within $1-2 \sigma$ errorbars.
    That is to say, after accounting for the 
    $Q^{'}_{*}-P$ dependence, our inference that hot Jupiters originate from multiple formation pathways with different timescales remains valid.

    It is worth noting that the best fit of the hybrid model obtained using $Q^{'}_{*}$ as \cite{2018AJ....155..165P} does not align very well with observations, as the inflection point appears earlier than 1 Gyr. 
    Moreover, the best fit of Hybrid model obtained using $Q^{'}_{*}$ as \cite{2018AJ....155..165P} exhibits a poorer agreement with observations compared to that obtained using a common $Q^{'}_{*}$ (Extended Data Fig. 6), with a larger AIC scores of -105.6 ($\rm \Delta AIC = 10.9$). 

    A better agreement with observations could potentially be achieved by modifying the coefficients ($A, B ,C$) of $Q^{'}_{*}-P$ (${\rm max}[\frac{A}{P^B_{\rm tide}}, C]$).
    However, this approach would increase the number of free parameters from 2 to 4, and the results would be highly sensitive to the sample orbiting young stars.
    Therefore, due to the limited sample size and (particularly) the lack of young data, we do not pursue this further in the present work.
\end{enumerate}

\subsection*{Comparison with results by adopting other secular chaos models}
The formation timescale of hot Jupiter via the secular chaos mechanism depends on the initial conditions of the planetary system \citep{2011ApJ...735..109W}, which remain unclear. 
To test how sensitive their conclusion is to this choice of secular chaos model and its initial conditions, we also adopted the secular chaos model from Teyssandier et al. (2019) \citep{2019MNRAS.486.2265T} as the `late' model and fitted it to the observational age-frequency trend.

Teyssandier et al. (2019) conducted numerical simulations and derived the distribution of the formation timescales until 2 Gyr (the maximum integration time) for transient eccentric warm Jupiters (see their fig. 8).
They also suggested that 10\% of transient warm Jupiters would form on longer timescales. Here we simply assume that these 10\% form uniformly in logarithmic time between 2 and 10 Gyr.
Due to to weak friction tides, the orbits of transient warm Jupiters will be circularized, leading to the formation of an hot Jupiters.
Based on the semimajor axis and eccentricity, we calculate their tidal circularization timescale. 
The formation timescales of hot Jupiters is the sum of the formation and circularization timescales of transient warm Jupiters.
The initial mass distribution of hot Jupiters is assumed to be the same as that of the innermost warm Jupiters, i.e., uniformly distributed between 0.3-1.5 $M_J$.
The initial orbital periods obey a distribution referring to their fig. 10.

We then perform numerical simulations for the tidal decays of hot Jupiters with various stellar tidal quality factor ($Q^{'}_{*}$).
Figure \ref{figFHJmodel12a2016_T2019} shows the best match of $F_{\rm HJ}$ from the hybrid model comparing to the observational results.
The Hybrid model is preferred comparing to the single `Early' model and single `Late' model with confidence levels of $\gtrsim 2$-$\sigma$, respectively.
The best-fits for the $f_{\rm Late}$ and $Q^{'}_*$ are $45^{+18}_{-16}\%$ and $5.6^{+0.2}_{-0.3}$, which are nearly identical to those when adopting the secular model from Hamers et al. (2017) \citep{2017MNRAS.464..688H} though with a larger AIC score of -111.9.

The above analyses suggest that our constraints on the origin and tidal evolution of hot Jupiters are not sensitive to the choice of secular chaos model.
}

{\subsection*{Influence of age uncertainty}
In this work, we have adopted the kinematic method, which could not directly provide the ages and their uncertainties of individual stars.
However, the age uncertainty may distort the frequency-age relation.
For example, if the age uncertainties are larger, the observed frequency-age trend could become flatter regardless of the underlying true pattern.

To evaluate whether the `early' model alone, after considering the effect of age uncertainties, can reproduce the observed broken age-frequency relation, we adopt the `early' model and simulate the true ages for stars with and without Jupiters by taking a $\log Q^{'}_{*} = 5.5$, which yields a declining slope consistent with observational trends beyond approximately 2 Gyr.
For each simulated star, we perturb its true age by drawing from a Gaussian distribution with relative uncertainties of 50\%, 100\%, 200\%, 300\%, 400\%, and 500\%, respectively. Using these perturbed ages, we recalculate the frequency–age relation.
This entire procedure is repeated 1,000 times to assess the statistical robustness of the results.

Figure \ref{figFHJAgeuncertainty} shows the median frequency-age relations using these perturbed ages with different uncertainties.
As expected, accounting for age uncertainties tends to flatten the declining trend. 
However, under the same level of uncertainty, this flattening effect caused by age uncertainties is similar for both young and old stars, and thus cannot produce a broken pattern as observed.
Moreover, this effect exhibits a saturation behaviour: when the uncertainty exceeds 300\%, the trends under different uncertainty levels tend to become indistinguishable.

A broken pattern may also be reproduced if the relative uncertainties are larger at younger ages but smaller at older ages, as shown in Figure \ref{figAgeuncertaintycomparion}. However, even with a 500\% uncertainty for stars younger than 2 Gyr and 0\% uncertainty for stars older than 2 Gyr, the degree of the observed broken pattern (first-flatten-then-decline) is still not achieved. 
%If the broken declining pattern is attributed to the effect of age uncertainties, a possible scenario is that the relative uncertainties are larger at younger ages. 
%However, as shown in Figure \ref{figFHJAgeuncertainty}, even with a 500\% uncertainty, the decline in occurrence rate with age remains steeper than that observed. 
To evaluate whether this effect can produce a broken trend as pronounced as the one observed, we consider an extreme case: the age uncertainty is set to 0\% after 2 Gyr, while being significantly larger before 2 Gyr. In this case, the declining slope beyond 2 Gyr matches the observed trend.
When the age uncertainty before 2 Gyr is set to 50\%, none of the 1,000 realizations produce a frequency–age trend that is flatter than the observed one (i.e., fitted $\gamma$ greater than or equal to the observed value). 
With a 100\% uncertainty, only 1 instance out of 1,000 realizations yields a flatter trend. 
Even when the uncertainty is increased to 500\%, only 59 instances out of 1,000 match the observed flattening.

%It is worthing noting that the above cases go beyond the maximum impact that age uncertainties alone can plausibly account for, since uncertainties as large as 100\%-500\% are unrealistic at young ages and unlikely to be zero at old ages. 
Therefore, the above analyses demonstrate the age uncertainties, even under such exaggerated conditions, are insufficient to reproduce a flatten-then-decline trend as strong as the one observed. 
This strongly suggests that other physical mechanisms must be contributing to the observed frequency–age relation.
}

\newpage

\begin{figure}[!t]
\centering
\includegraphics[width=\textwidth]{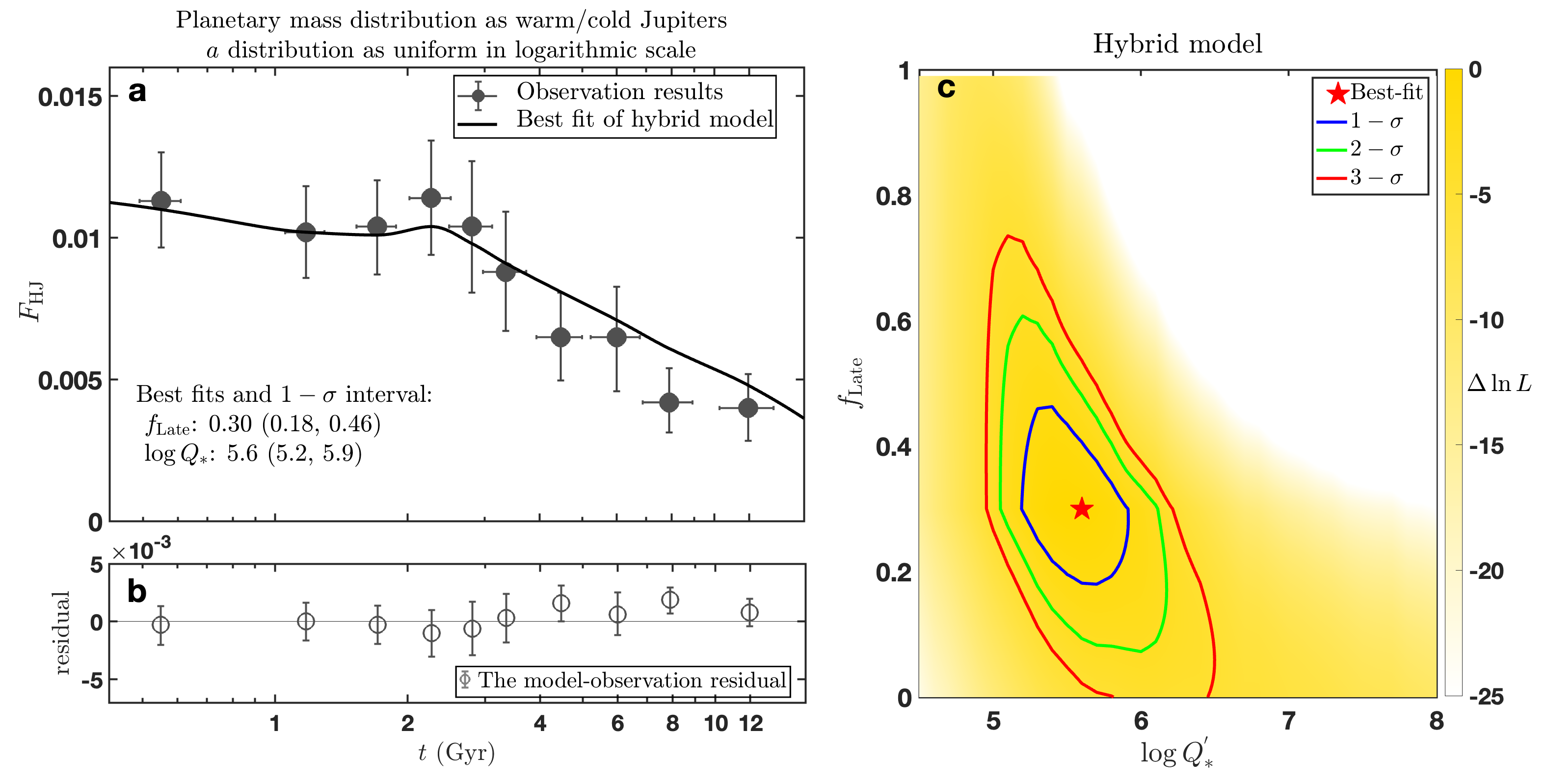}
\caption{{Similar to Extended Data Fig. 6 but the initial $a-$distribution of hot Jupiters in simulations is set as uniform in logarithmic scale.}
\label{figFHJmodel12auniformMCJ}.}
\end{figure}

\begin{figure}[!t]
\centering
\includegraphics[width=\textwidth]{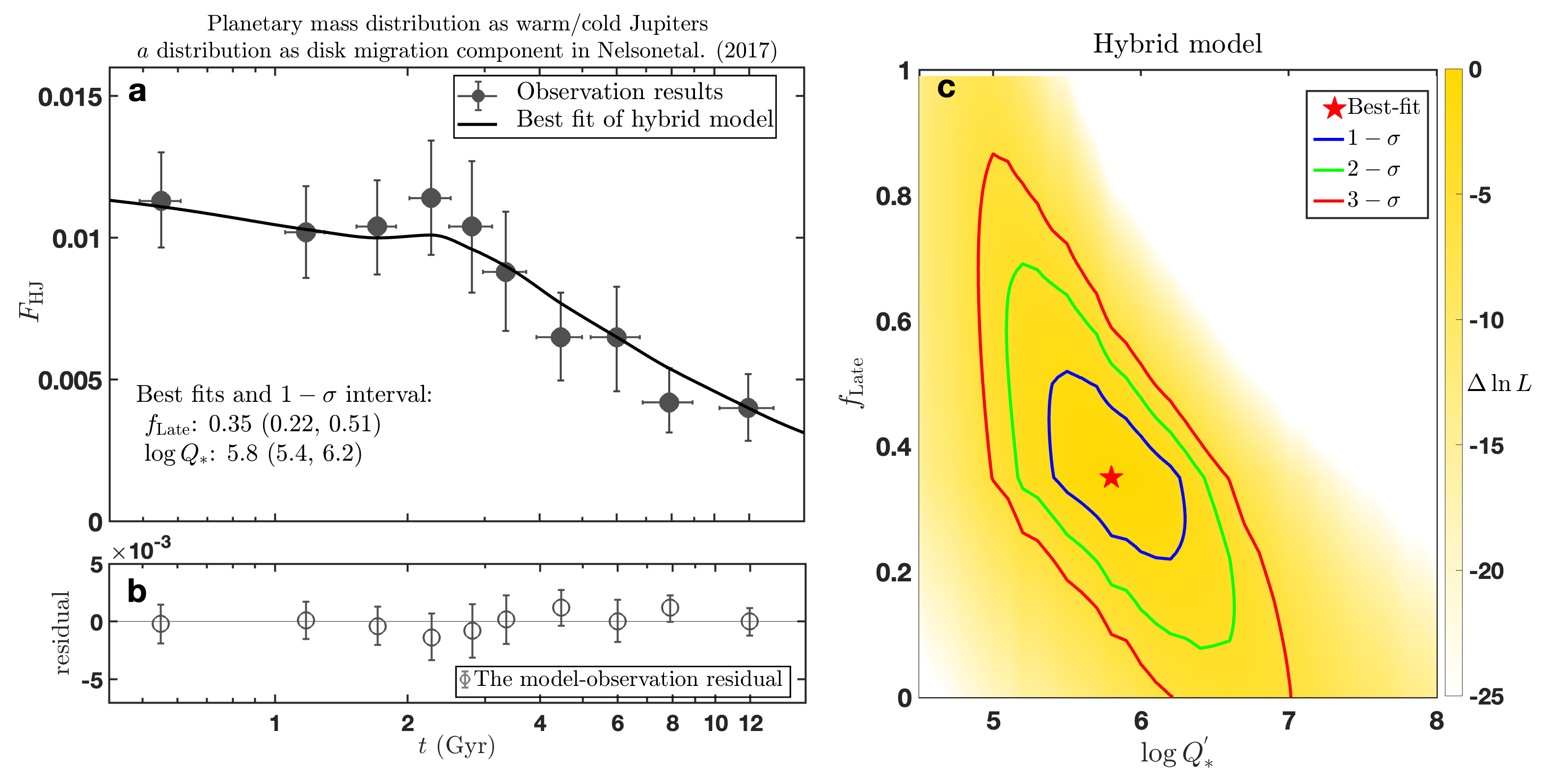}
\caption{{Similar to Extended Data Fig. 6 but the initial planetary mass in simulations is set as the initial $a-$distribution of hot Jupiters in simulations is set as the disk migration
component in Nelson et al. (2017).}
\label{figFHJmodel12aNelsondiskMCJ}}
\end{figure}

\begin{figure}[!t]
\centering
\includegraphics[width=\textwidth]{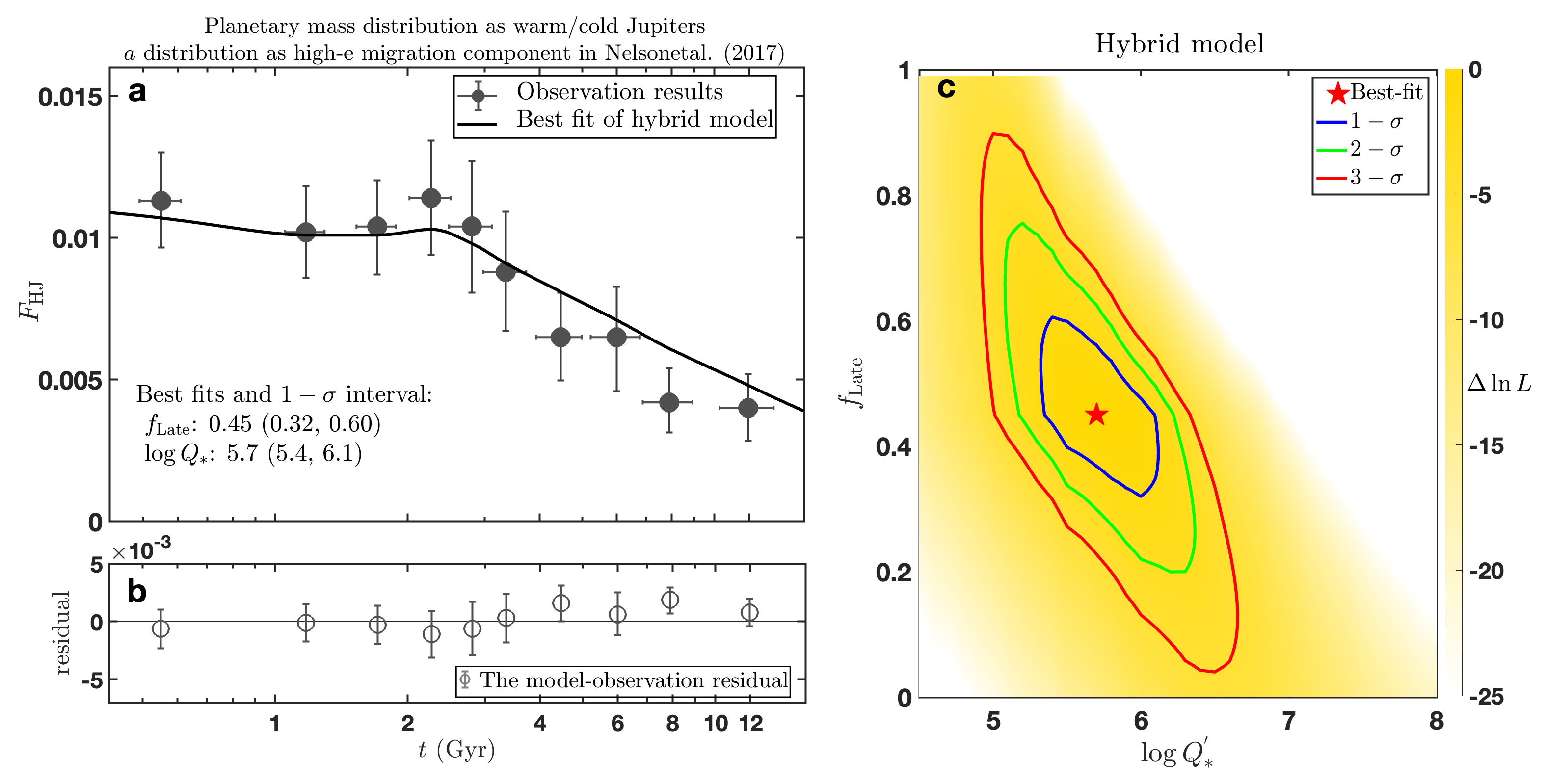}
\caption{{Similar to Extended Data Fig. 6 but the initial planetary mass in simulations is set as the initial $a-$distribution of hot Jupiters in simulations is set as the high-eccentricity migration
component in Nelson et al. (2017).}
\label{figFHJmodel12aNelsonhigheMCJ}}
\end{figure}

\begin{figure*}[!t]
\centering
\includegraphics[width=0.5\textwidth]{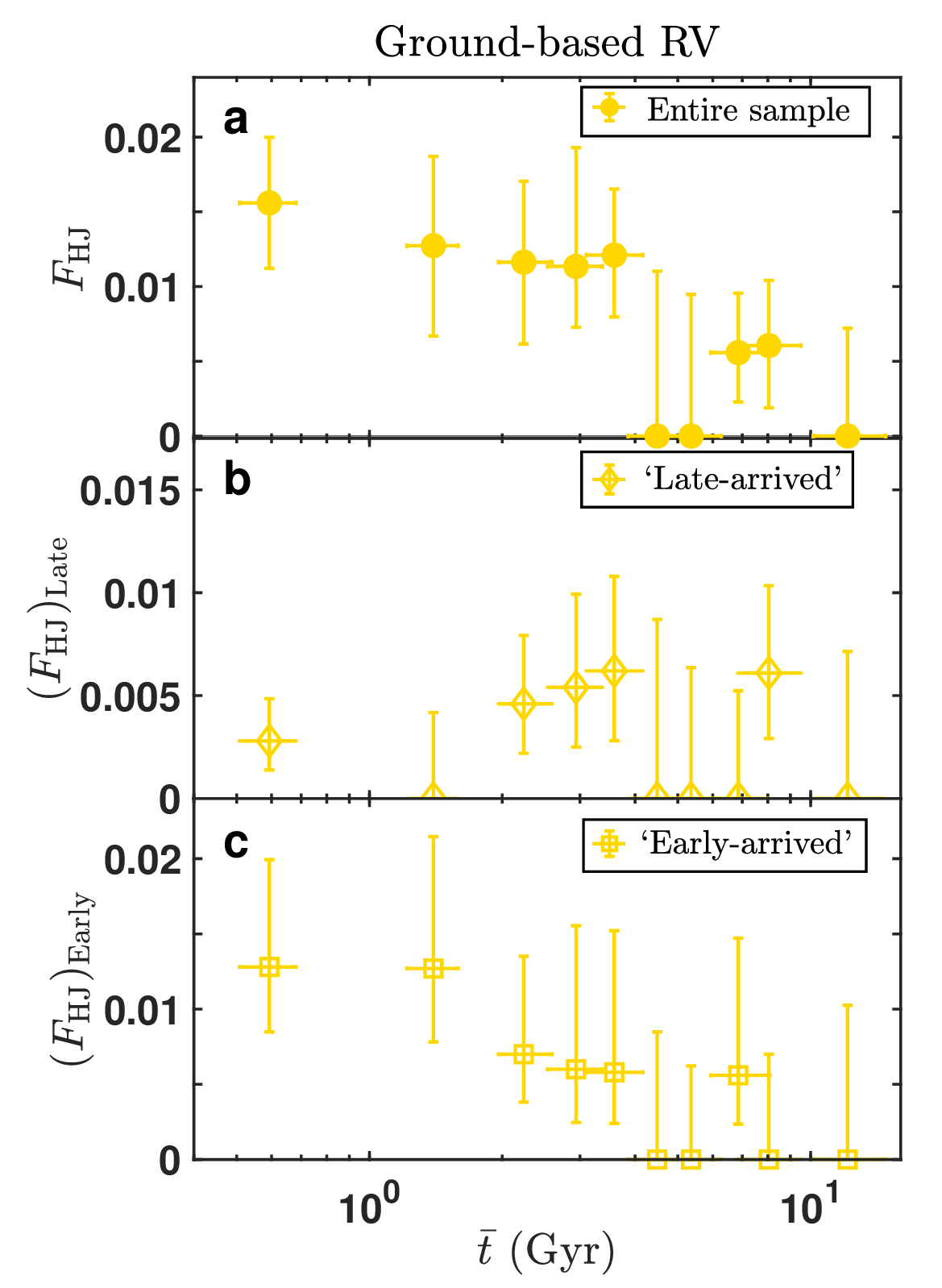}
\caption{{Similar to Figure 1 in the main text but here we only consider the Ground-based RV subsample.}
\label{figfHJAgeRV}}
\end{figure*}

\begin{figure*}[!t]
\centering
\includegraphics[width=\textwidth]{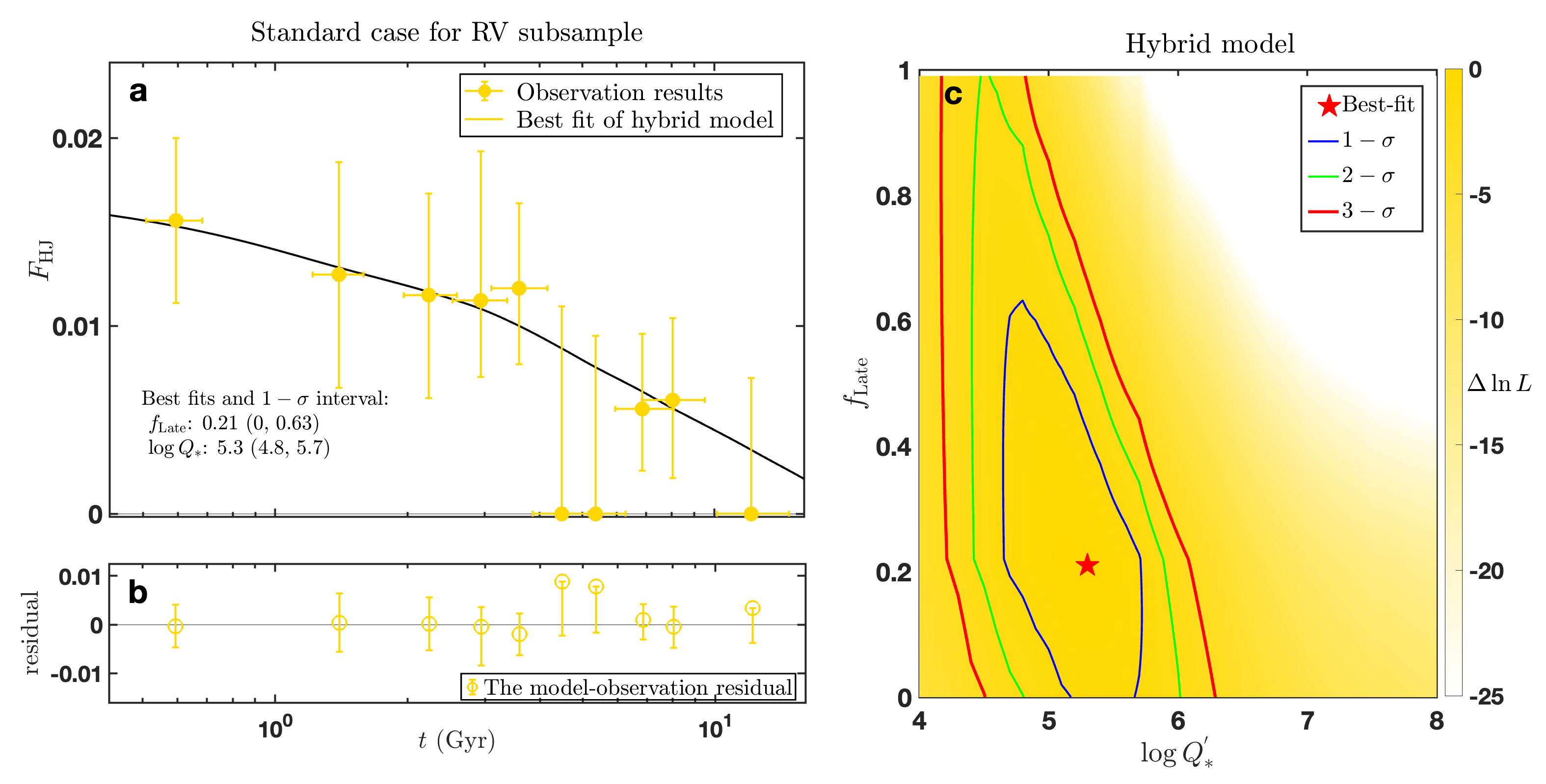}
\caption{{Similar to Extended Data Fig. 6 but here we only consider the Ground-based Radial velocity (RV) subsample.}
\label{figFHJmodel12a2016MCJRV}}
\end{figure*}

\begin{figure*}[!t]
\centering
\includegraphics[width=0.5\textwidth]{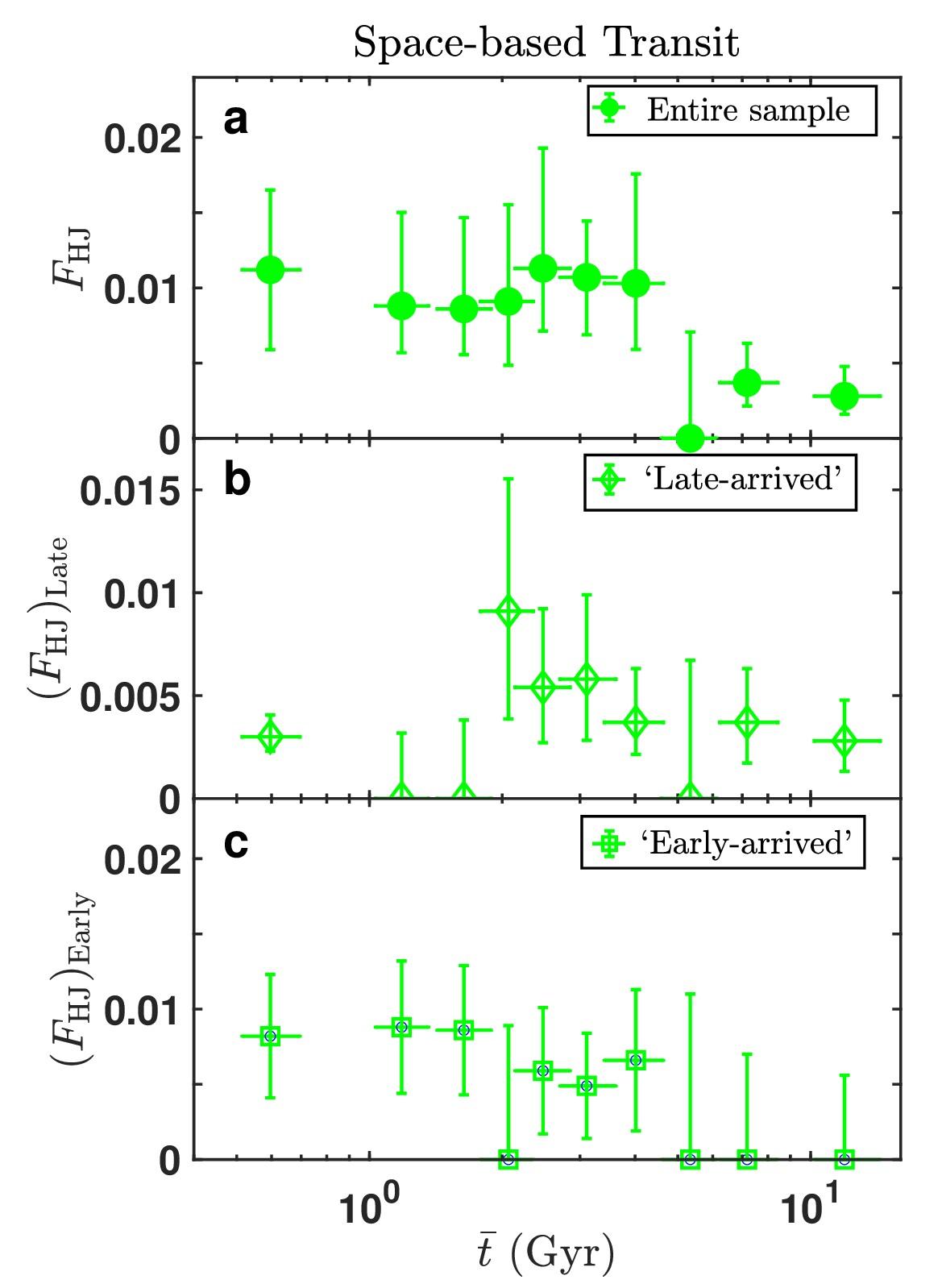}
\caption{{Similar to Figure 1 in the main text but here we only consider the Space-based Transit subsample.}}
\label{figfHJAgeST}
\end{figure*}

\begin{figure*}[!t]
\centering
\includegraphics[width=\textwidth]{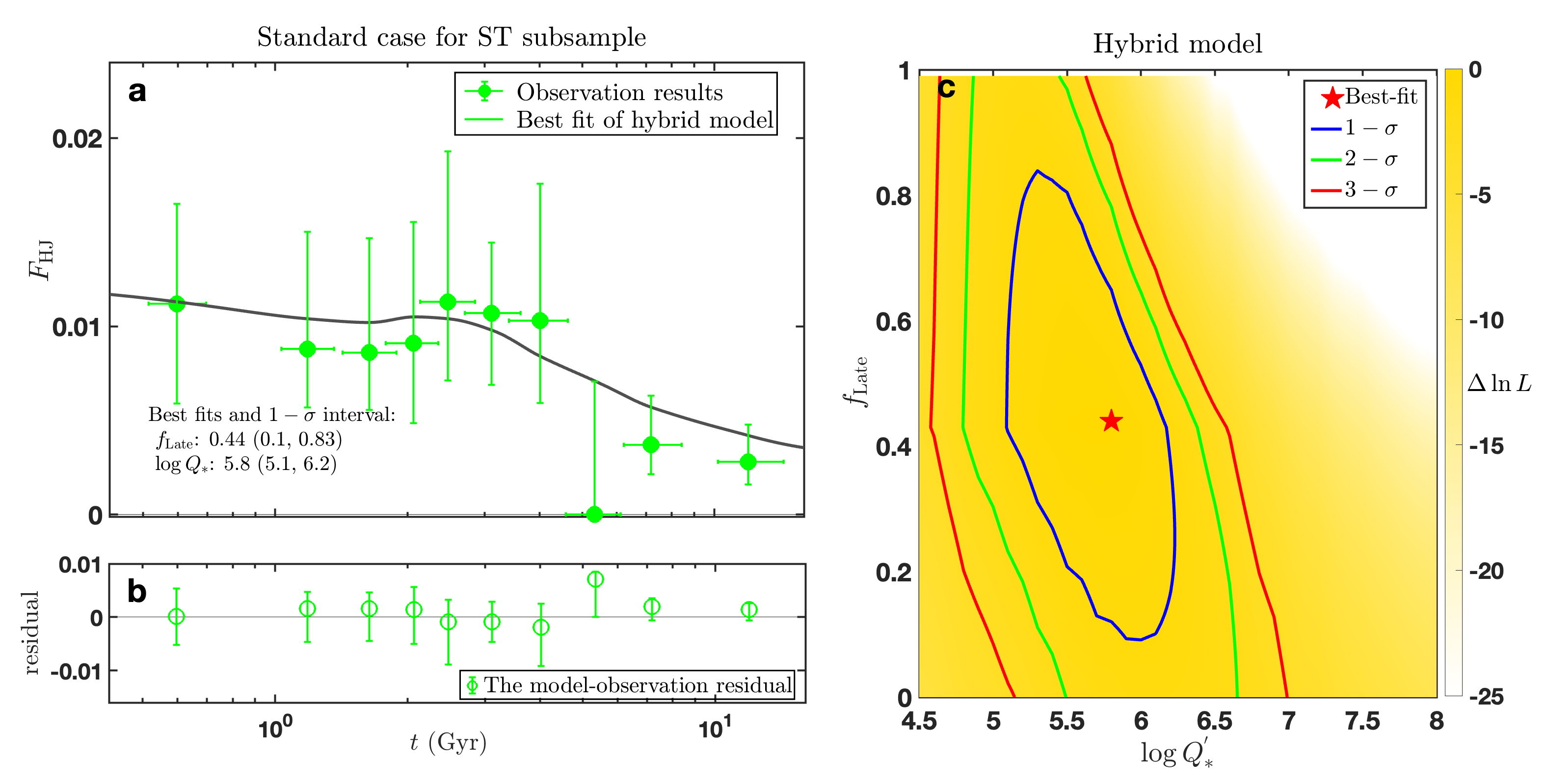}
\caption{{Similar to Extended Data Fig. 6 but here we only consider the Space-based Transit (ST) subsample.}
\label{figFHJmodel12a2016MCJST}}
\end{figure*}

\begin{figure*}[!t]
\centering
\includegraphics[width=0.5\textwidth]{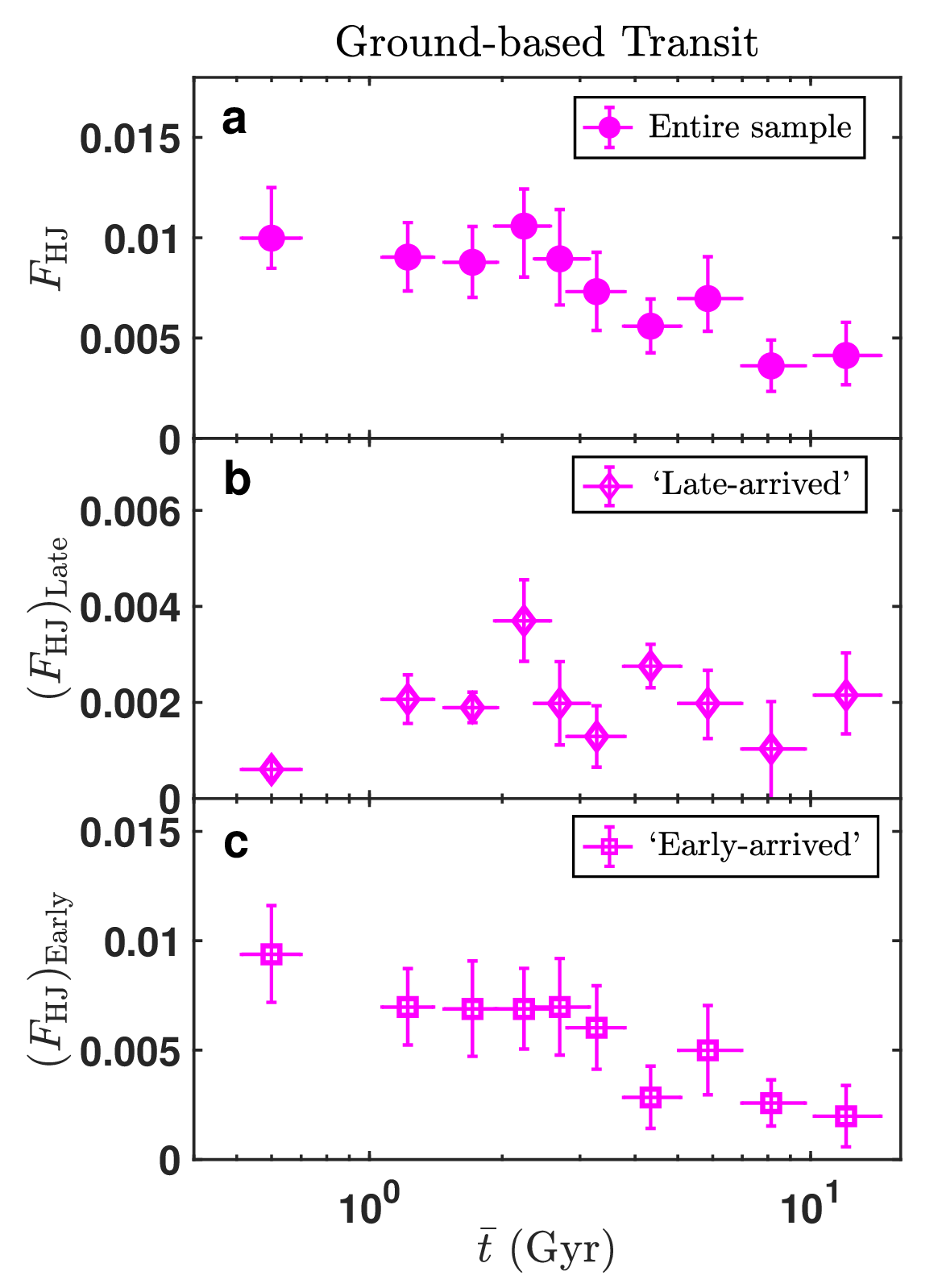}
\caption{{Similar to Figure 1 in the main text but here we only consider the Ground-based Transit subsample.}}
\label{figfHJAgeGT}
\end{figure*}

\begin{figure*}[!t]
\centering
\includegraphics[width=\textwidth]{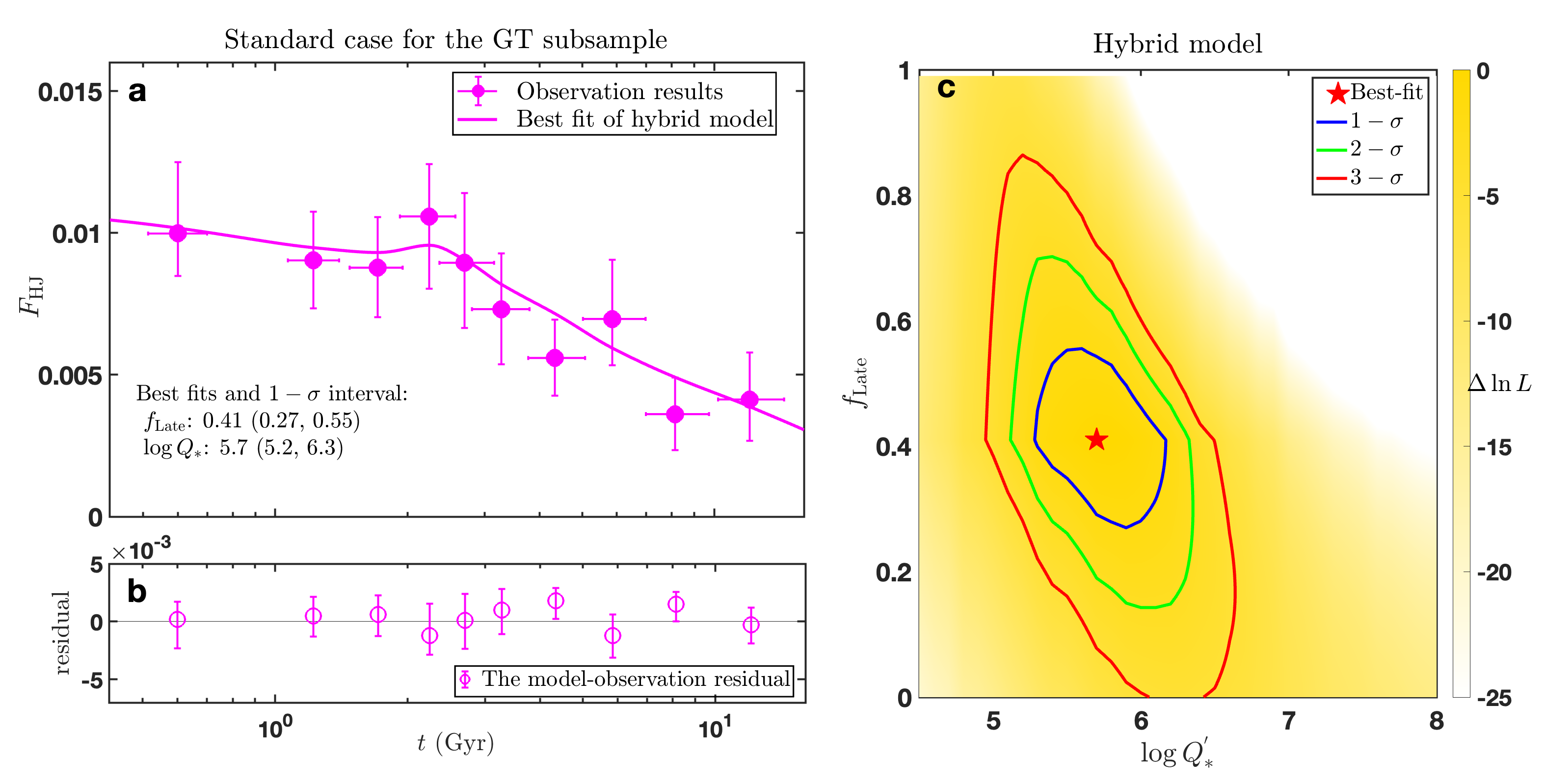}
\caption{Similar to Extended Data Fig. 6 but here we only consider the Ground-based Transit (GT) subsample.
\label{figFHJmodel12a2016MCJGT}}
\end{figure*}

\begin{figure*}[!t]
\centering
\includegraphics[width=\textwidth]{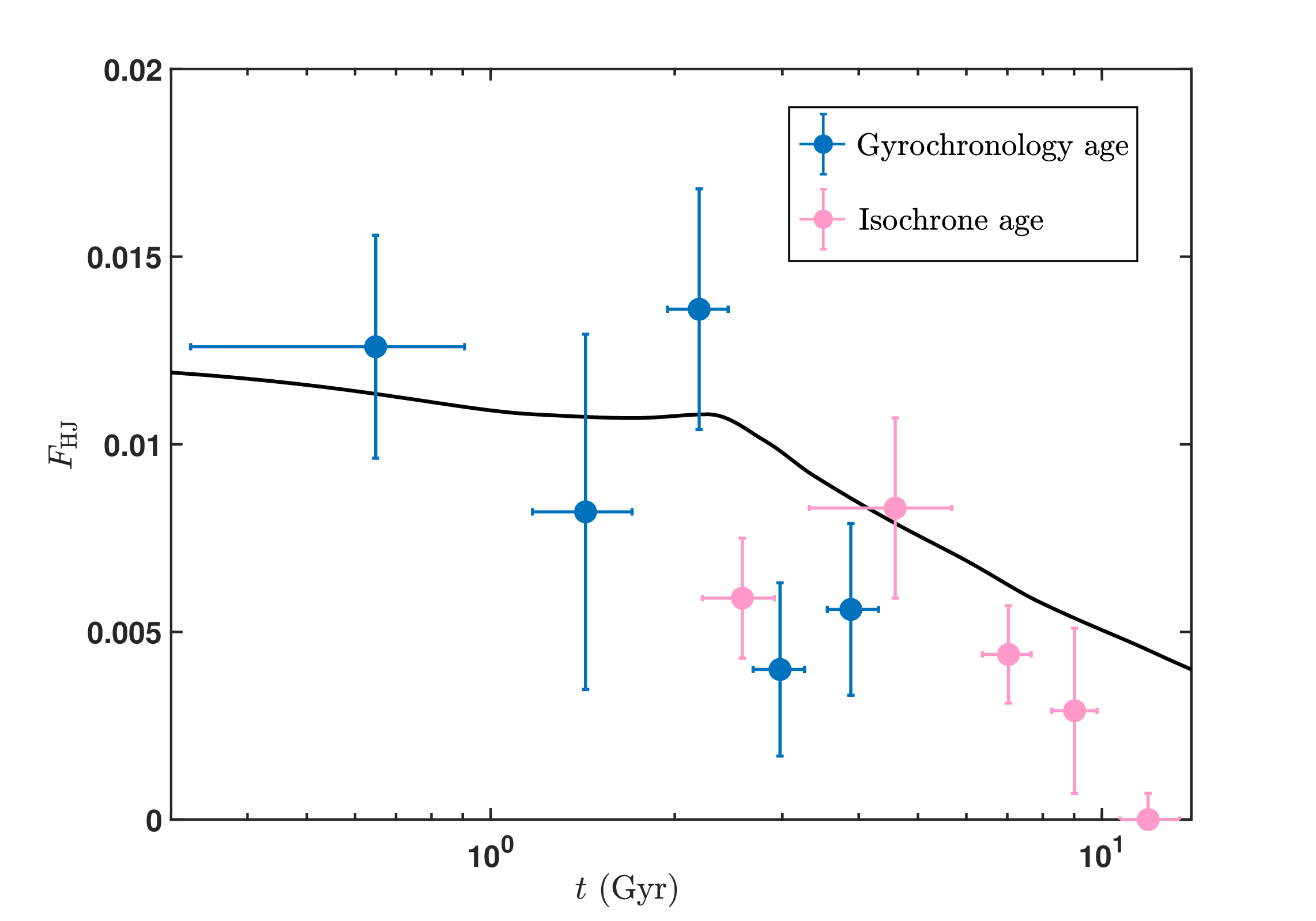}
\caption{ Frequency of hot Jupiters $F_{\rm HJ}$ as a function of Gyrochronology age (dark blue) and Isochrone age (pink) after renormalizing to Solar metallicity.
The solid line denotes the best-fit of the hybrid model derived from the kinematic results.
\label{figFHJAge_Rotation_Isochrone}}
\end{figure*}

\begin{figure*}
\centering
\includegraphics[width=\textwidth]{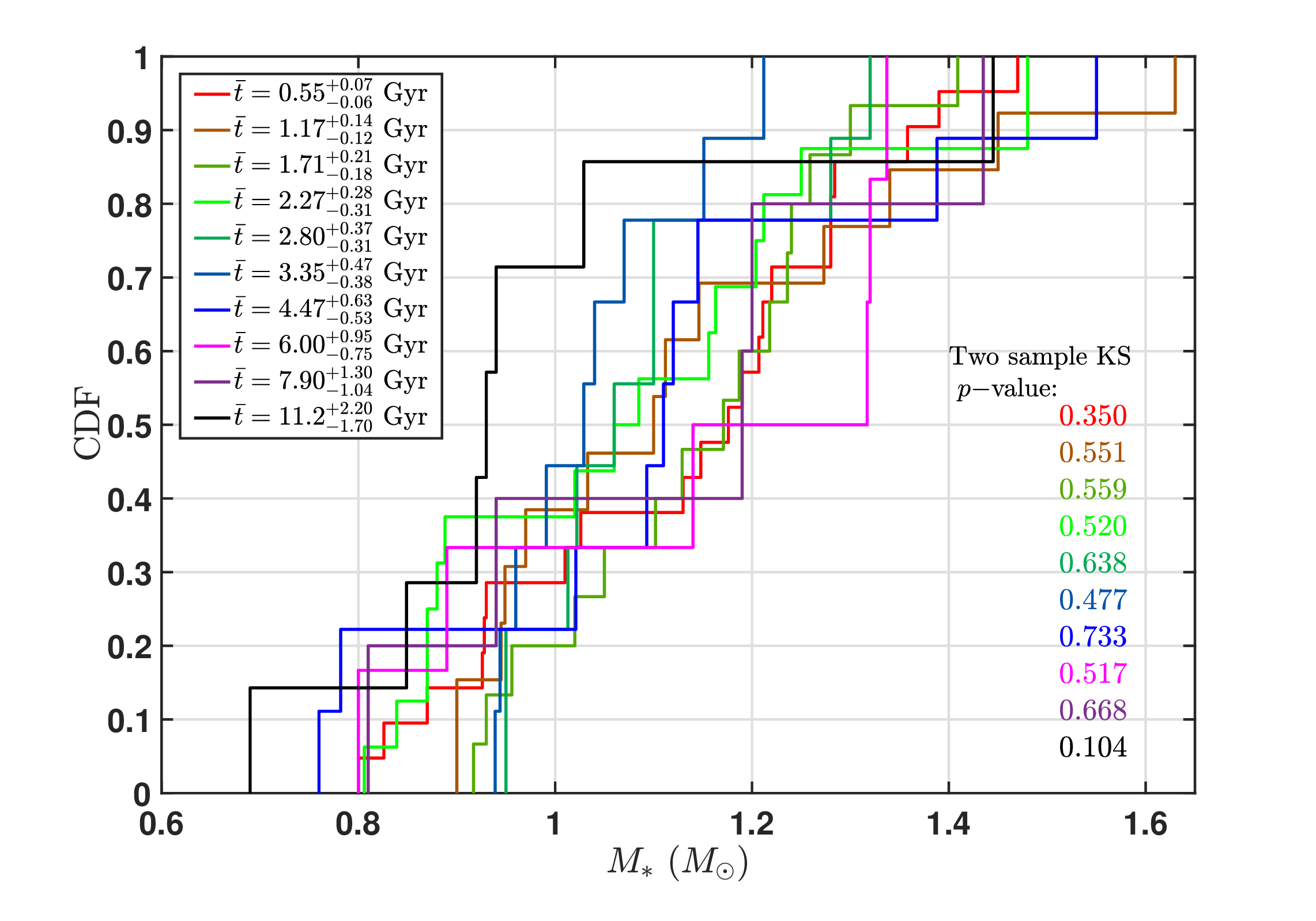}
\caption{ The cumulative distributions of the stellar masses for hot Jupiter host stars of different kinematic ages. 
In the bottom-right corner, we
print the two sample KS test $p-$values for the distributions of each bin compared to the total sample.}  
\label{figMass_Age_HJhosts}
\end{figure*}

\begin{figure*}
\centering
\includegraphics[width=\textwidth]{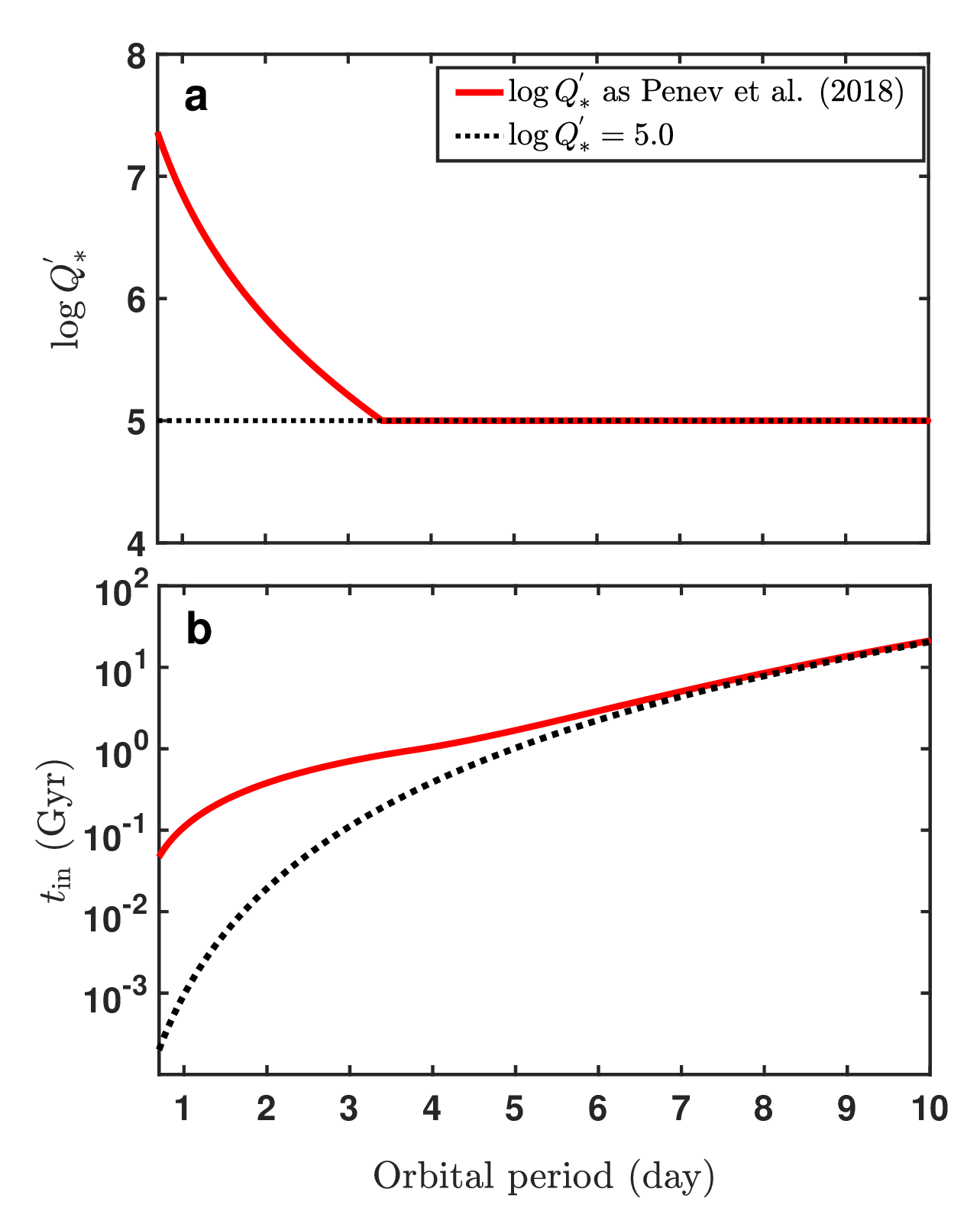}
\caption{ Top: The stellar tidal factor $Q^{'}_{*}$ (Top) as a function of orbital period ($P$) referring to \cite{2018AJ....155..165P}. 
Bottom: The in-sprial timescales $t_{\rm in}$ as a function of $P$ by considering the above $Q^{'}_{*}-P$ dependence.
The stellar and planetary periods are taken as the median values for the `Early' model population in the standard case.
For comparison, we also plot the result when taking a constant $\log Q^{'}_{*} = 5.0$.}
\label{figQs_tin_P_Penev2018}
\end{figure*}

\begin{figure}[!t]
\centering
\includegraphics[width=\textwidth]{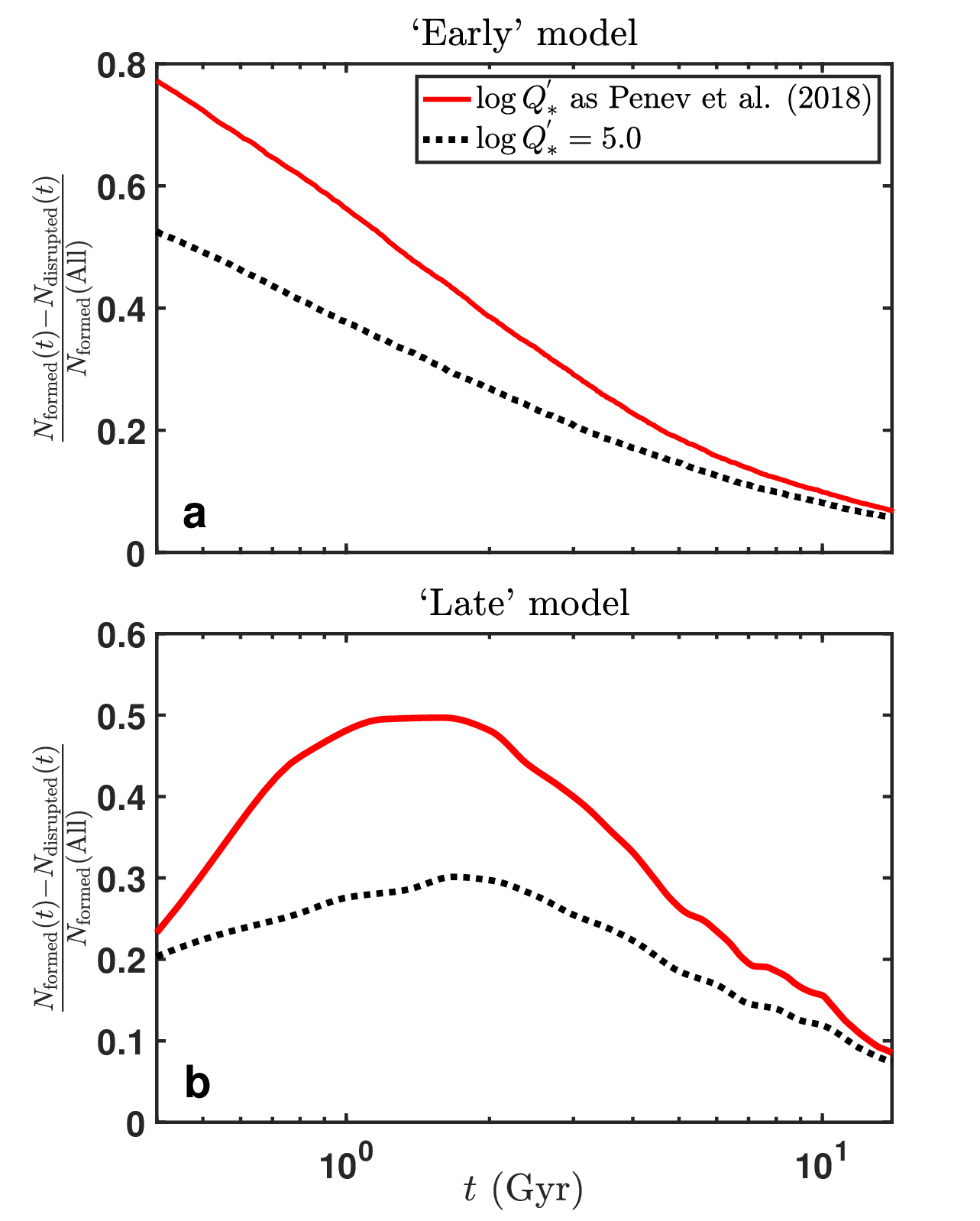}
\caption{ The number ratio of the left (formed $-$ tidally disrupted)  hot Jupiters before $t$ over the formed hot Jupiters of all times as a function of age for the `Early' model (Top) and `Late' model (Bottom), respectively.
The initial conditions are set as the standard case, i.e., the a-distribution of hot Jupiters is set as the results inferred from Kepler data \citep{2016A&A...587A..64S} and the initial planetary mass is set as that of cold Jupiters. 
The modified stellar tidal quality factor $Q^{'}_{*}$ is set as Equation \ref{eqQsPorb}.
For comparison, we also plot the result when taking a constant $\log Q^{'}_{*} = 5.0$.
\label{figHJTSofDM_QsPorb}}
\end{figure}

\begin{figure}[!t]
\centering
\includegraphics[width=\textwidth]{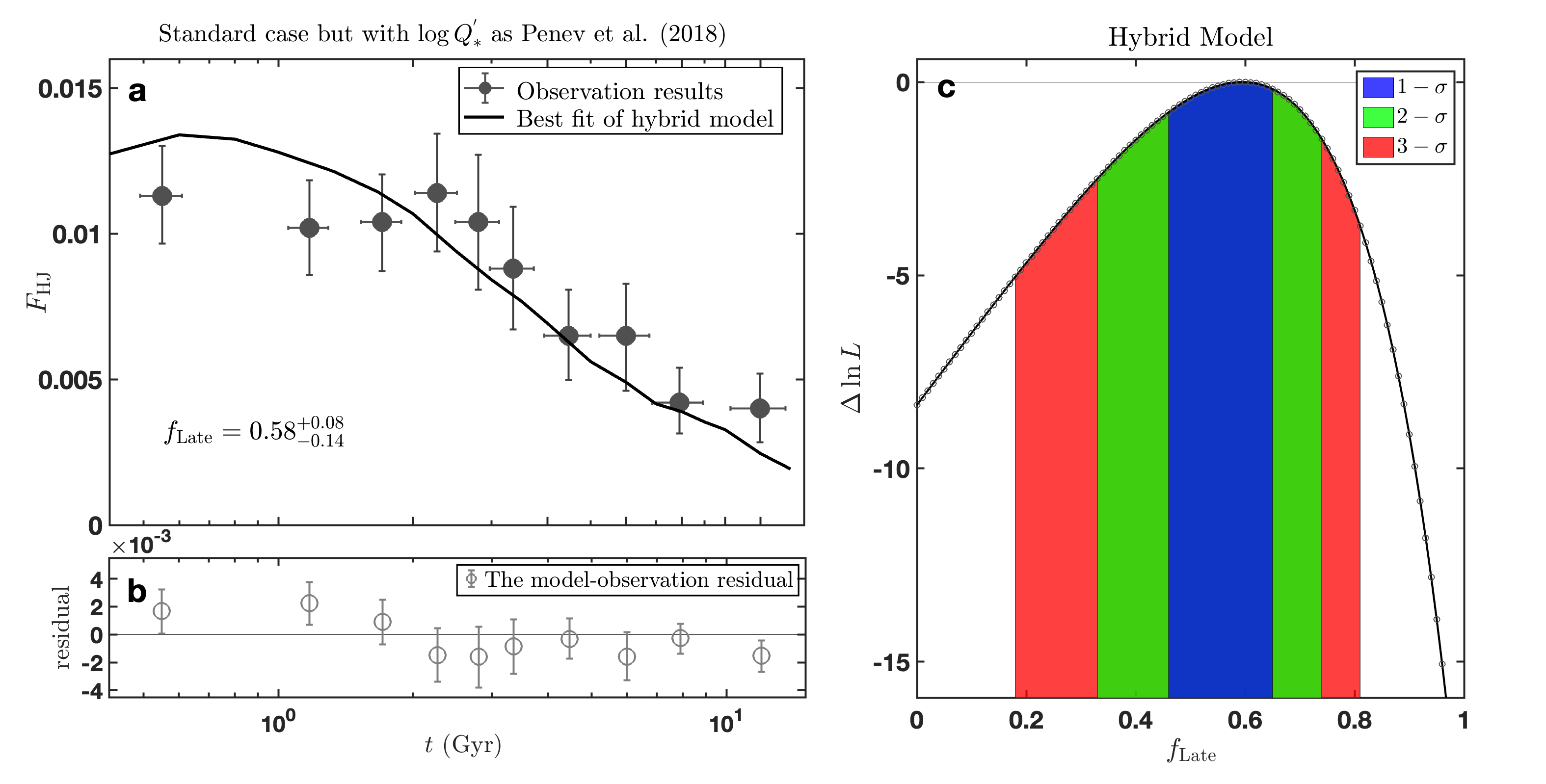}
\caption{ Similar to Extended Data Fig. 6 but the $Q^{'}_{*}$ is set as the results derived in \cite{2018AJ....155..165P}, which varies with orbital period (Eq. \ref{eqQsPorb}).
The right panel displays the relative likelihood in logarithm as a function of $f_{\rm Late}$.
\label{figFHJmodel12standard_QsPorb}}
\end{figure}

\begin{figure}[!t]
\centering
\includegraphics[width=\textwidth]{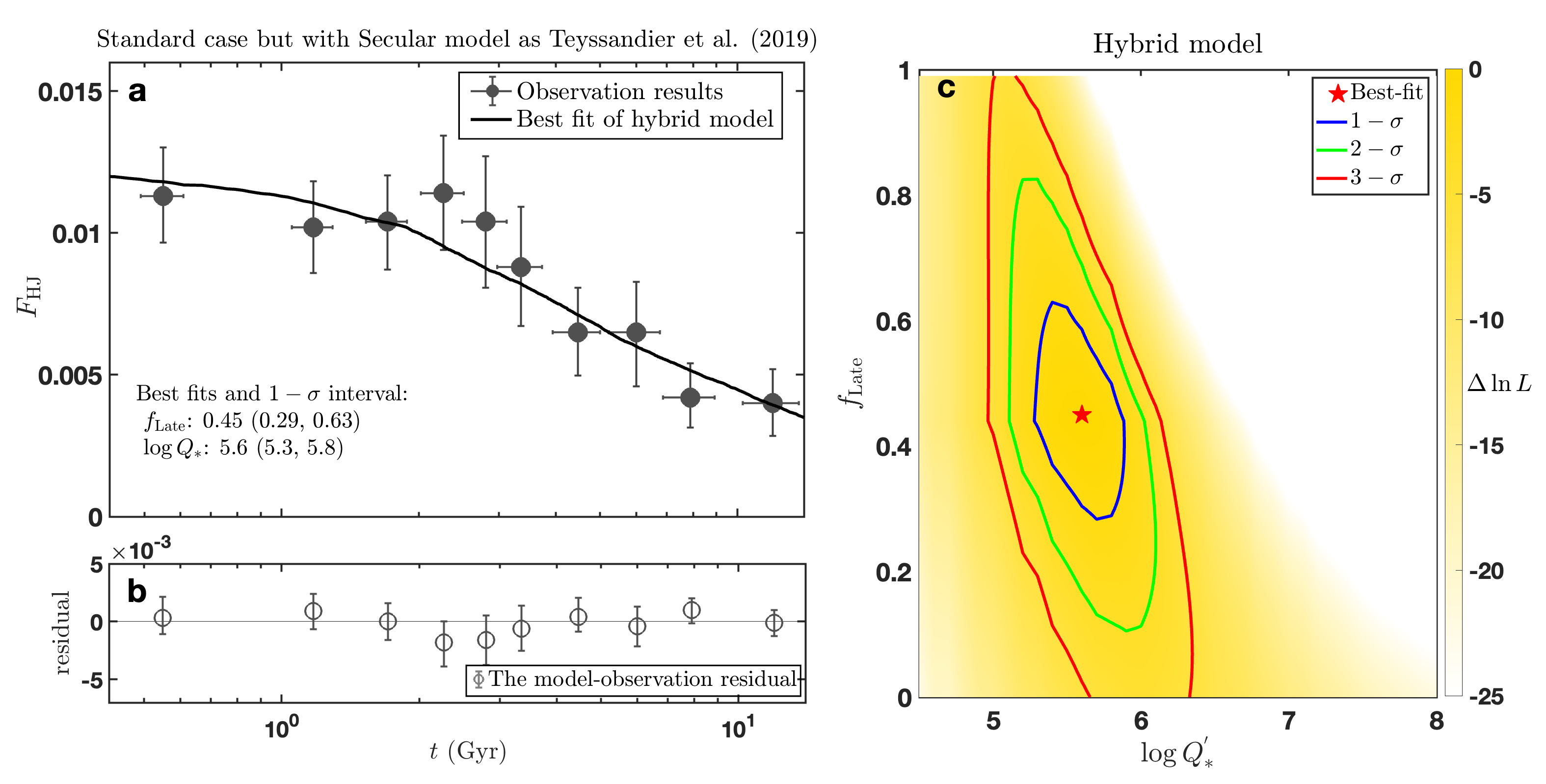}
\caption{{ Similar to Extended Data Fig. 6 but the `late' model is adopt as the secular chaos model from \citep{2019MNRAS.486.2265T}.}
\label{figFHJmodel12a2016_T2019}}
\end{figure}

\begin{figure}[!t]
\centering
\includegraphics[width=0.95\textwidth]{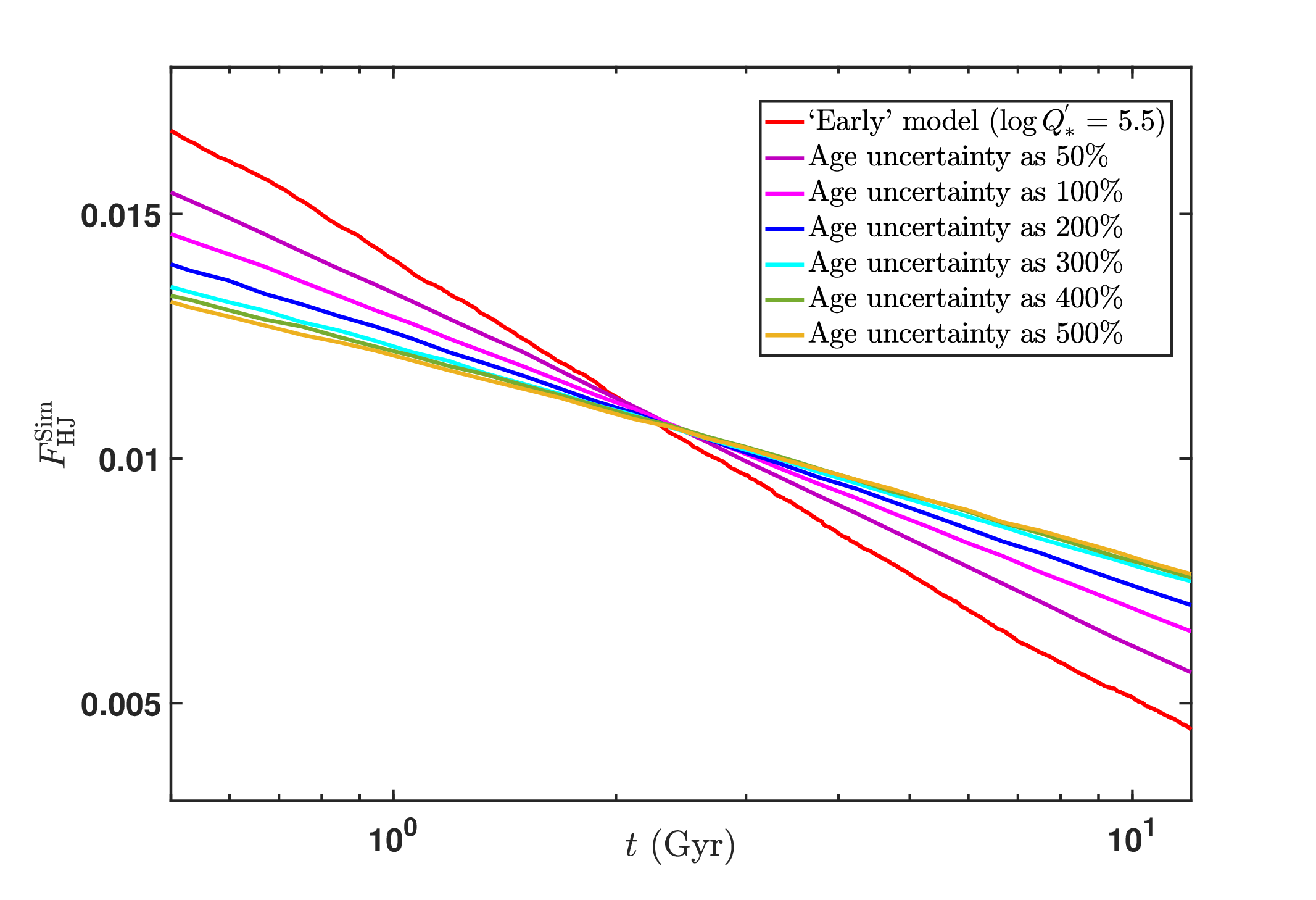}
\caption{The median frequency-age relations using these perturbed ages under different level of age uncertainties (in different color) for the ‘Early’ model.
\label{figFHJAgeuncertainty}}
\end{figure}

\begin{figure}[!t]
\centering
\includegraphics[width=0.95\textwidth]{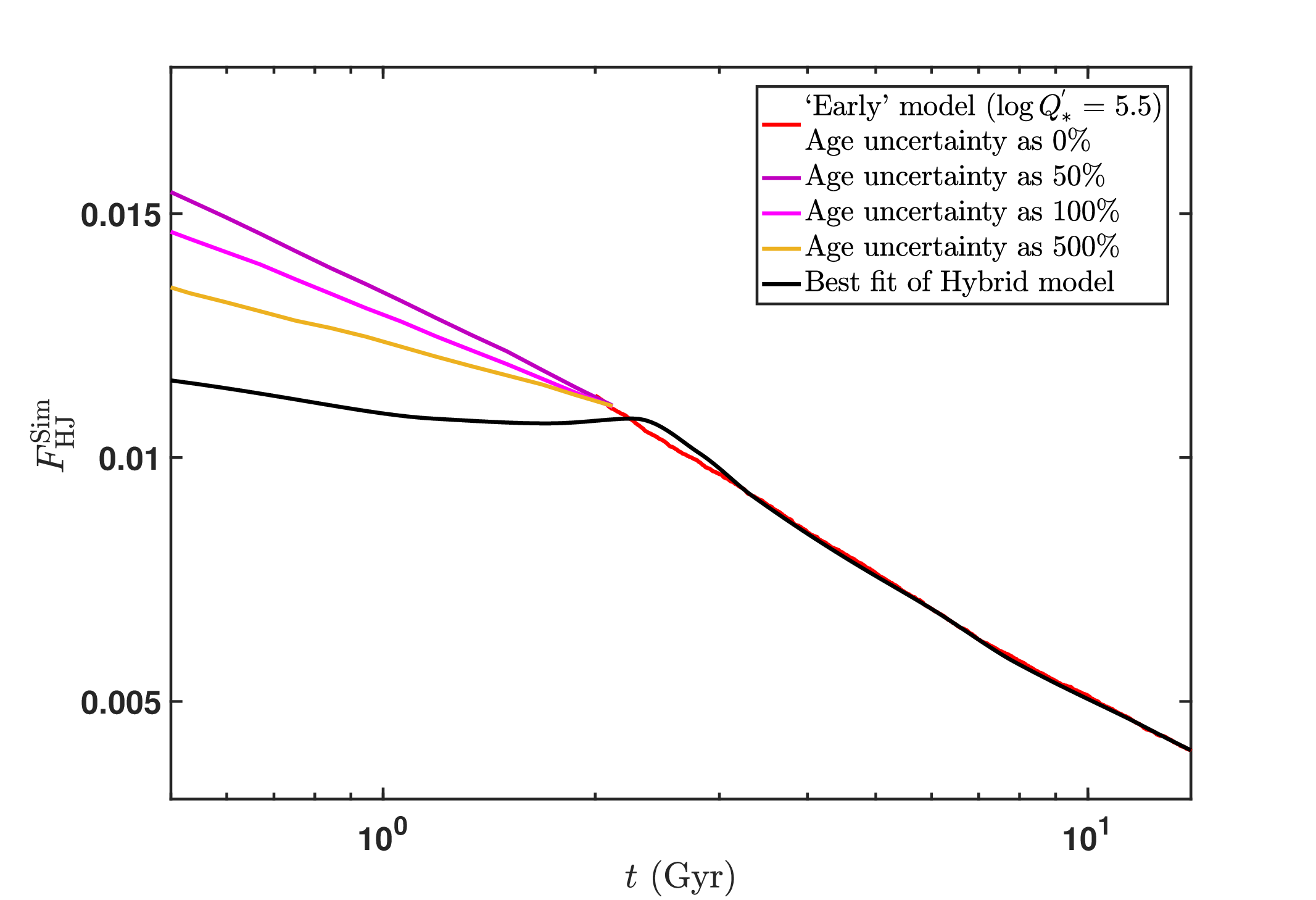}
\caption{An extreme case for the median frequency-age relations using these perturbed ages for the ‘Early’ model.
The age uncertainty is set to 0\% after 2 Gyr, while being significantly larger before 2 Gyr.
\label{figAgeuncertaintycomparion}}
\end{figure}

\clearpage
\bibliography{sn-bibliography.bib}

\end{document}